# Discrete intrinsic localized modes in a microelectromechanical resonator

Authors: Adarsh Ganesan, Cuong Do and Ashwin A. Seshia

**Intrinsic Localized Modes (ILMs) or Discrete Breathers (DBs) are produced through a non-linear vibration localization phenomenon. While Anderson localization is due to lattice defects, the nonlinearity of lattices provides the basis for ILM excitation. Over the past two decades, these ILMs have been realized in a wide range of physical systems including photonic crystals, nonlinear atomic lattices, anti-ferromagnets, coupled Josephson junction arrays and coupled cantilevers. This paper brings out the feasibility of exciting ILMs in a standalone mechanical resonator. Through piezoelectric driving and optical visualization, various intriguing features of ILMs have been recorded. The ILMs in our system are observed as spectral bushes and their frequencies are much lower than that of the drive frequency. The excitation of ILMs is mediated through large amplitude instability of a mechanical mode. The spatial prevalence of discrete ILM excitations is at antinodes of that mode. Further, the ILMs have been observed to be time-variant and various events including attraction-repulsion (or splitting-merging) of ILMs and hopping occur during the time evolution of ILMs.**

Intrinsic localized modes (ILM) or Discrete Breathers (DBs) have been a subject of interest for the last two decades ever since the pioneering generalization by Takeno and co-workers [1-3]. Such nonlinear vibration energy localization phenomenon has had much impact in condensed matter physics [4-6] and led to the understanding and control of physical processes [7-10] involving high density vibration energy. The key features concerning ILMs are nonlinearity [1-3] and discreteness [11-12]. Also, these ILMs are not externally imposed as in the case of Anderson localization [13-14] rather due to intrinsic nonlinearity. More recently, the presence of ILMs in a wide range of physical systems including photonic crystals [15], nonlinear atomic lattices [16], anti-ferromagnets [17] and coupled Josephson junction arrays [18] has been demonstrated. Particularly, in the mechanical domain, the ILMs have been experimentally realized through the elastic coupling of cantilevers [19].

This paper brings out the feasibility of exciting travelling ILMs in a standalone microelectromechanical resonator. Using the combination of Laser Doppler Vibrometry (LDV) and piezoelectric driving, we observed the two-dimensional transport of ILMs on mechanical structure and many interesting features of ILMs are recorded. This seemingly simple device (and other similar devices) can therefore form an experimental testbed for the future experimental investigations of ILMs. Such devices can be fabricated using conventional semiconductor micromachining techniques.

Also, the complementary visualization of ILM transport within a single resonator element at microscopic resolution can lead to intriguing and surprising concepts concerning the behaviour of ILMs.

**Results and Discussion**

**Experimental Method**

For the experimental realization of ILMs, a 1-D extensional-compressional mechanical mode (Figure S1) (eigen-frequency $3.86\ MHz$) of free-free beam structure of dimensions $1100\ \mu m \times 350\ \mu m \times 11\ \mu m$ (Figure 1A) is considered. A $0.5\ \mu m$ thick AlN layer is harnessed for the piezoelectric excitation of the structure (Figure S1). An electrical signal derived from the Agilent Waveform Generator 335ARB1U is fed through one of the split electrodes for driving the mechanical mode. Through optical visualization using the Polytec Laser Doppler Vibrometer MSA-400, an insightful assessment of ILM behaviour was carried out. Due to the capture area limitation with our vibrometry setup, only one half of the device is imaged. The displacements averaged over the capture area are automatically computed using the 'Scan' option available in our LDV measurement setup. The measurements are carried out at a spatial resolution of $35\ \mu m$ and $125\ Hz$ spectral resolution which leads to a reasonable temporal resolution of 1 minute. Higher temporal resolution can be reached by compromising the spatial or spectral resolution to probe fast moving ILMs.

**Observation of Intrinsic Localized Modes**

At the drive power level $-10\ dBm$, the frequency spectrum consists of just the drive tone. In contrast, at $10\ dBm$, the low frequency ILMs and a sub-harmonic tone is also observed along with the drive tone (Figure 1B). This sub-harmonic tone results from an auto-parametric excitation through the intrinsic coupling between the driven mode and sub-harmonic mode. Such parametric resonances have been previously encountered in a number of experiments [20-24]. The displacement profiles corresponding to the drive tone $\omega_d$ and the parametrically excited sub-harmonic tone $\omega_d/2$ are shown in figure 1C. The elevated amplitudes associated with the parametric amplification process has now been shown to result in the excitation of discrete ILMs. Figure 1D shows the presence of spectral bushes around $35 - 75\ kHz$ and these correspond to the ILMs. The RMS displacement profile (Figure 1D) corresponding to these frequencies validates the spatial discreteness of ILMs.

It is now interesting to note that the spatial average frequency spectrum (Figure 2A) at the drive power level presents an ensemble of three sub-processes: auto-parametric excitation of sub-harmonic mode (Figure 2a1); excitation of low frequency ILMs (Figures 2a2 and 2a3) and the individual inherent coupling of ILMs with sub-harmonic (Figure 2a4) and driven modes (Figure 2a5). The coupling results in the splitting of sub-harmonic and driven modes at the location of ILM excitation (Figure 2a2) and the resultant spectral lines are spaced away by twice the frequency of ILMs (Figures 2a4 and 2a5). The elevated amplitudes of the sub-harmonic mode lead to the excitation of ILMs. Particularly, since the maximal displacements are associated with the antinodes of this mode, the prevalence of ILM excitations is at these spatial locations. The displacement profiles corresponding to a large set of ILM experiments (Supplementary movies S1-S4) validate this nature (Figure 2b2). In order to look at the quantitative traits of ILM excitation, the displacements at the frequencies $\omega_d$, $\omega_d + \omega_i$, $\omega_d - \omega_i$, $\frac{\omega_d}{2}$, $\frac{\omega_d}{2} + \omega_i$, $\frac{\omega_d}{2} - \omega_i$ and $\omega_i$ are measured (Figures 2c1-2c7). Soon after the parametric excitation of sub-harmonic tone i.e. for $S_{in} > -3\ dBm$, the displacements at $\frac{\omega_d}{2}$ increases (Figure 2c4). In addition, there is also a slight increase in the displacements at $\frac{\omega_d}{2} + \omega_i$ and $\frac{\omega_d}{2} - \omega_i$ (Figures 2c5-2c6). Now, for $S_{in} > 4\ dBm$, the ILM excitation onsets. This results in the increase of displacements at $\omega_i$ (Figure 2c7), $\omega_d + \omega_i$, $\omega_d - \omega_i$ (Figures 2c2-2c3), $\frac{\omega_d}{2} + \omega_i$, $\frac{\omega_d}{2} - \omega_i$ (Figures 2c5-2c6).

Now, we investigate the temporal dependence of ILMs. To this end, the displacements in the spectral range $35 - 75\ kHz$ were imaged every minute for about 30-60 minutes from the time when the drive signal is introduced (Supplementary movies S1-S12). This entire process was carried out several times. (The entire set of experimental data is provided in the supplementary material S2.) The RMS displacement patterns (supplementary S2) provide the evidence for variations in spatial locations corresponding to ILMs. During these measurements, a number of interesting features of ILMs were recorded and a few are presented below.

*1 – Spatial stabilization –* In run 2, from $32^{th}$ minute, there is an evidence for locking of ILMs at a spatial location $a1$ for a long period of atleast 20 minutes (Figure 3-A).

*2 – Hopping –* In run 3, from $35^{th}$ to $37^{th}$ minute, the ILMs are seen to hop from the points $b1$ and $b2$ to $b3$ (Figure 3-B).

*3 – Attraction-repulsion of breathers –* In run 4, from $7^{th}$ to $9^{th}$ minute, the ILMs at the points $c1$ and $c3$ tend to coalesce and separate due to attraction and repulsion and as a result, the displacement at $c3$ increases and decreases and corresponding decrease and increase in the displacement is observed at $c4$ (Figure 3-C). This behaviour may also be conceived as 'ILM Splitting' [20].

To demonstrate the universality of ILMs and their features in any mechanical mode, the experiments were also carried out with other eigen-modes of the free-free beam. To this end, the drive frequencies $51\ kHz$ and $264\ kHz$ are applied at the drive power level $20\ dBm$. The resultant vibration patterns are presented in the figures 4a1 and 4b1. Here, the modes are directly driven at the elevated amplitudes which in turn results in the excitation of low frequency ILMs (Figures 4a2 and 4b2). Further, the auxiliary aspects including ILM - driven mode coupling (Figures 4a3, 4b3 and 4c3) and prevalence of ILMs at the antinodes (Figures 4a4, 4b4 and 4c4) are also qualitatively reproduced in these modes as well. Interestingly, the frequency ranges for ILMs are also observed to be modal dependent (Figures 4a2, 4b2 and 4c2). That is, the modes corresponding to the frequencies $51\ kHz$, $268\ kHz$ and $1.93\ MHz$ lead to the excitations of ILMs with nominal frequencies close to $2.88\ kHz$, $104.1\ kHz$ and $57.6\ kHz$ respectively. Interestingly, the observed trend does not exhibit monotonous behaviour.

**Summary**

We have experimentally observed both the discreteness and nonlinearity of ILMs in a micromechanical resonator. Unlike the ILMs in coupled cantilevers [19], the ILMs in our system are observed as spectral bushes and their frequencies are much smaller than that of the drive. The excitation of ILMs in our mechanical resonator is mediated via the elevated amplitudes of mechanical mode. Our results also showcase different compulsive features of ILMs including hopping, attraction-repulsion and single and spatial stabilization. Such features can be individual subjects for future experimental investigations. The various physical aspects of ILMs reported in this letter also motivate the development of a predictive first-principles model.

**Acknowledgements**

Funding from the Cambridge Trusts and the Engineering and Physical Sciences Research Council is gratefully acknowledged.

**Authors' contributions**

AG and CD designed the device and performed the experiments; AG and AAS analyzed the results and wrote the manuscript; AAS supervised the research.

**Additional information**

There are no competing financial interests.

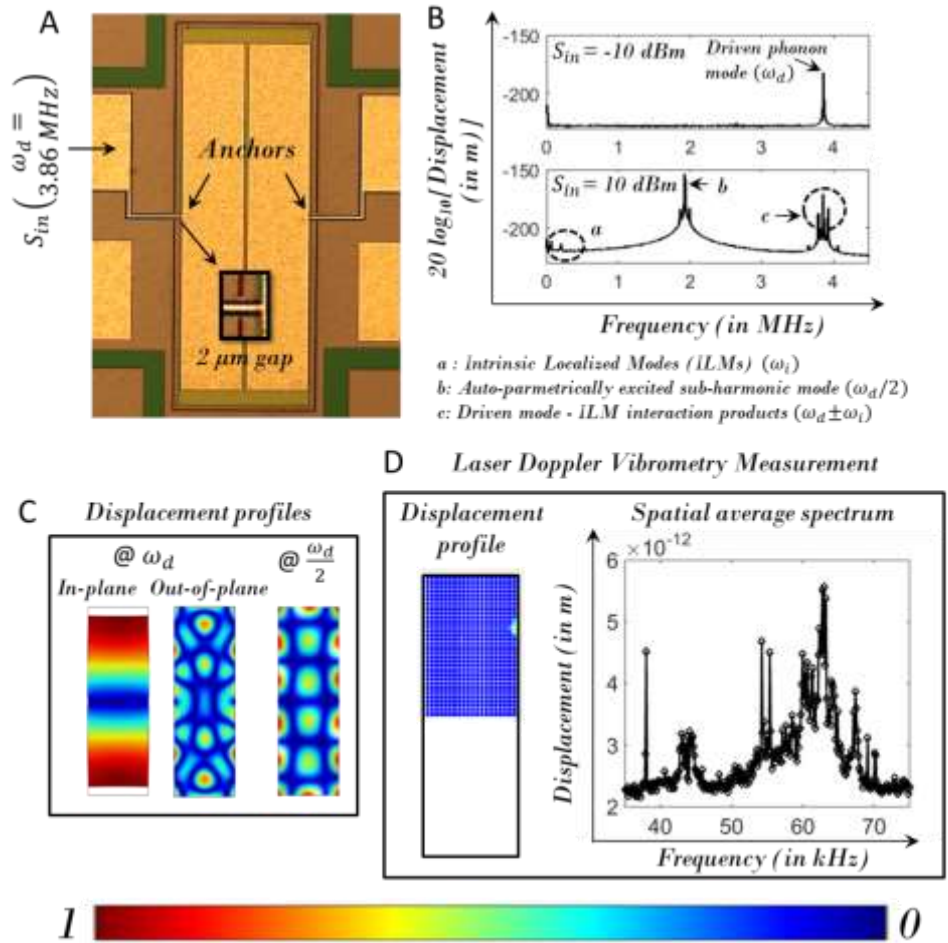

Figure 1: **Observation of Intrinsic Localized Modes.** A: Signal $S_{in}(\omega_d = 3.86\ MHz)$ is applied on a free-free beam microstructure; B: The spatial averaged spectral responses for $S_{in} = -10\ dBm$ and $S_{in} = 10\ dBm$; C: The displacement profiles at the frequencies $\omega_d = 3.86\ MHz$ and $\frac{\omega_d}{2} = 1.93\ MHz$; D: Left: RMS displacement profile (averaged over the spectral range $35 - 75\ kHz$); E: Right: The spatial average frequency spectrum over a specific one minute interval. Note: The displacement profiles are normalized to the maximum displacement in the structure as color-coded.

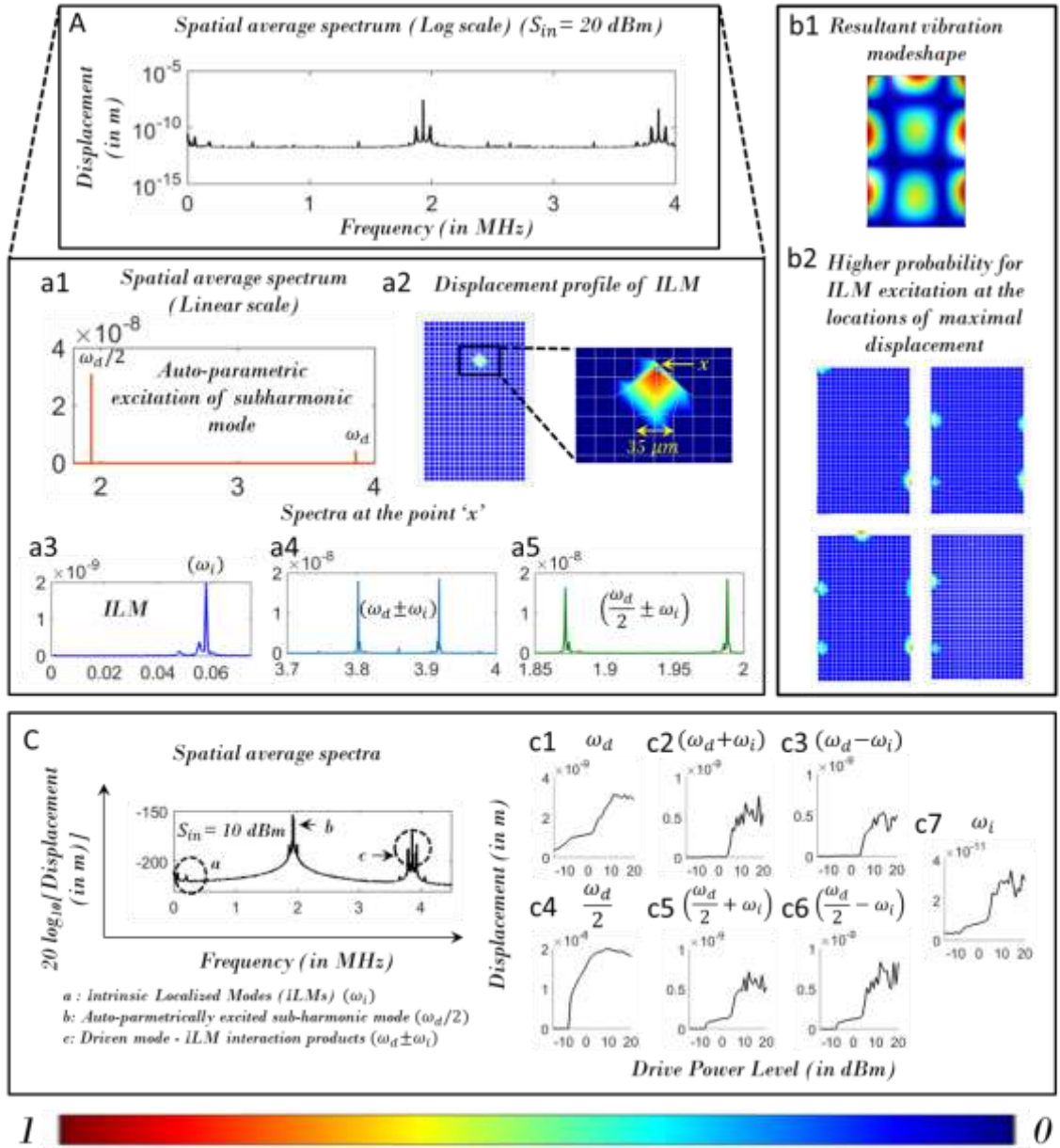

Figure 2: **Spatio-spectral features of Intrinsic Localized Modes.** A: Spatial average spectrum (Log scale) at the drive power level $S_{in}(\omega_d = 3.86\ MHz) = 20\ dBm$; a1: Spatial average spectrum (Linear scale) indicating the significant excitation of sub-harmonic mode via auto-parametric resonance; a2: Displacement profile of ILM at frequency $57.6\ kHz$ indicating the elevated displacement at the point $x$; a3-a4: Spectrum at the point $x$, zoomed into the frequency ranges $0 - 70\ kHz$; $3.7 - 4\ MHz$ and $1.85 - 2\ MHz$ respectively; b1: Resultant vibration pattern (averaged over the spectral range $0 - 5\ MHz$) of the beam; b2: The RMS displacement profiles (averaged over the spectral range $35 - 75\ kHz$) at the $12^{th}$, $14^{th}$ and $16^{th}$ minutes of run 2 and at the $10^{th}$ minute of run 3 indicating the spatial prevalence of ILMs; C: The spatial averaged spectral responses for $S_{in} = 10\ dBm$; c1-c7: The displacements at $\omega_d$, $\omega_d + \omega_i$, $\omega_d - \omega_i$, $\frac{\omega_d}{2}$, $\frac{\omega_d}{2} + \omega_i$,

$\frac{\omega_d}{2} - \omega_i$ and $\omega_i$. Note: The displacement profiles are normalized to the maximum displacement in the structure- i.e. Red corresponds to 1 and blue corresponds to 0.

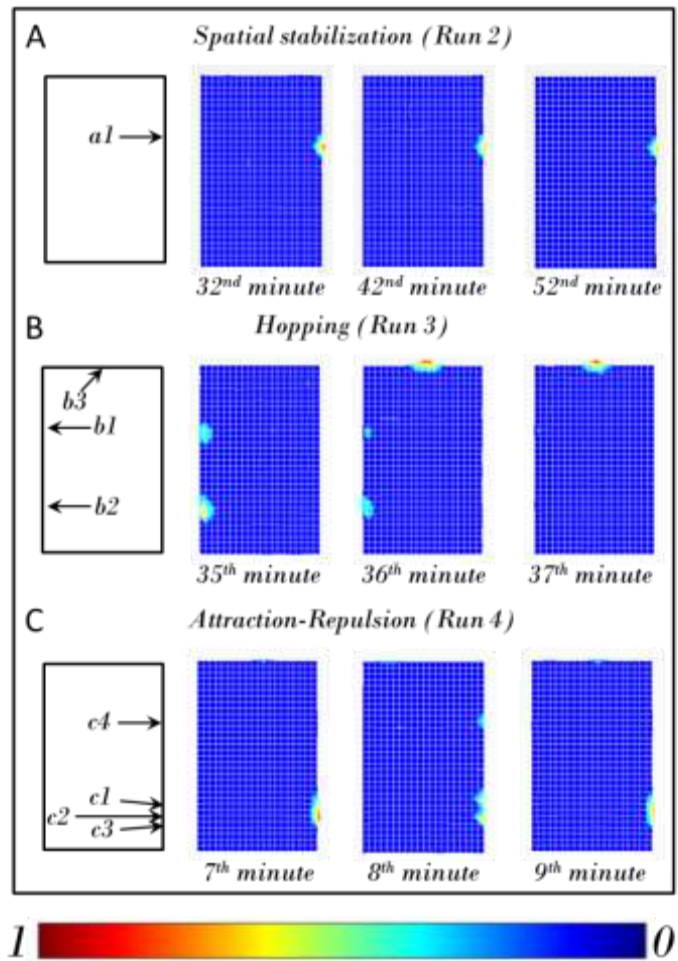

Figure 3: **Temporal dependence of Intrinsic Localized Modes.** A correspond to spatial stabilization feature; B correspond to the attraction-repulsion feature; C correspond to the hopping feature. Note 1: The RMS displacement profiles (averaged over the spectral range $35 - 75\ kHz$) are normalized to the maximum displacement in the structure. Note 2: The RMS displacement profiles are selected from the supplementary section S2.

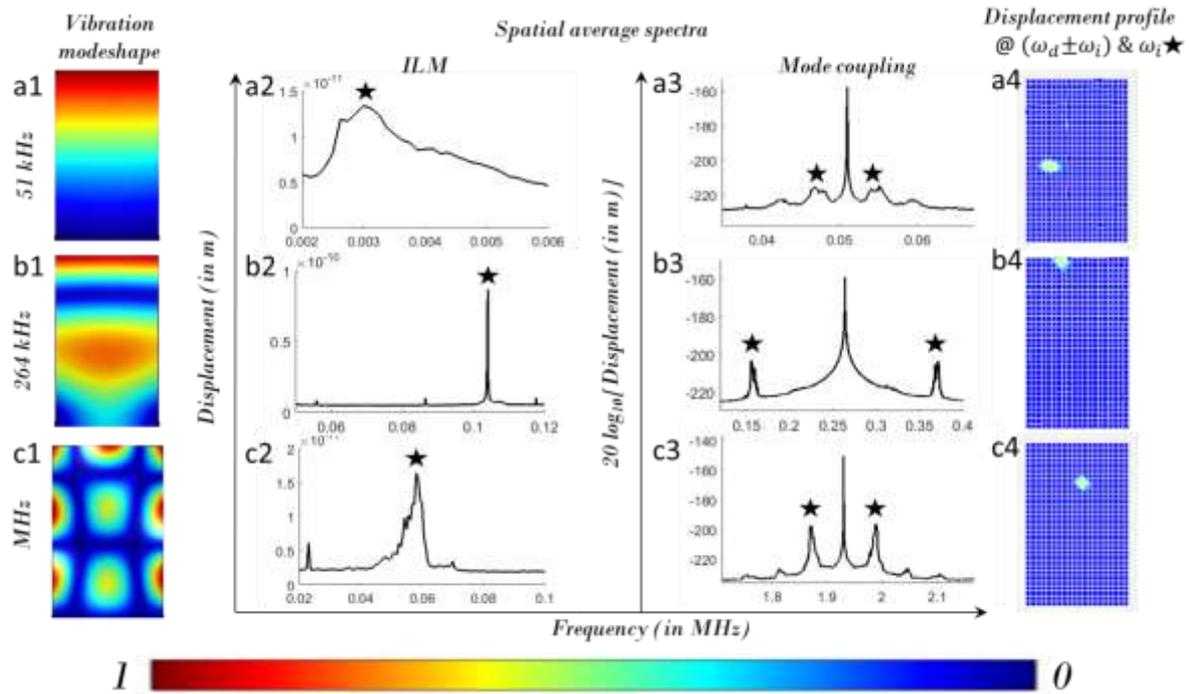

Figure 4: **Modal dependence of Intrinsic Localized Modes.** a1-a4: The vibration pattern, spatial average spectra corresponding to ILM and driven mode coupling regions and displacement profile at $\omega_i$, $\omega_d \pm \omega_i$ indicated by stars for $S_{in}(\omega_d = 51\ kHz) = 20\ dBm$; b1-b4: The resultant vibration pattern (averaged over the spectral range $0 - 2\ MHz$), spatial average spectra corresponding to ILM and driven mode coupling regions and displacement profile at $\omega_i$, $\omega_d \pm \omega_i$ indicated by stars for $S_{in}(\omega_d = 264\ kHz) = 20\ dBm$; c1-c4: The resultant vibration pattern (averaged over the spectral range $0 - 5\ MHz$), spatial average spectra corresponding to ILM and driven mode coupling regions and displacement profile at $\omega_i$ and $\frac{\omega_d}{2} \pm \omega_i$ indicated by stars for $S_{in}(\omega_d = 3.86\ MHz) = 20\ dBm$. Note: The displacement profiles are normalized to the maximum displacement in the structure- i.e. Red corresponds to 1 and blue corresponds to 0.

**Supplementary Information**

**Discrete intrinsic localized modes in a microelectromechanical resonator**

Authors: Adarsh Ganesan[1], Cuong Do[1], Ashwin Seshia[1]

1. Nanoscience Centre, University of Cambridge, Cambridge, UK

**Supplementary section S1**

Operation of piezoelectrically driven micromechanical resonator

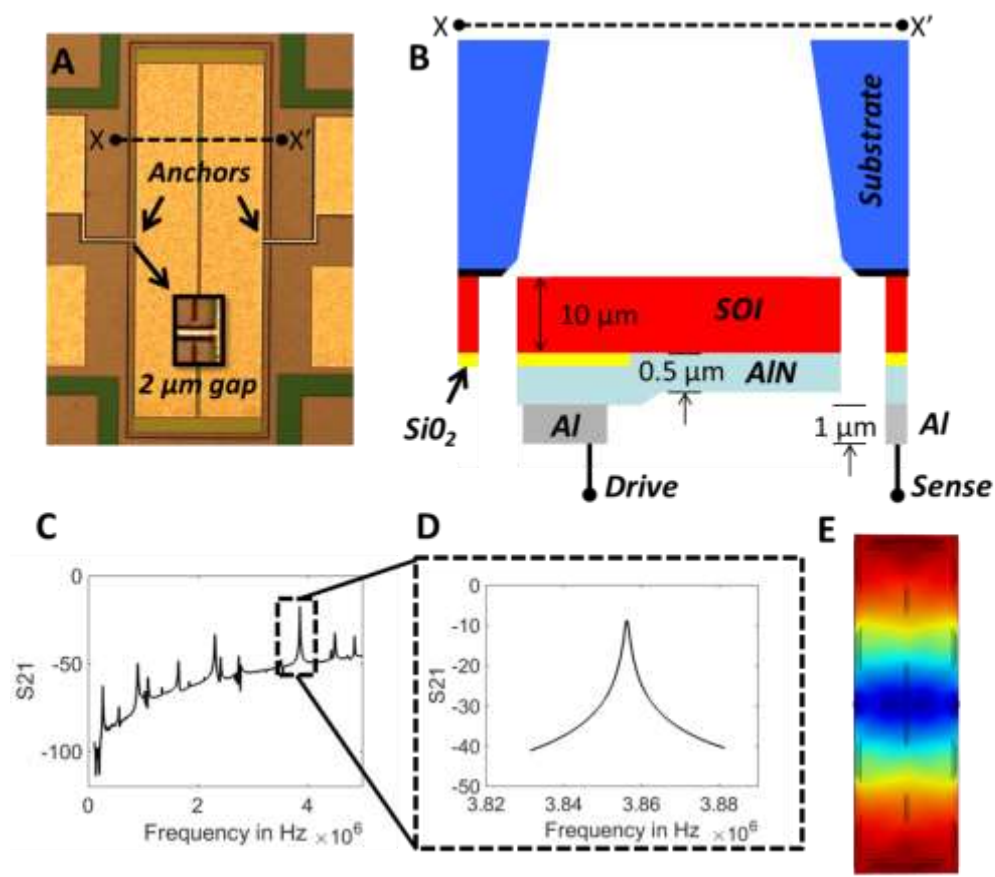

**Figure S1: Operation of piezoelectrically driven micromechanical resonator**: **A**: Free-free beam topology with 2 μm air gap for in-plane mode excitation; **B**: 1 μm thick Al electrodes patterned on 0.5 μm thick AlN piezoelectric film which is in-turn patterned on SOI substrate; the 10 μm thick SOI layer is then released through back-side etch to realize mechanical functionality; **C**: The scattering parameter S21 denoting forward transmission gain across a broad range of frequencies from 0-4.5 MHz; **D**: across a smaller range of frequencies 3.83-3.88 MHz; **E**: Resonant mode shape of ~3.86 MHz vibrations from eigen-frequency analysis in COMSOL and Doppler shift vibrometry.

**Supplementary section S2**

Frequency responses and RMS surface displacement profiles obtained using Laser Doppler Vibrometry at different drive conditions

Run 1

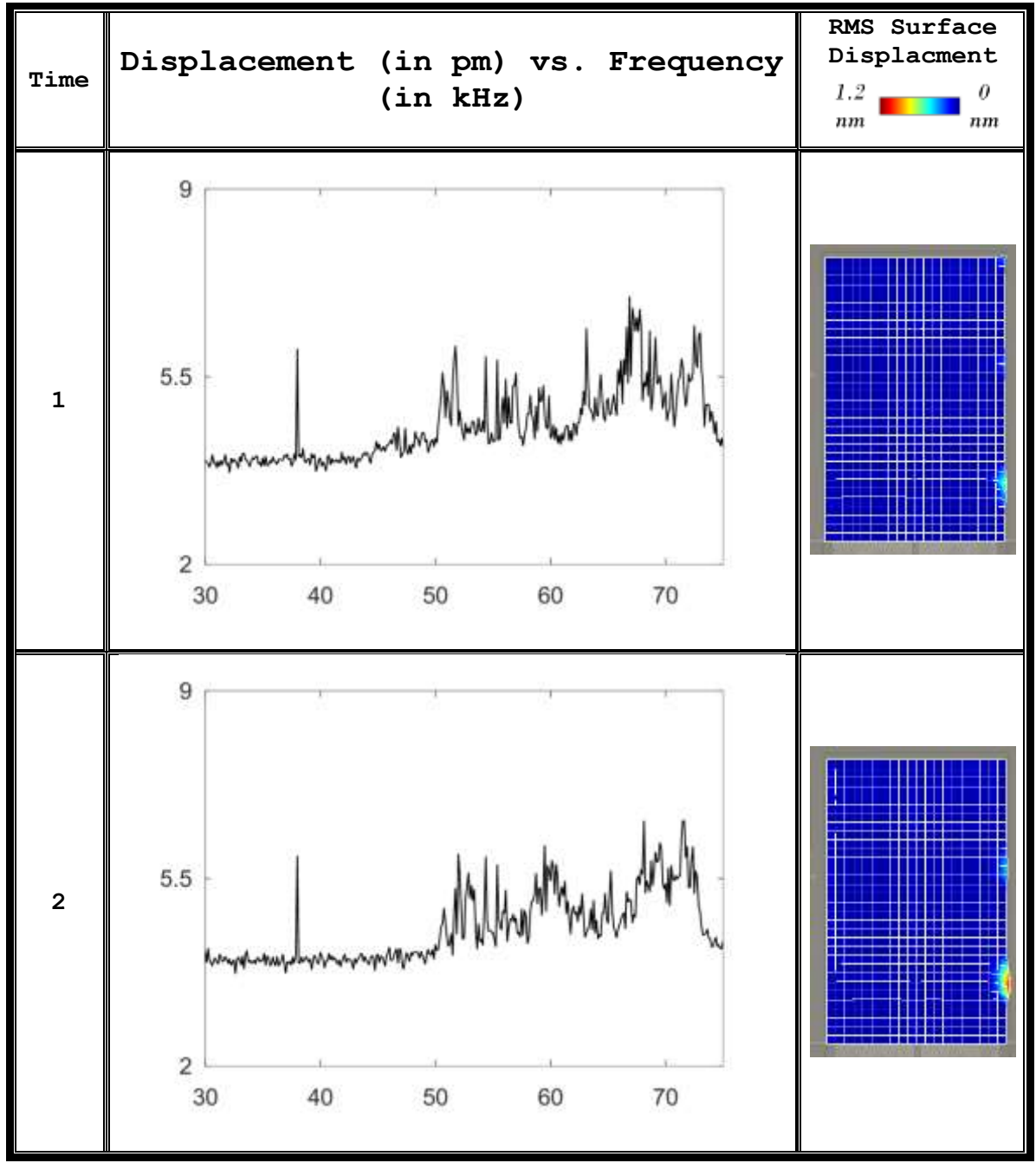

| 3 | 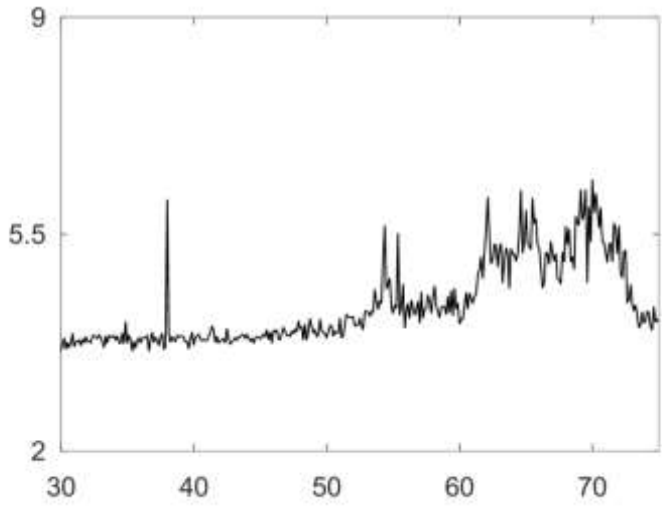 | 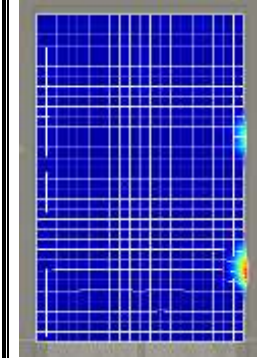 |
|---|---|---|
| 4 | 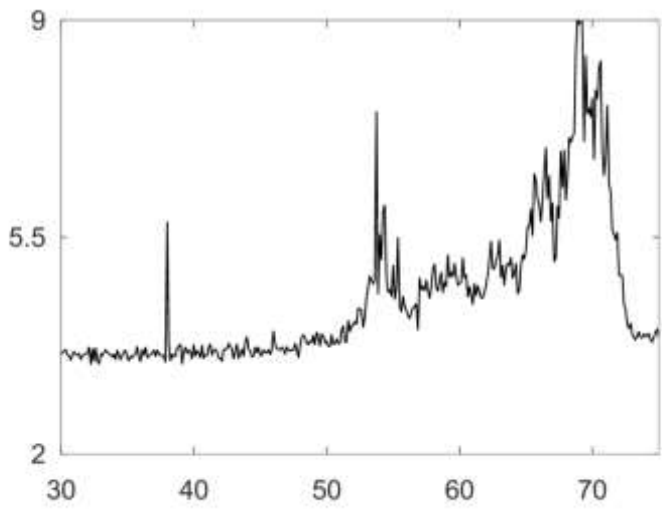 | 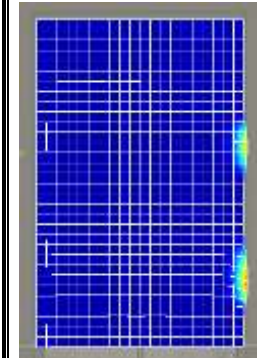 |
| 5 | 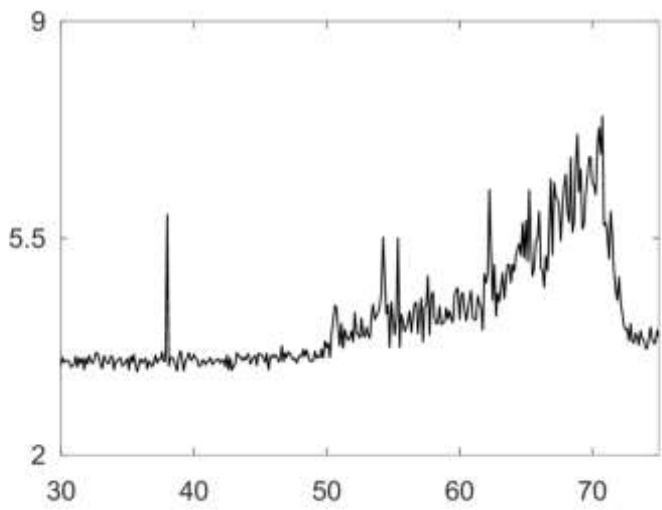 | 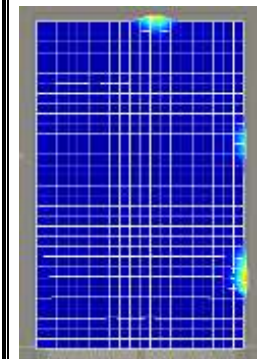 |

| 6 | 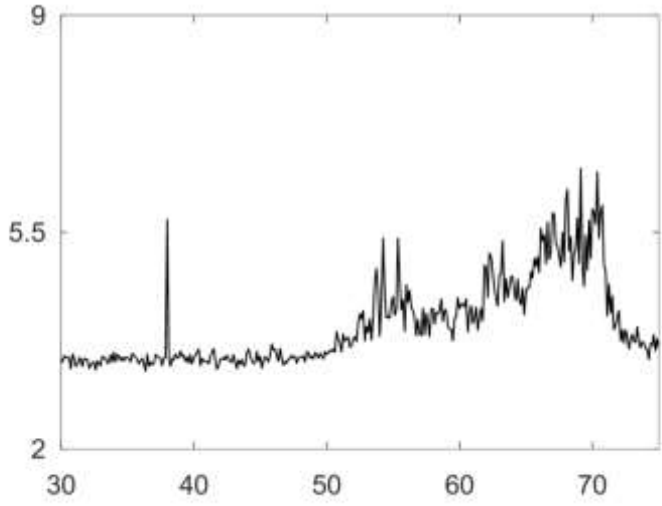 | 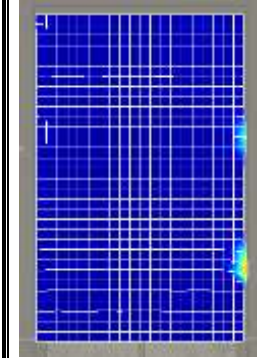 |
|---|---|---|
| 7 | 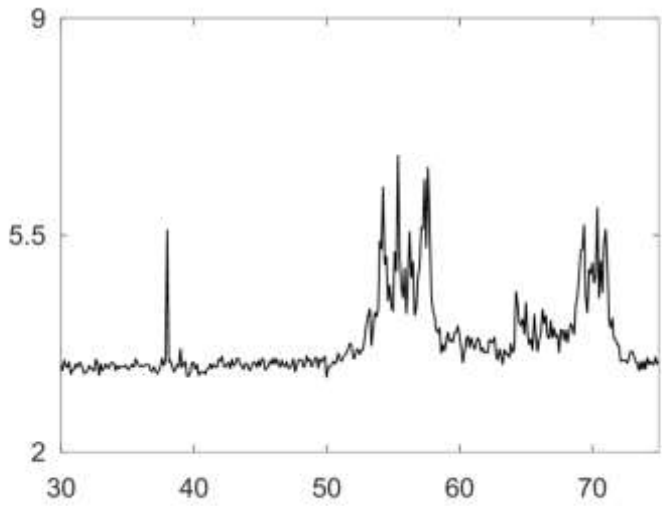 | 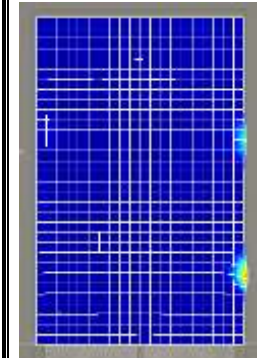 |
| 8 | 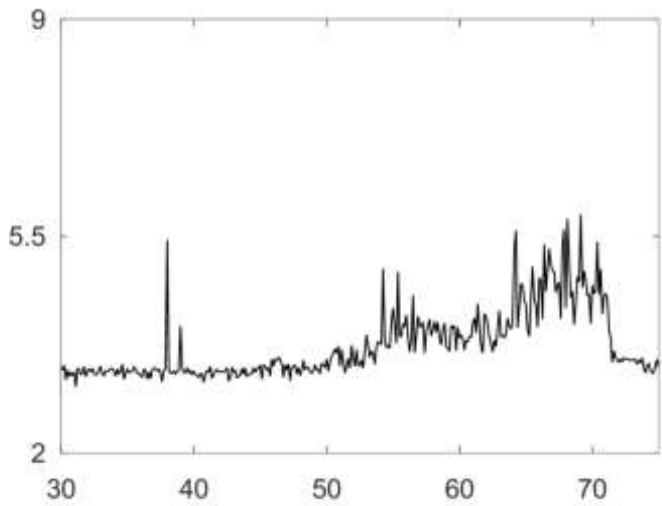 | 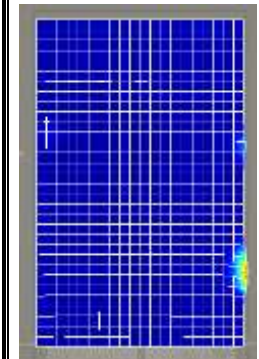 |

| 9 | 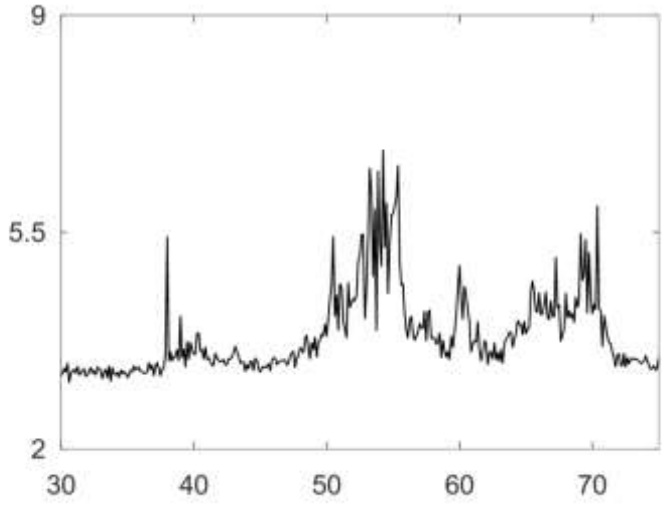 | 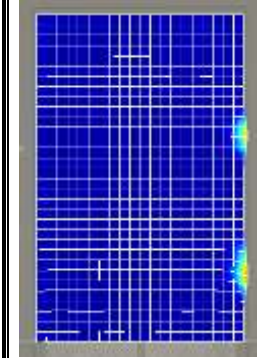 |
|---|---|---|
| 10 | 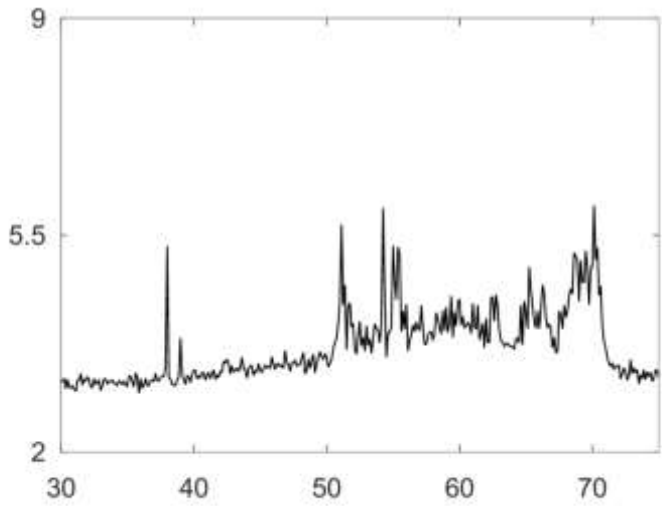 | 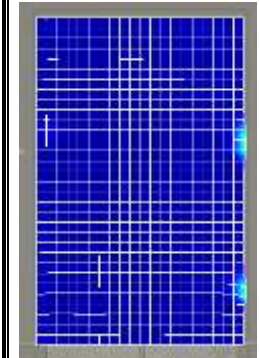 |
| 11 | 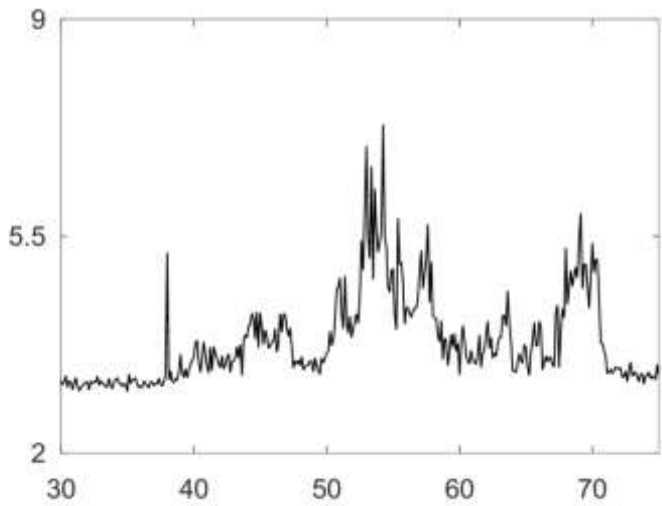 | 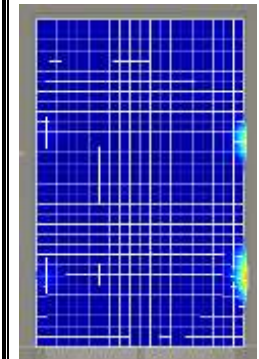 |
| 12 | 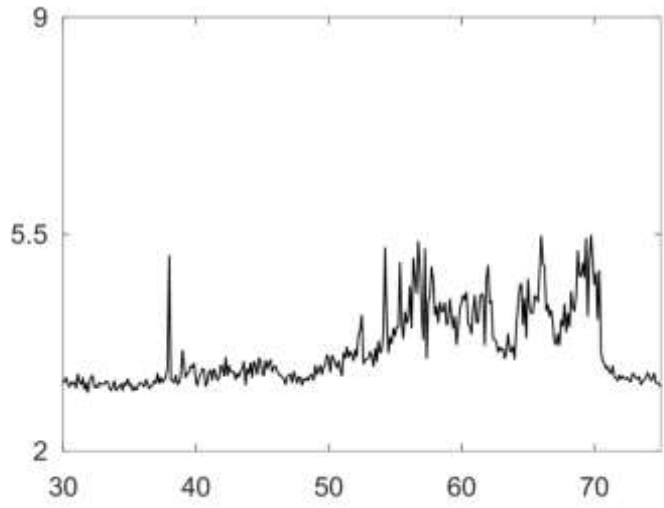 | 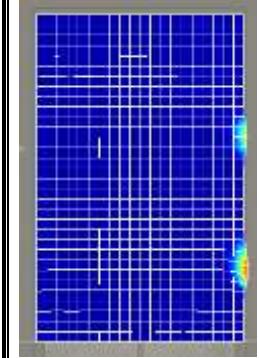 |
| 13 | 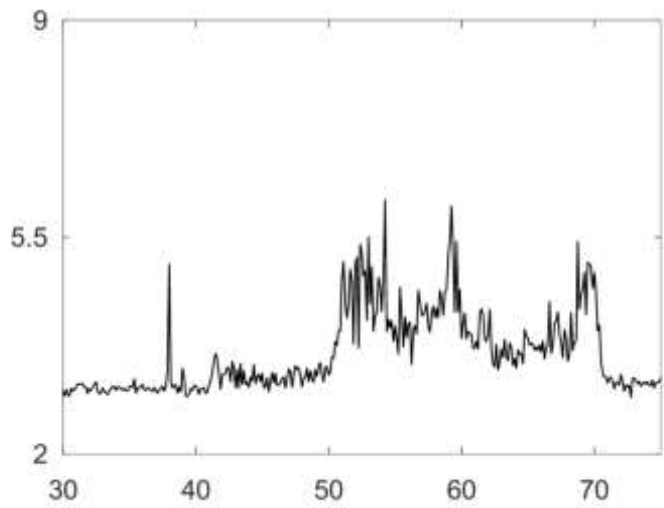 | 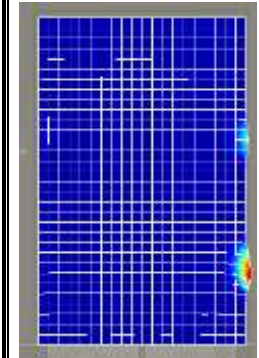 |
| 14 | 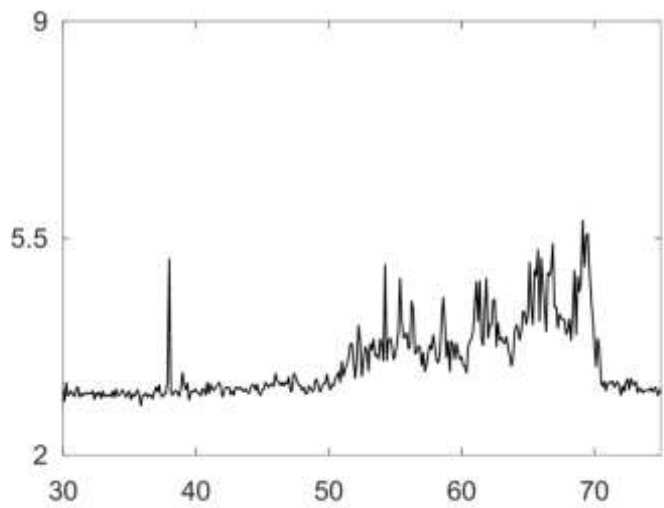 | 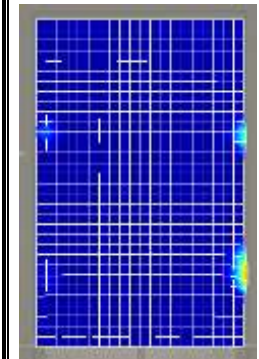 |
| 15 | 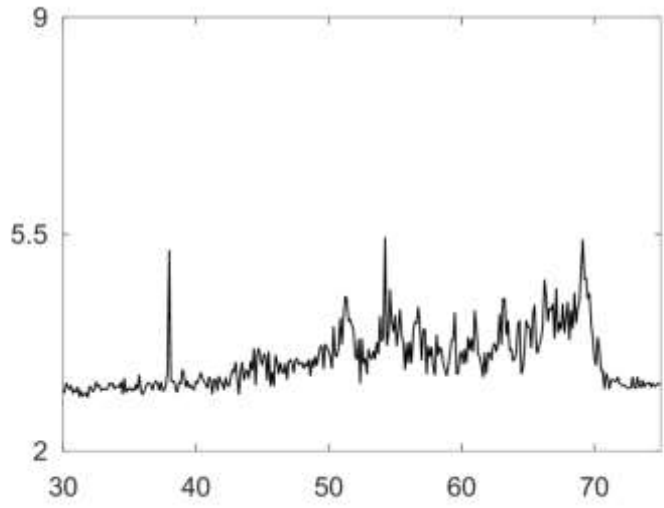 | 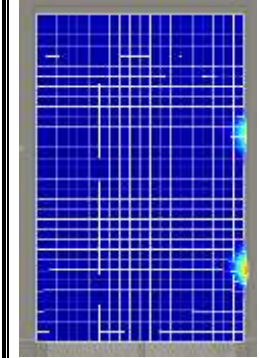 |
| 16 | 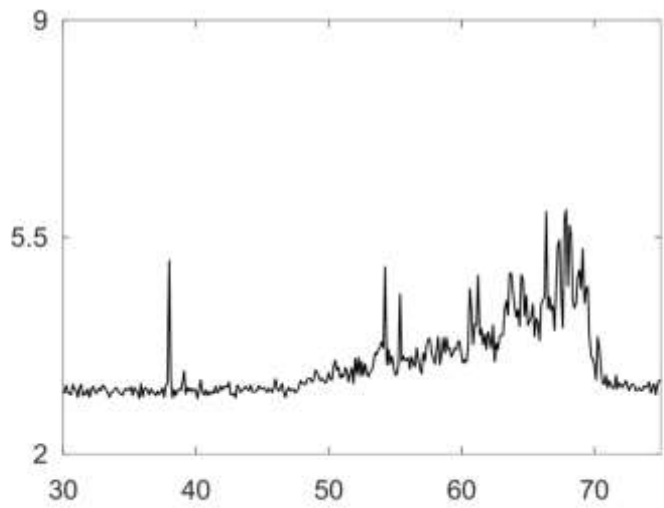 | 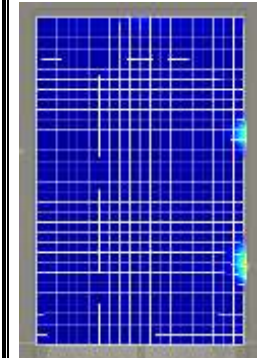 |
| 17 | 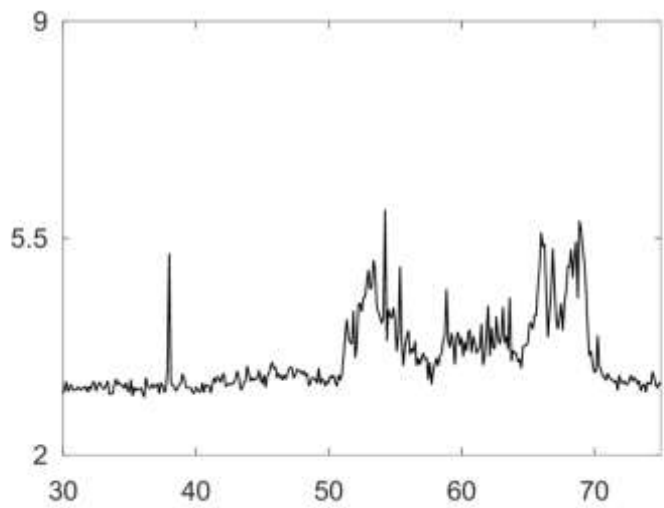 | 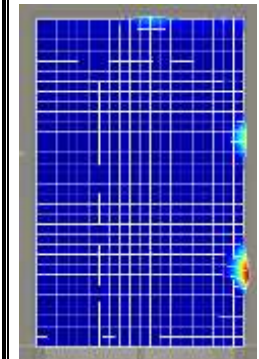 |

| 18 | 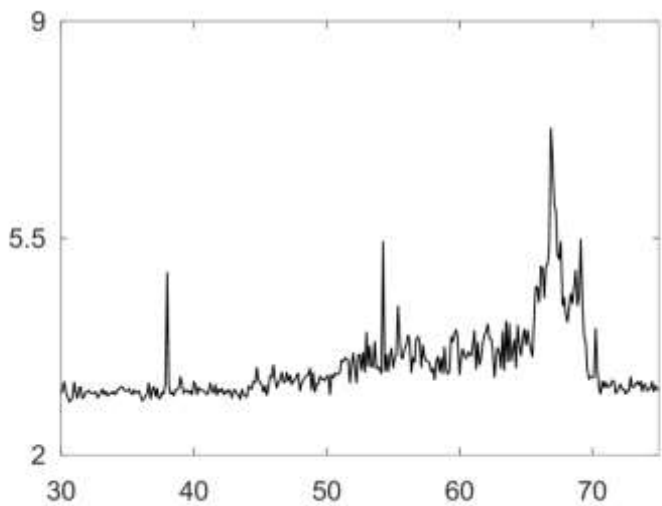 | 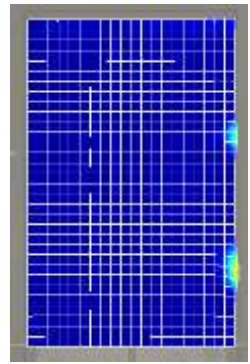 |
|---|---|---|
| 19 | 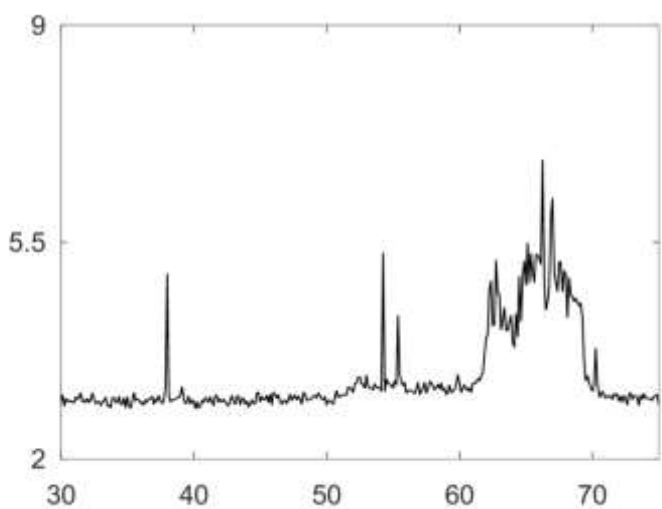 | 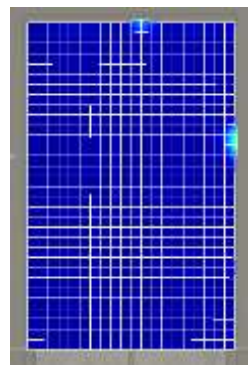 |
| 20 | 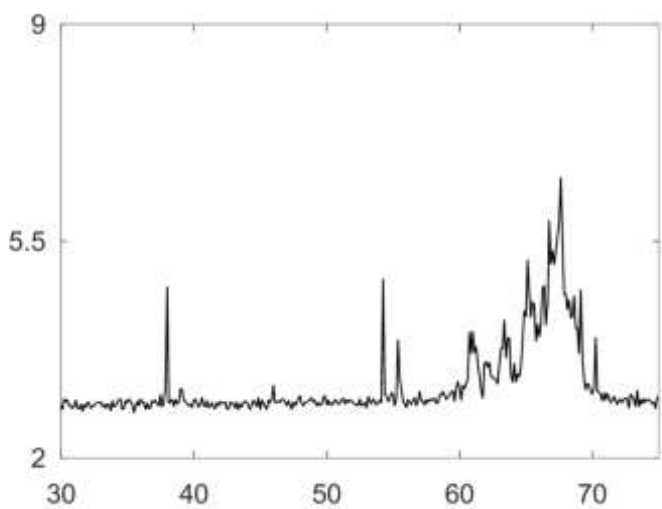 | 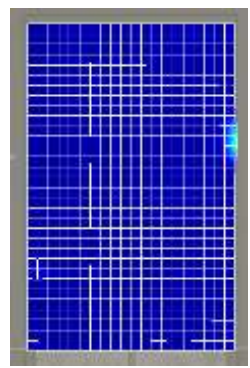 |

| 21 | 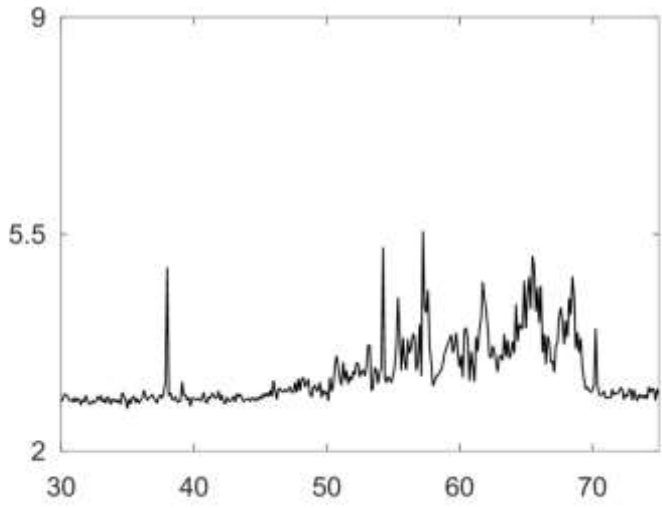 | 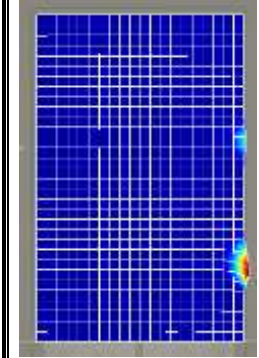 |
| --- | --- | --- |
| 22 | 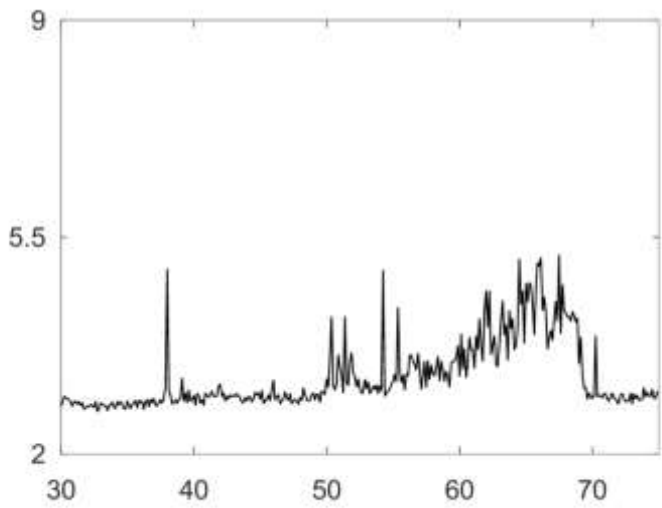 | 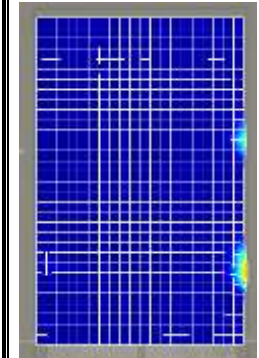 |
| 23 | 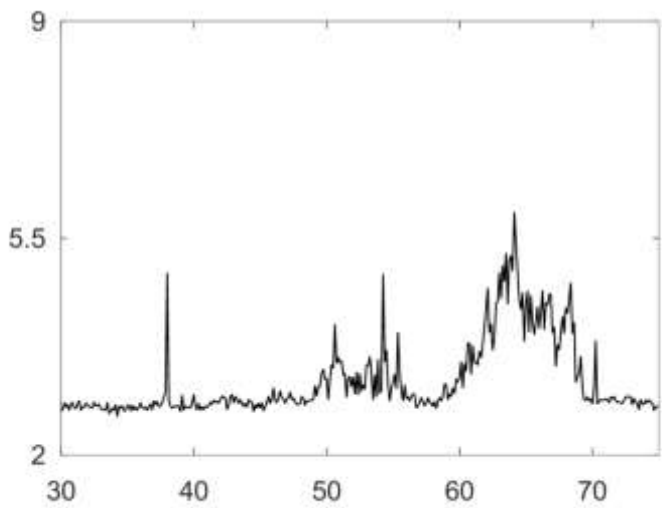 | 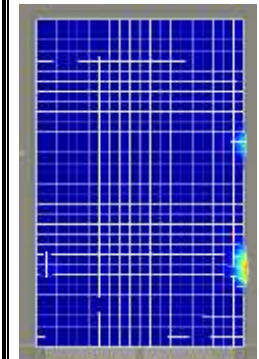 |
| 24 | 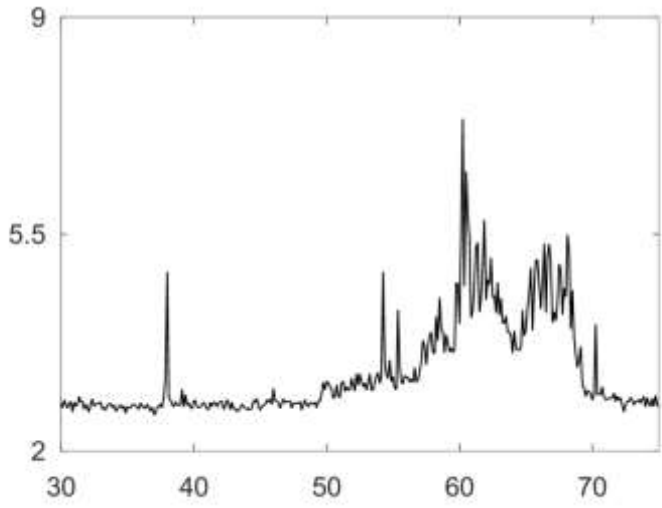 | 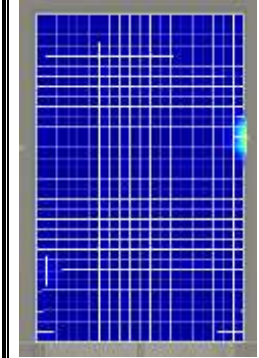 |
| 25 | 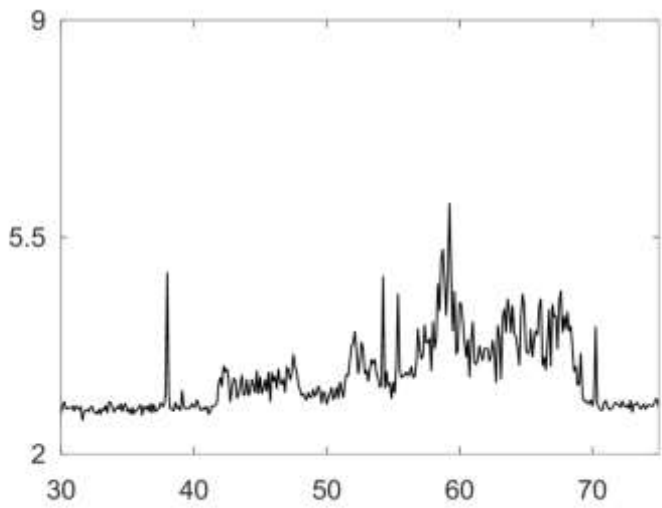 | 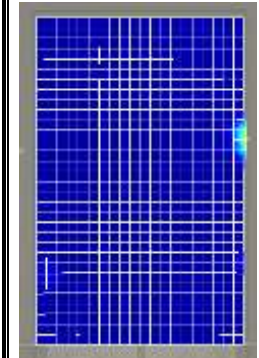 |
| 26 | 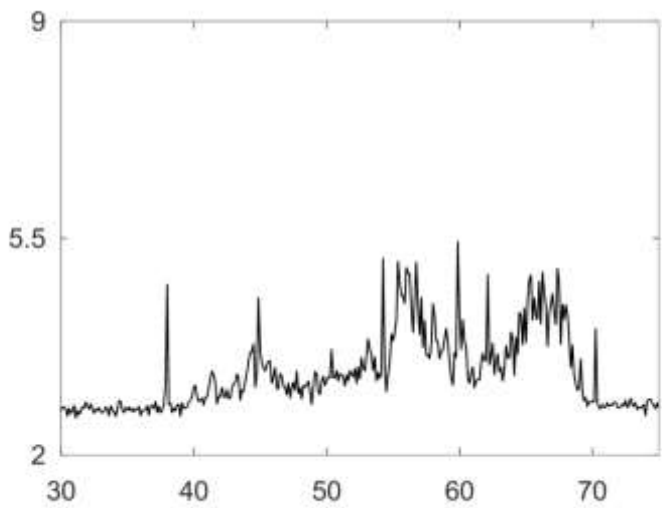 | 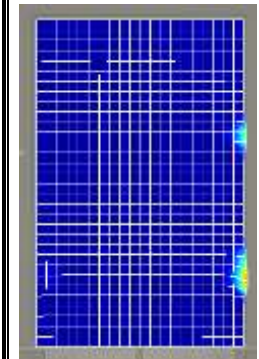 |

| 27 | 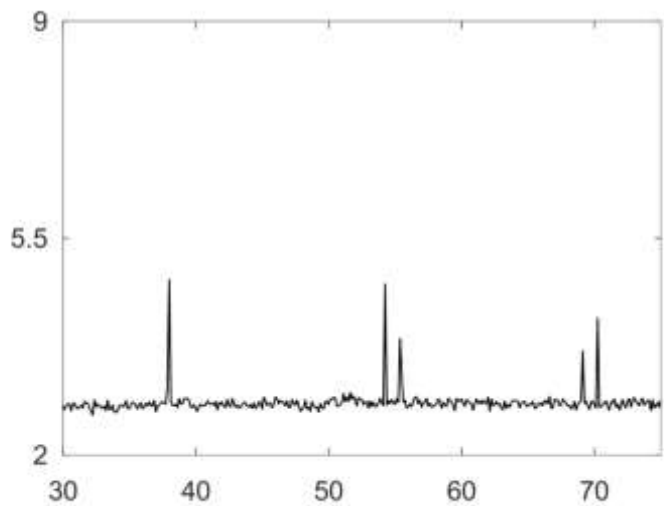 | 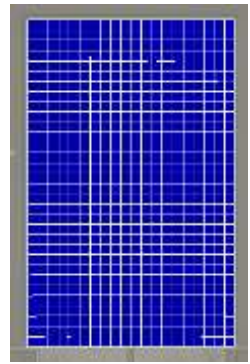 |
|---|---|---|
| 28 | 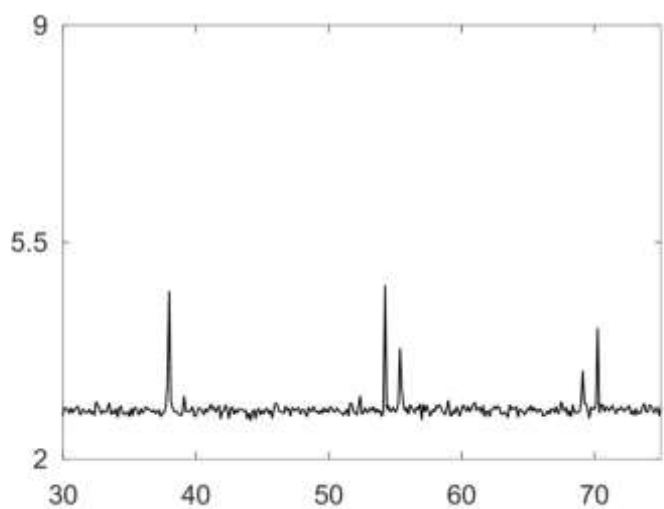 | 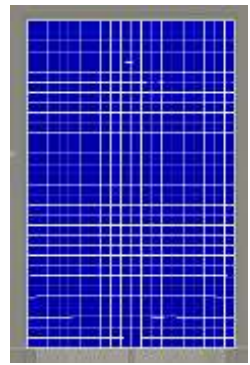 |
| 29 | 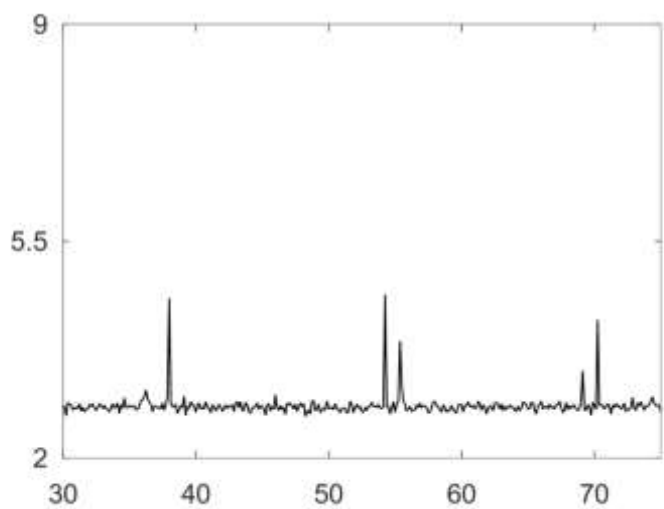 | 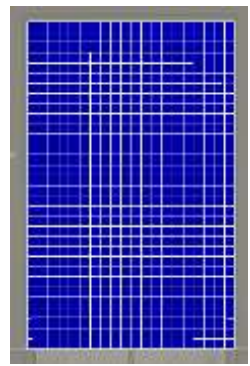 |

| Time | Displacement (in pm) vs. Frequency (in kHz) | RMS Surface Displacment |
|---|---|---|
| 30 | 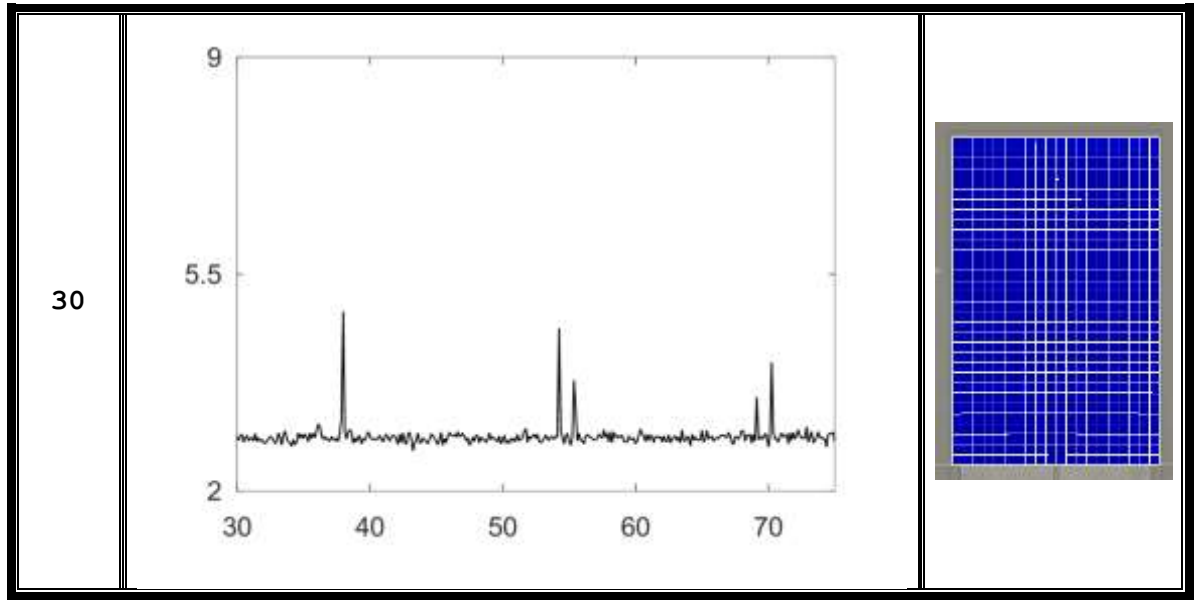 | |

Run 2

| Time | Displacement (in pm) vs. Frequency (in kHz) | RMS Surface Displacment 1.2 nm — 0 nm |
|---|---|---|
| 1 | 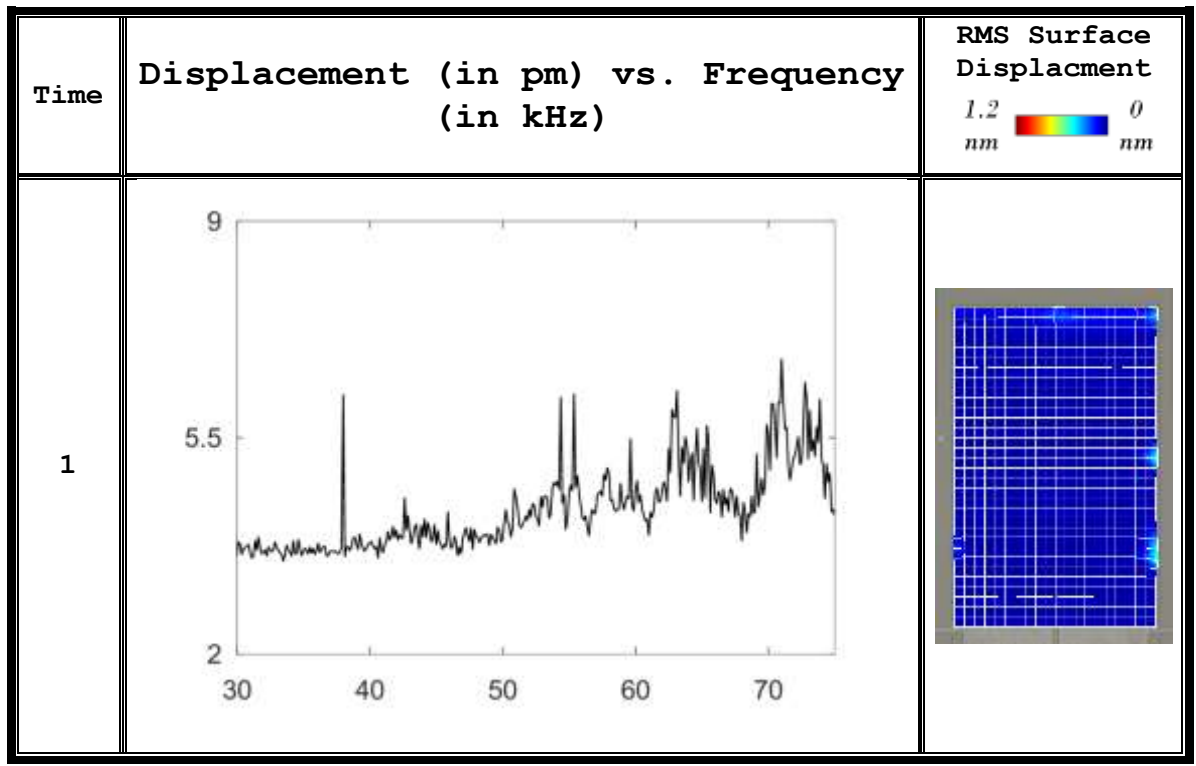 | |

| 2 | 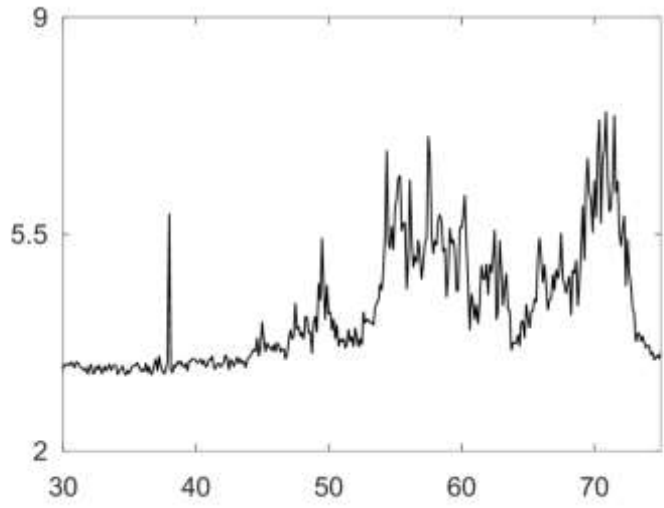 | 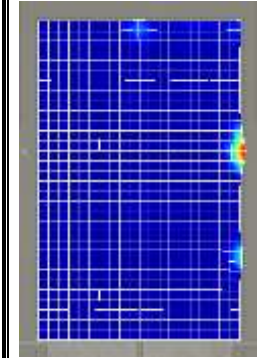 |
|---|---|---|
| 3 | 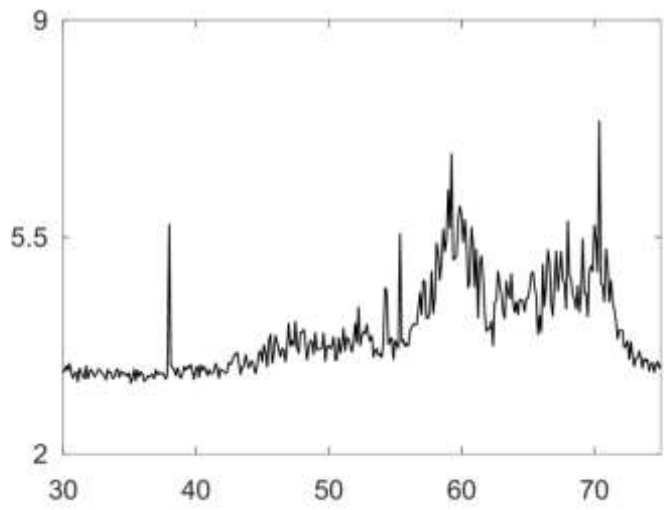 | 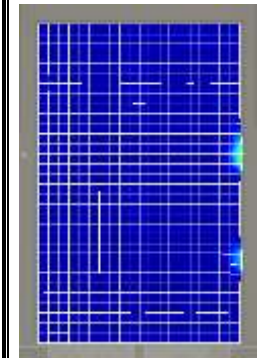 |
| 4 | 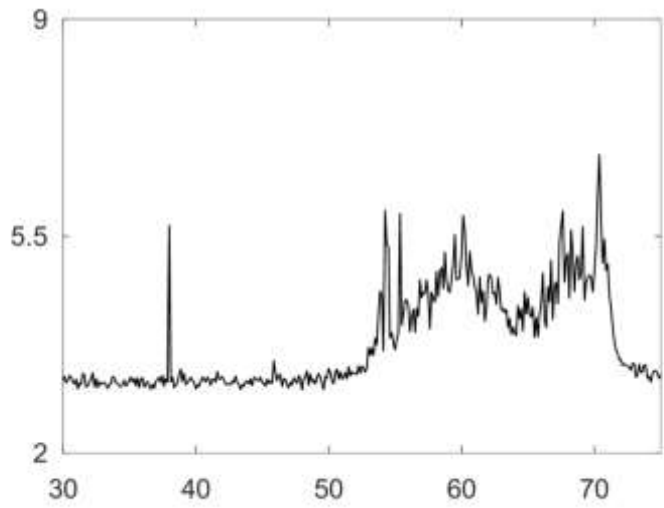 | 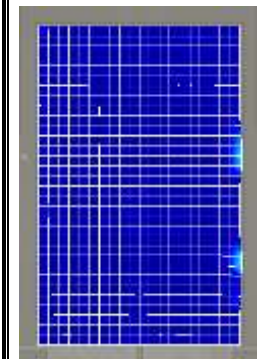 |

| 5 | 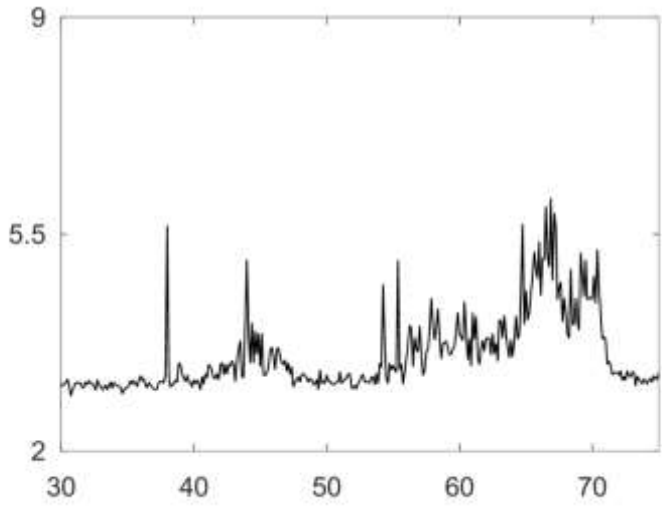 | 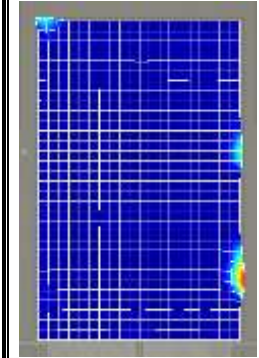 |
|---|---|---|
| 6 | 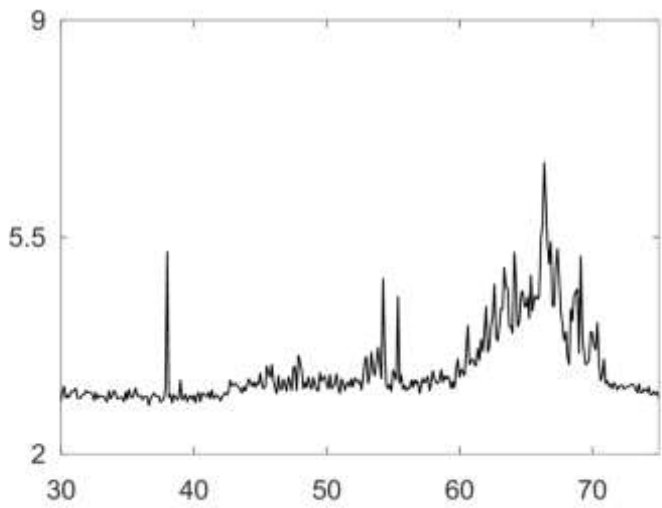 | 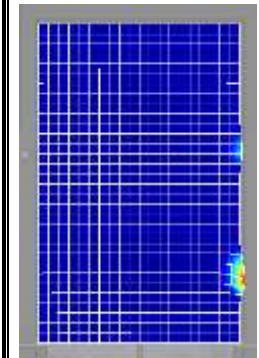 |
| 7 | 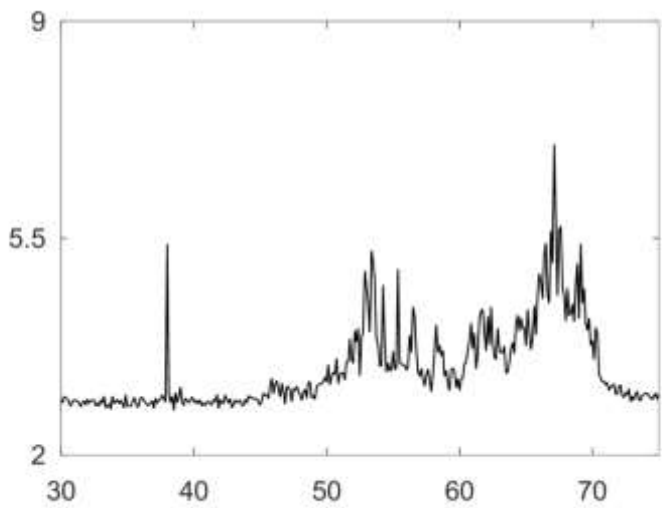 | 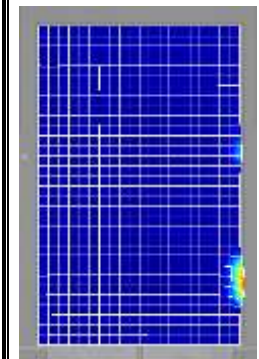 |

| 8 | 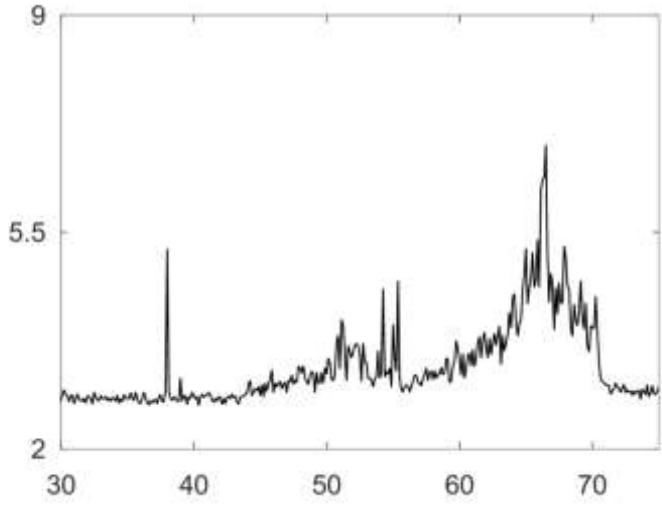 | 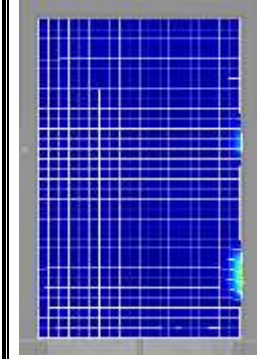 |
| --- | --- | --- |
| 9 | 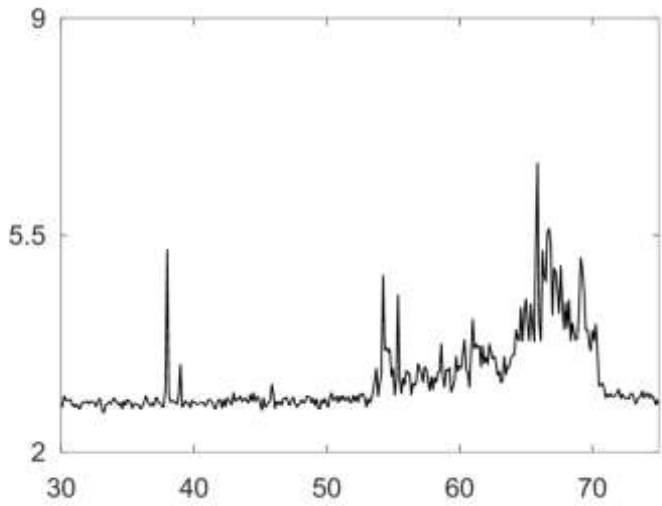 | 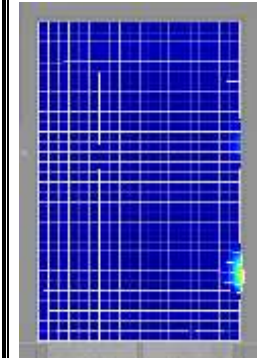 |
| 10 | 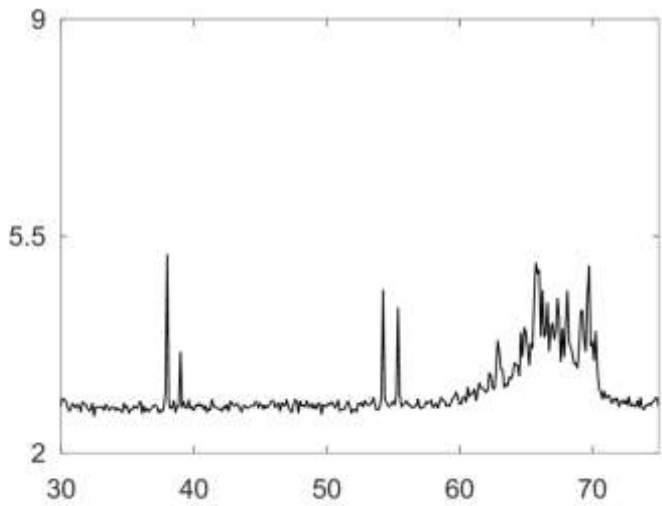 | 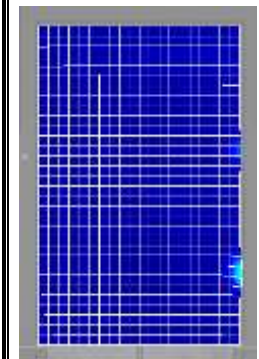 |

| 11 | 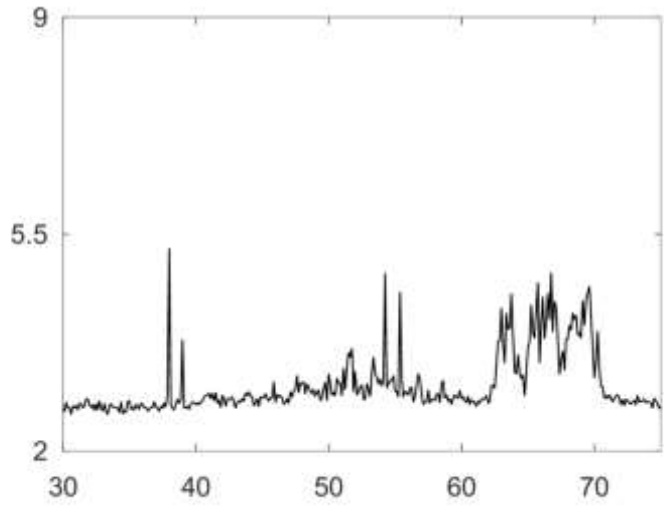 | 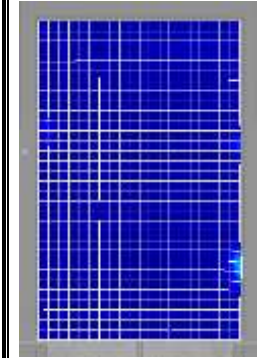 |
| --- | --- | --- |
| 12 | 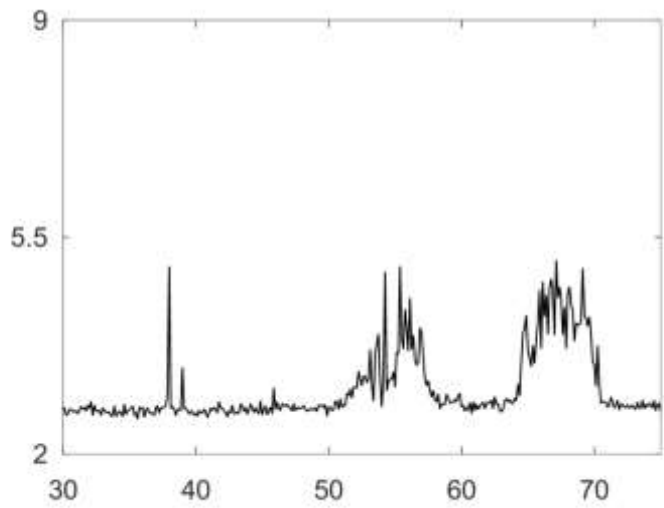 | 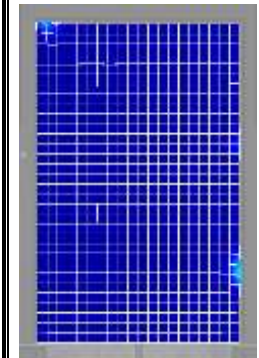 |
| 13 | 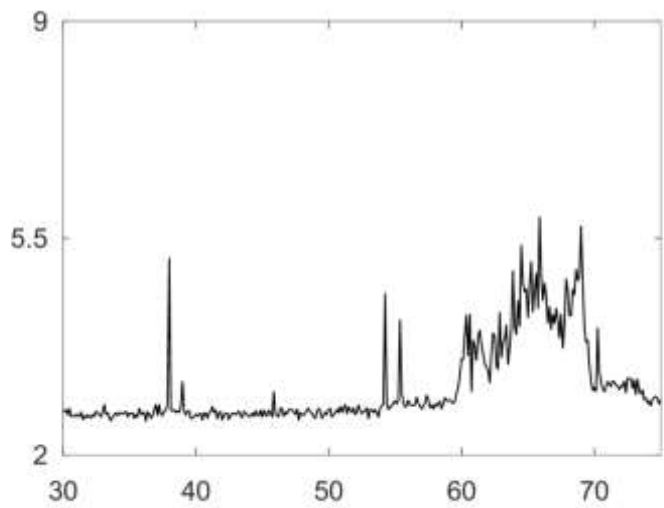 | 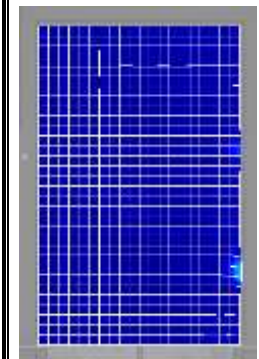 |

| 14 | 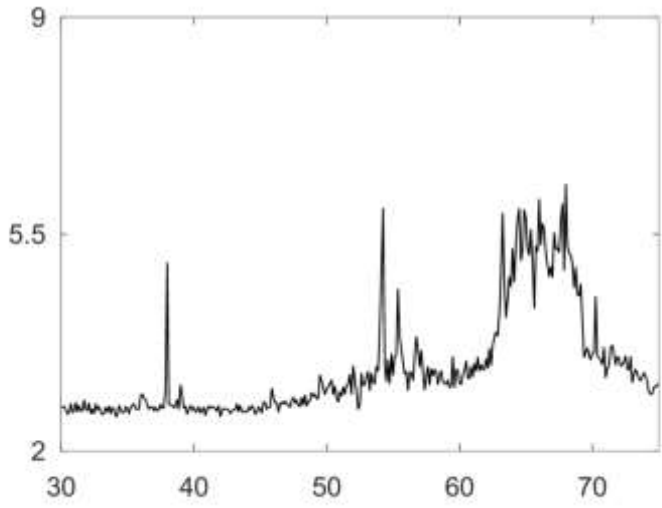 | 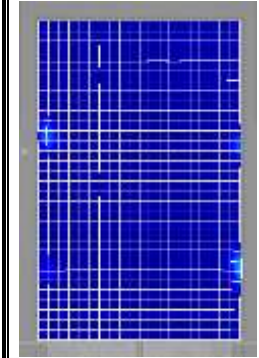 |
|---|---|---|
| 15 | 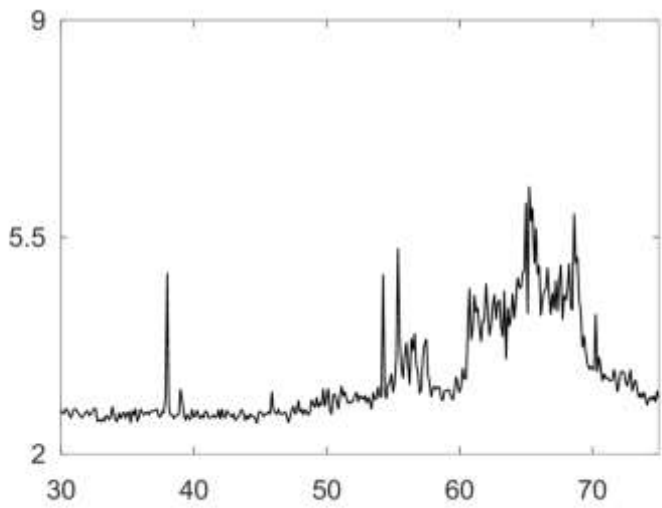 | 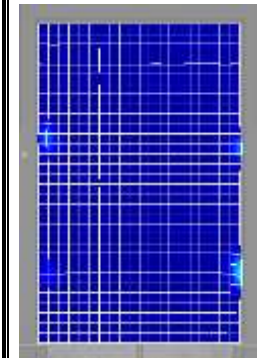 |
| 16 | 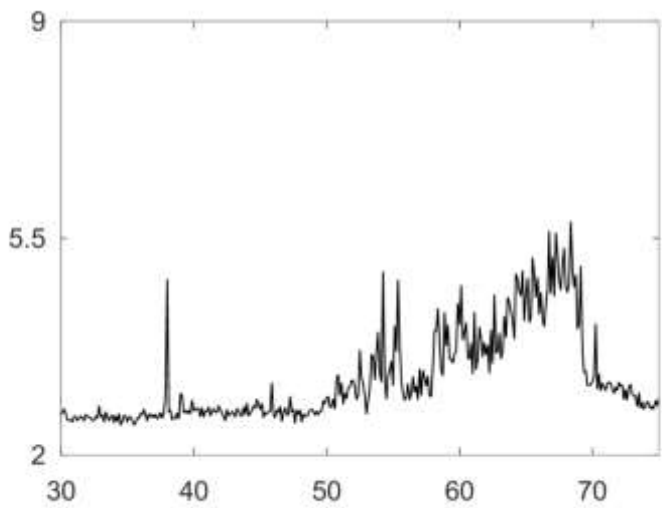 | 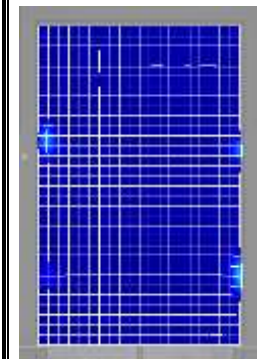 |

| 17 | 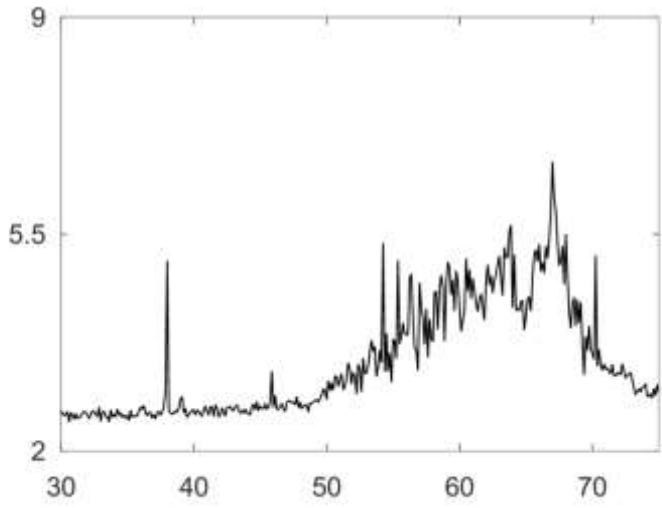 | 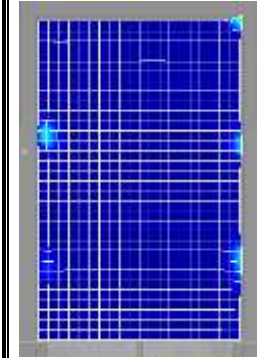 |
| --- | --- | --- |
| 18 | 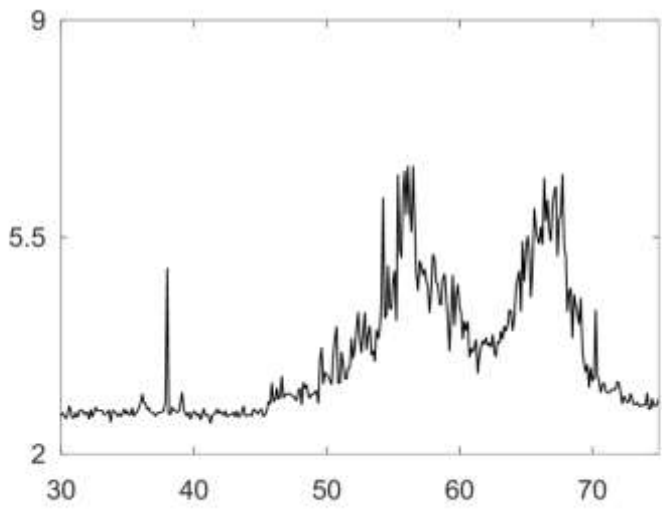 | 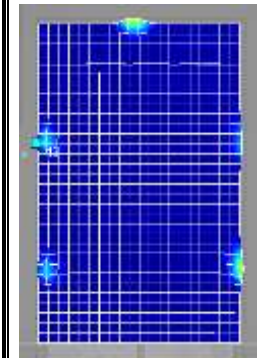 |
| 19 | 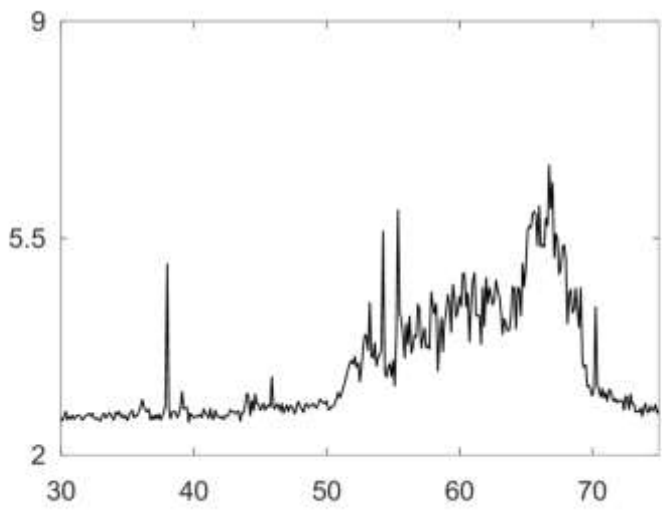 | 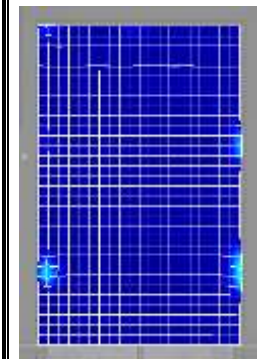 |

| 20 | 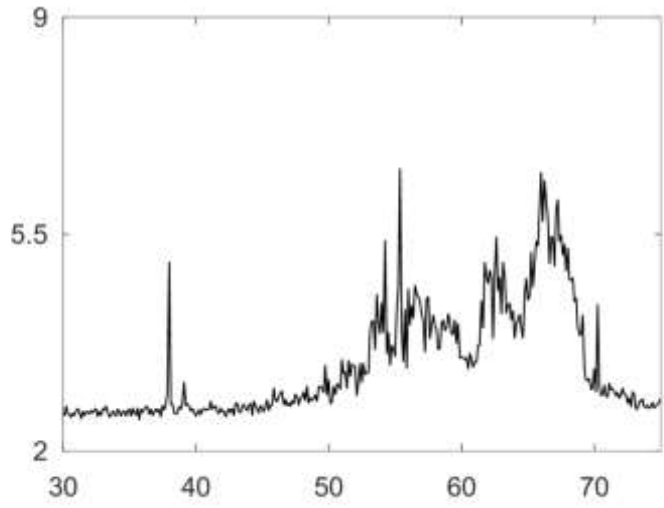 | 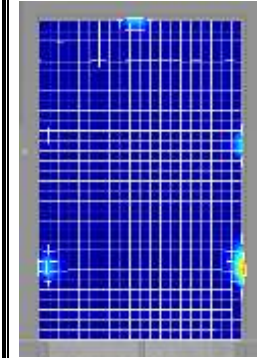 |
|---|---|---|
| 21 | 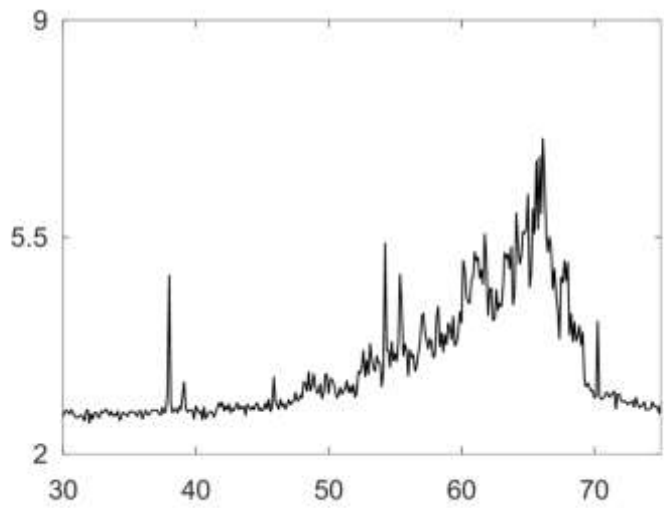 | 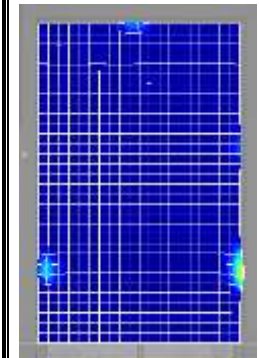 |
| 22 | 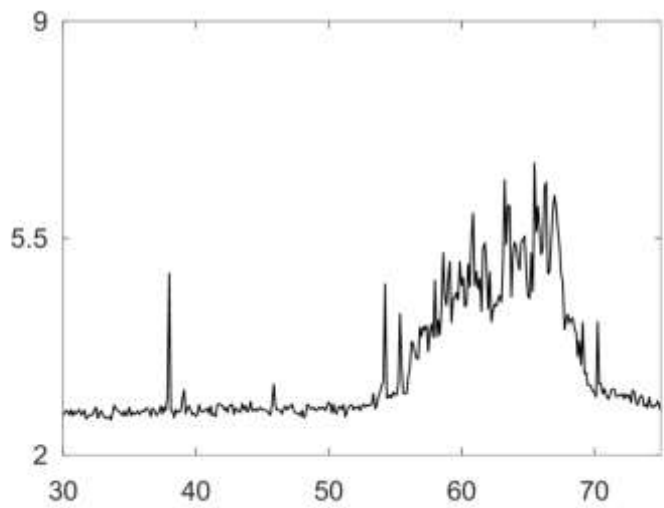 | 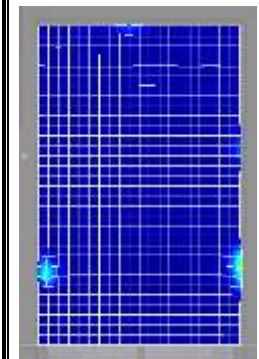 |

| 23 | 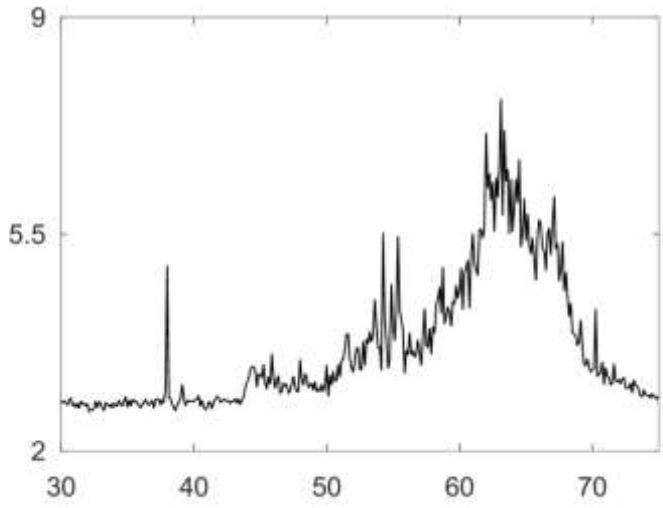 | 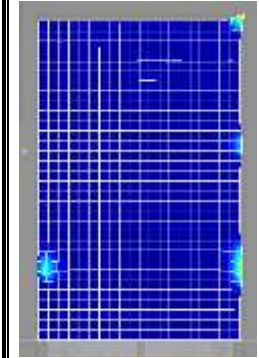 |
|---|---|---|
| 24 | 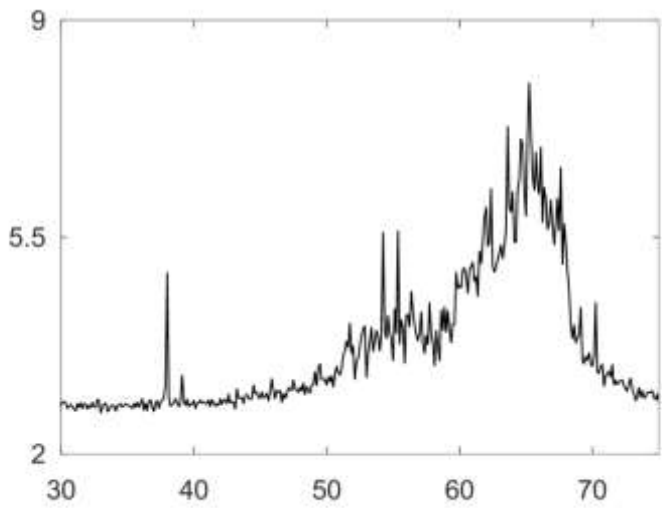 | 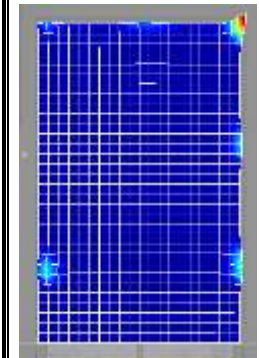 |
| 25 | 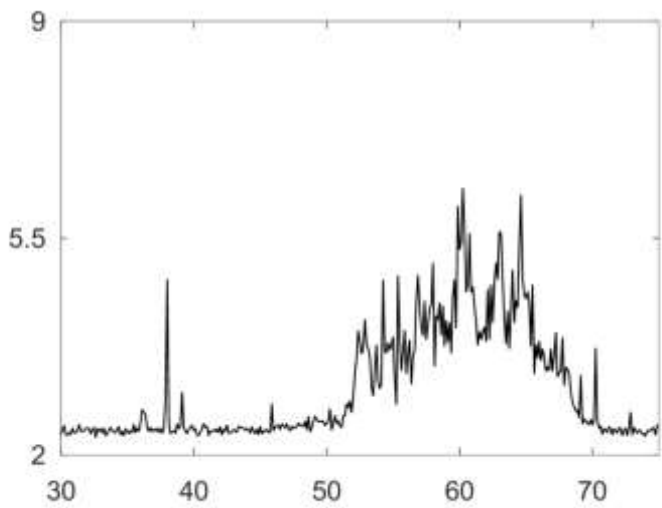 | 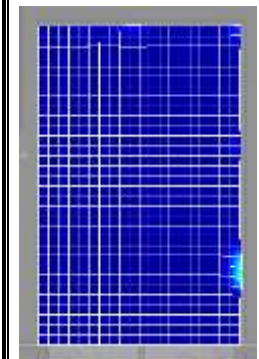 |

| 26 | 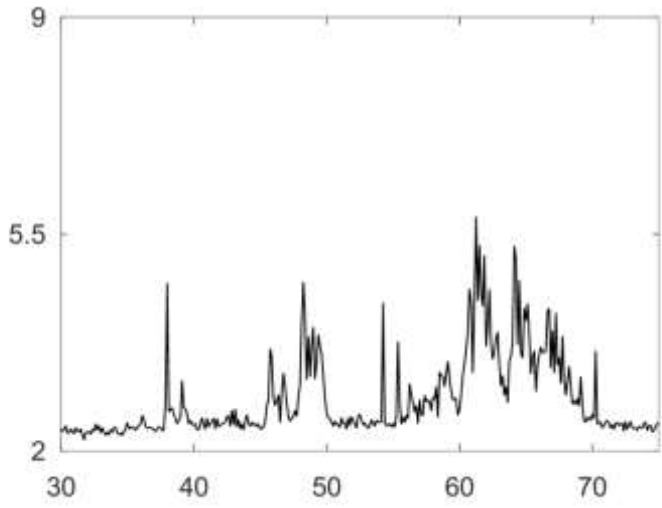 | 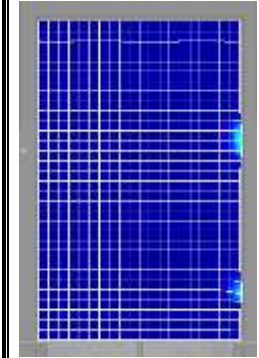 |
|---|---|---|
| 27 | 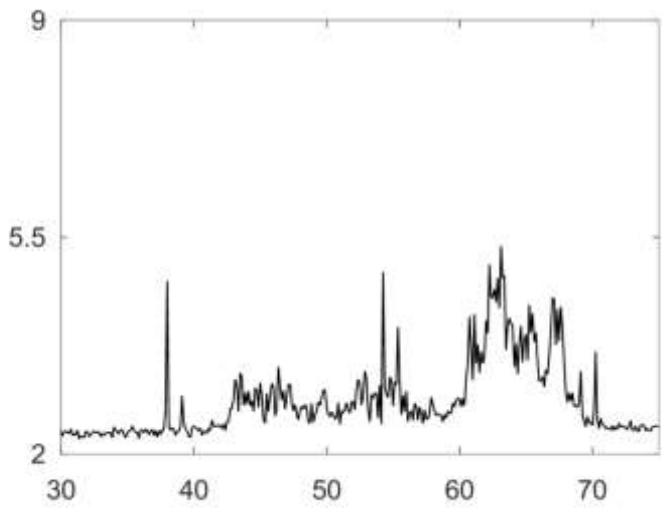 | 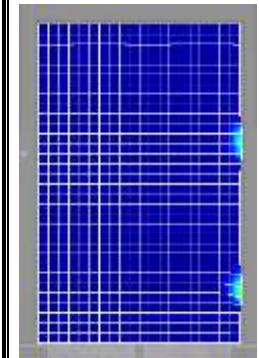 |
| 28 | 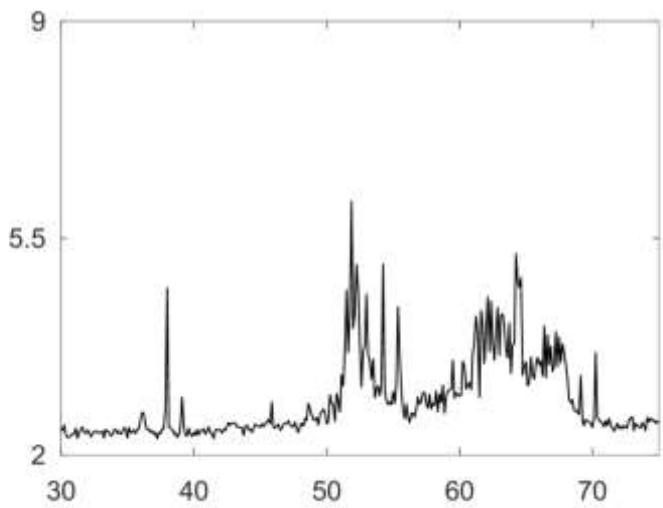 | 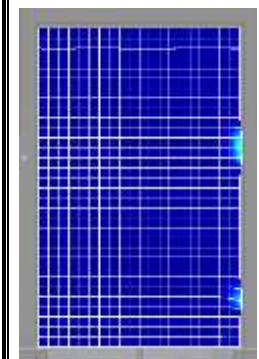 |

| 29 | 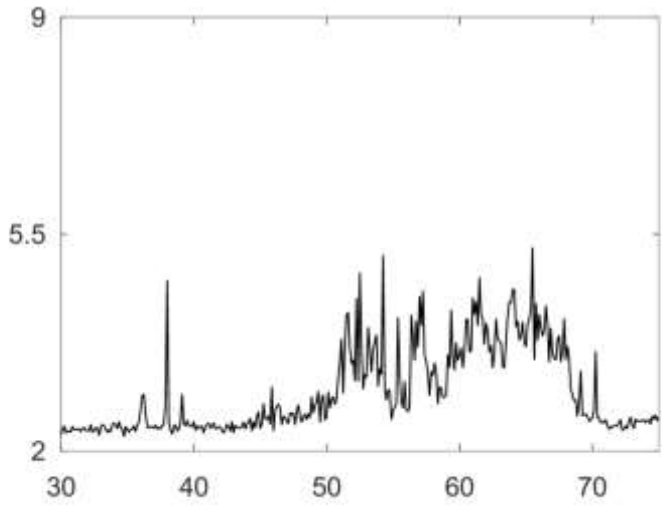 | 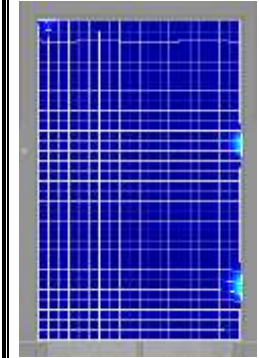 |
| 30 | 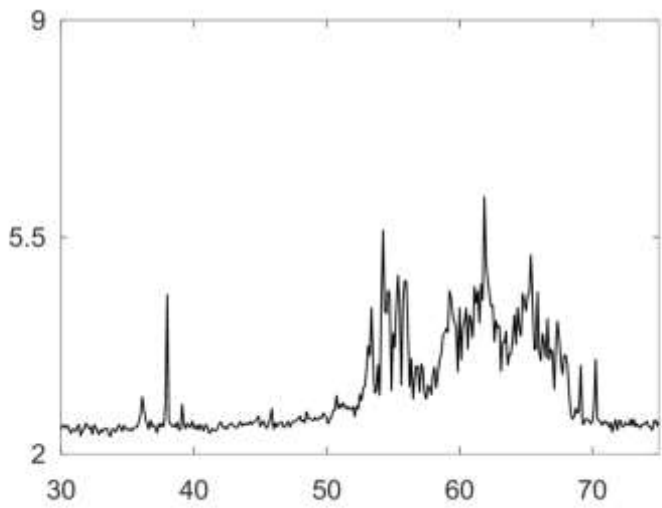 | 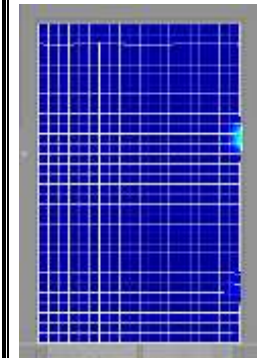 |
| 31 | 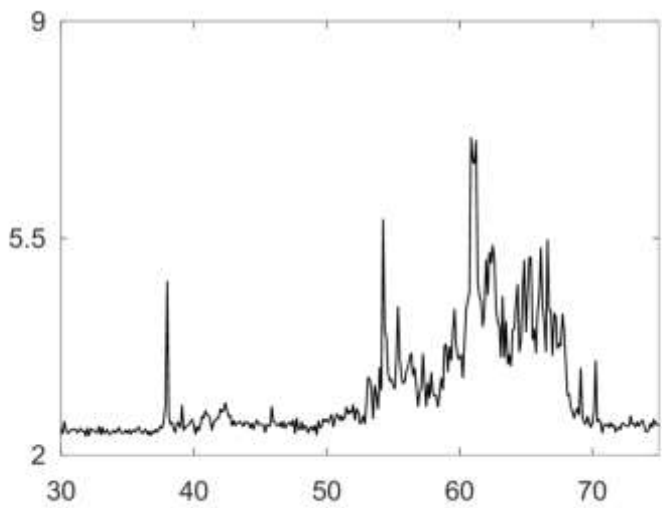 | 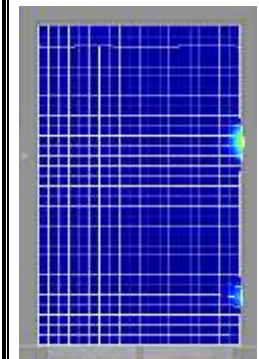 |

| 32 | 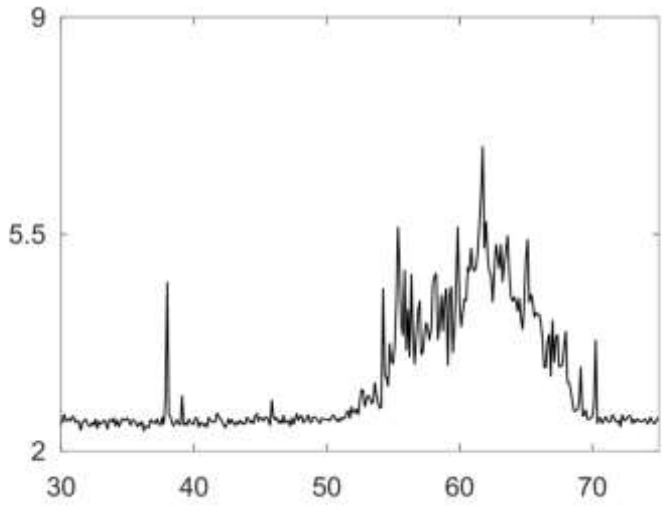 | 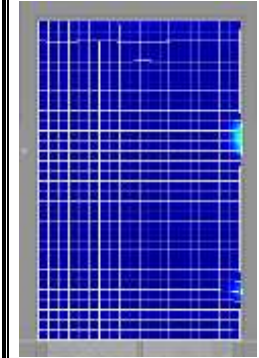 |
|---|---|---|
| 33 | 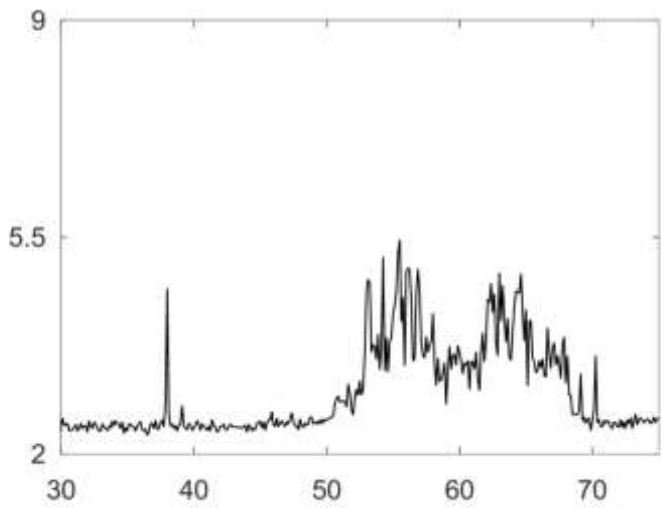 | 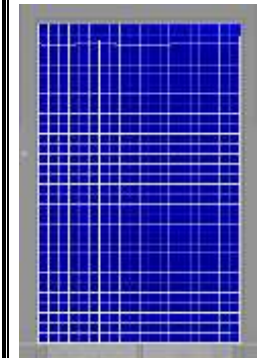 |
| 34 | 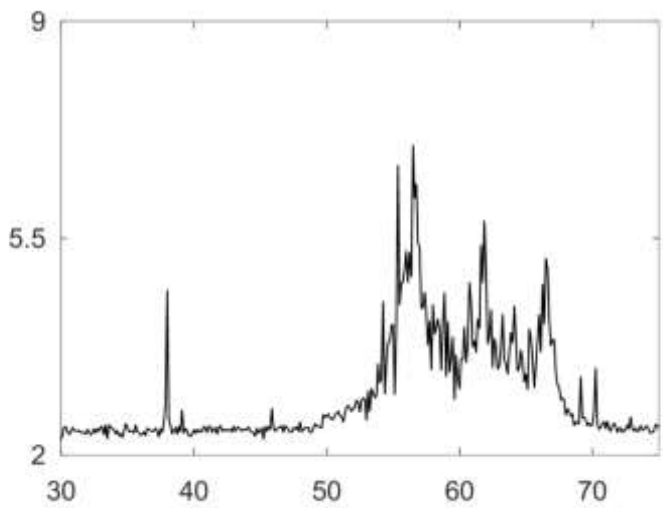 | 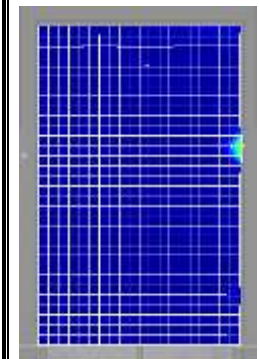 |

| 35 | 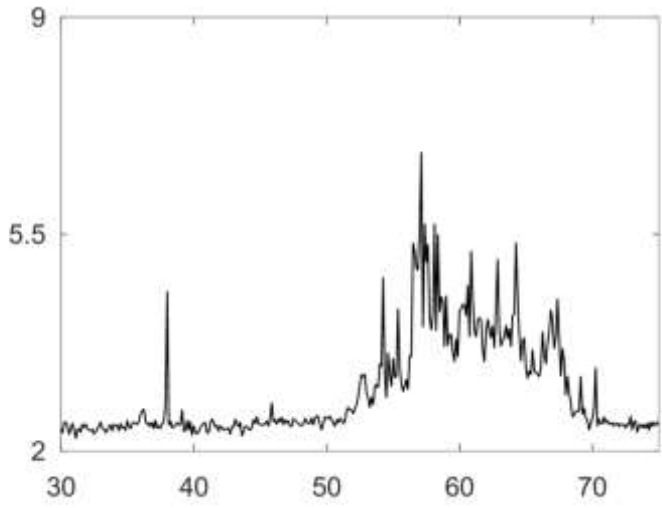 | 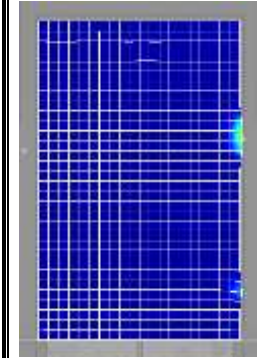 |
| --- | --- | --- |
| 36 | 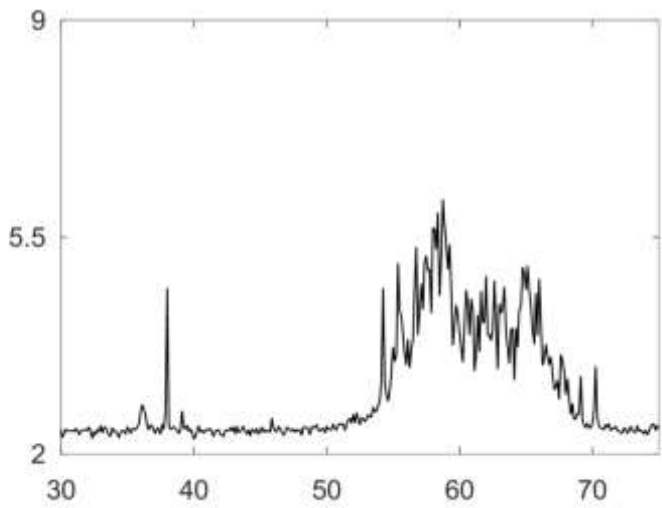 | 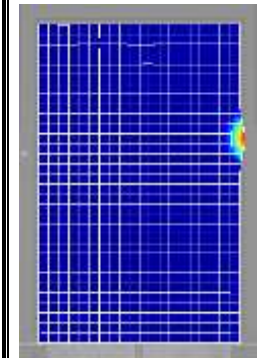 |
| 37 | 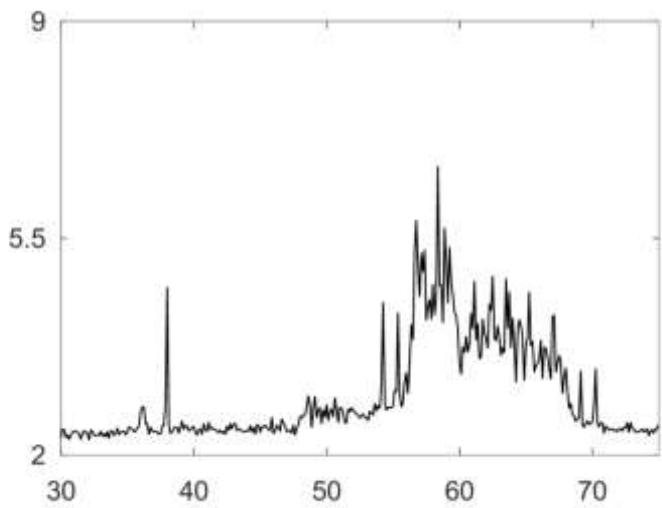 | 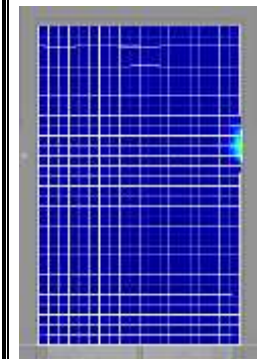 |

| 38 | 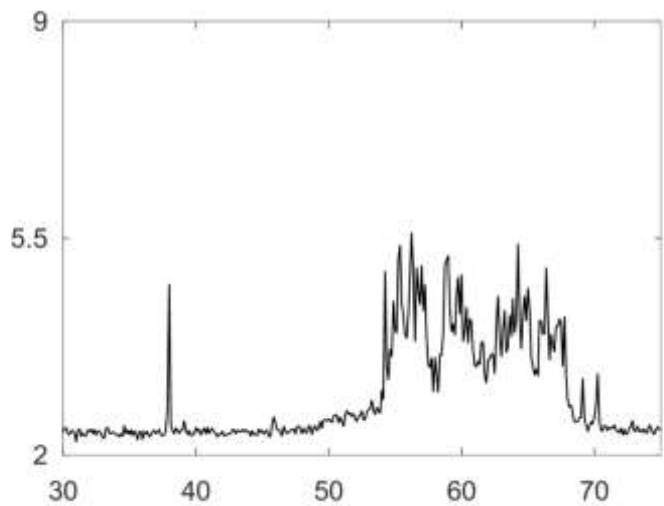 | 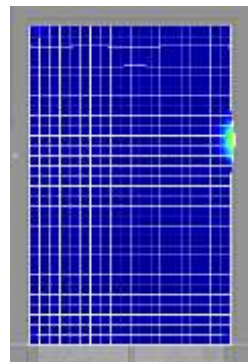 |
| --- | --- | --- |
| 39 | 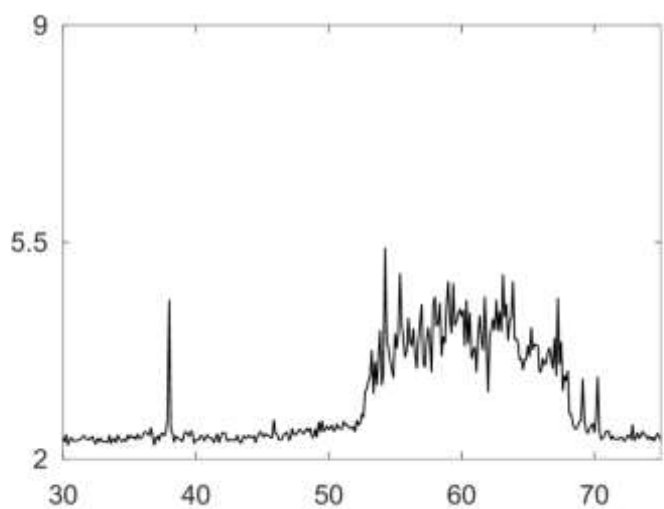 | 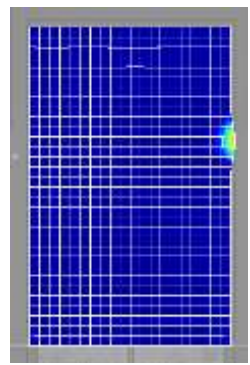 |
| 40 | 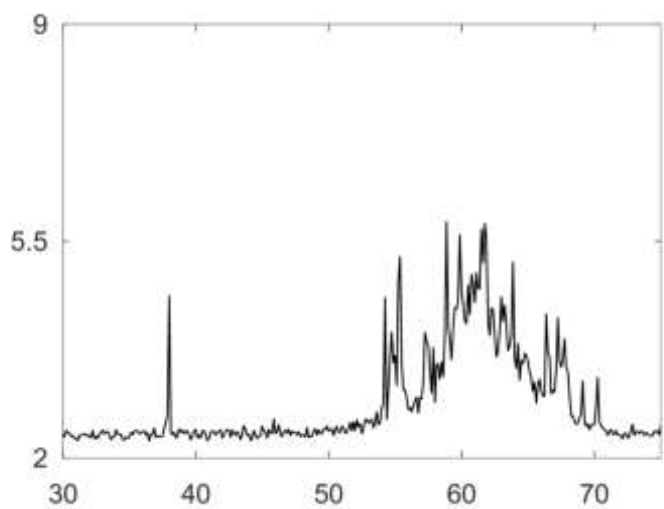 | 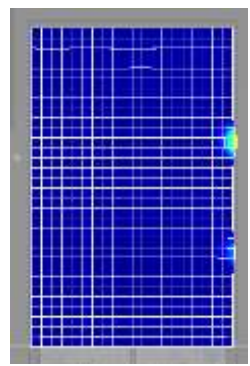 |

| 41 | 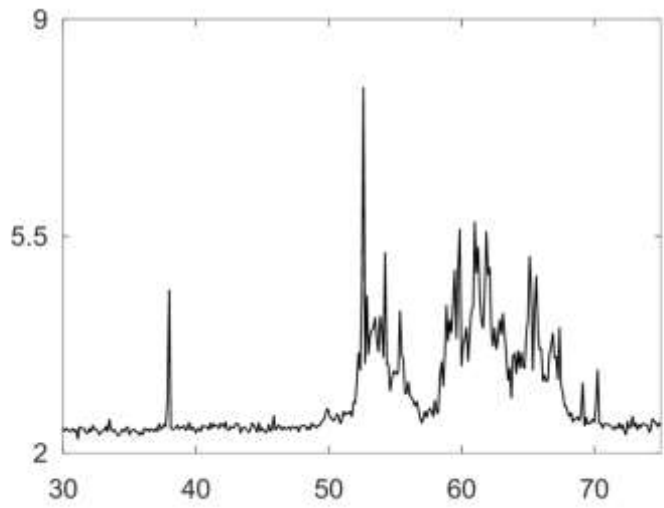 | 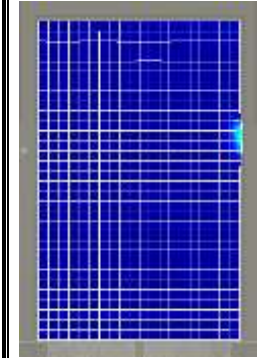 |
|---|---|---|
| 42 | 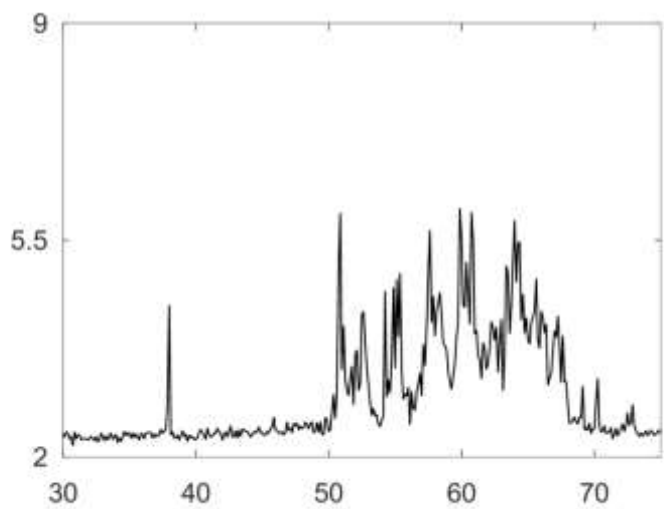 | 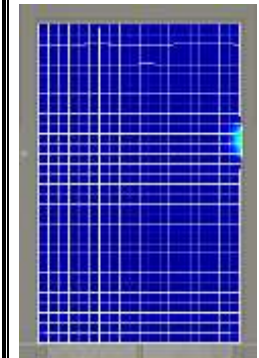 |
| 43 | 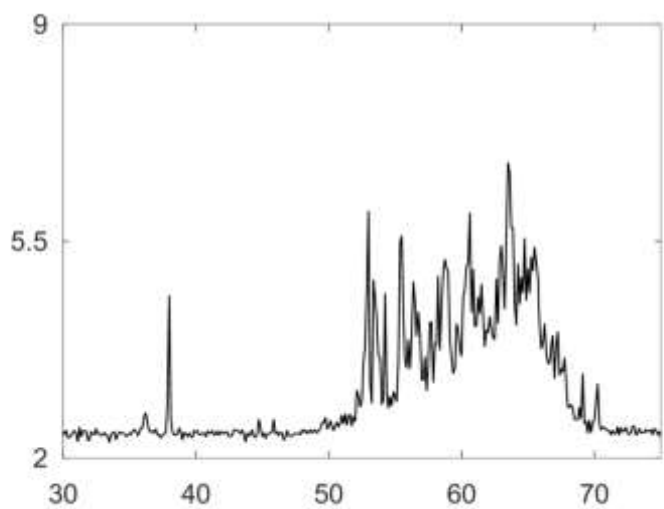 | 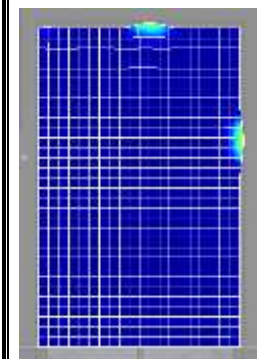 |

| 44 | 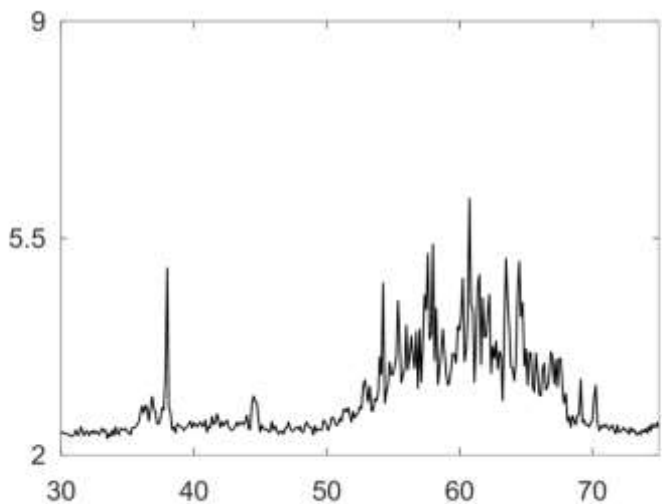 | 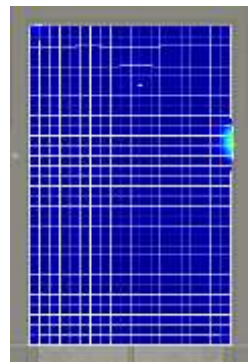 |
|---|---|---|
| 45 | 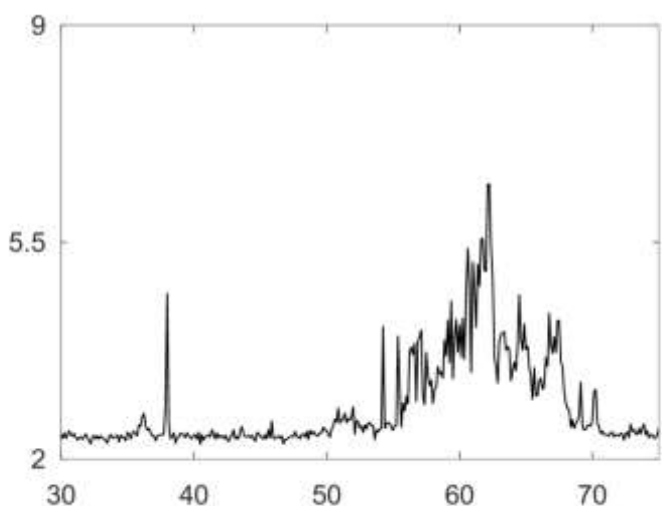 | 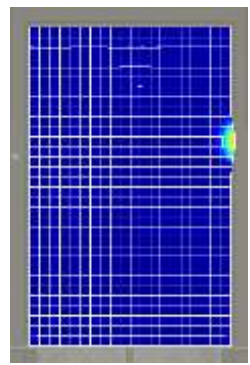 |
| 46 | 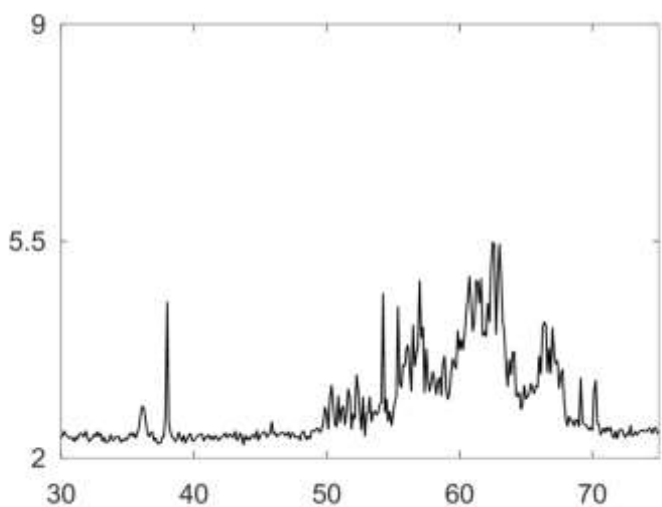 | 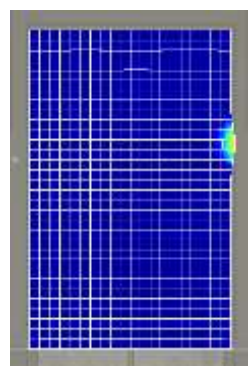 |

| 47 | 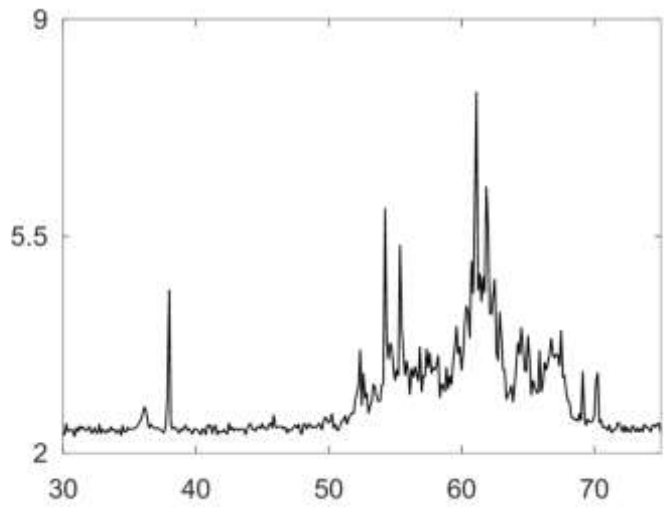 | 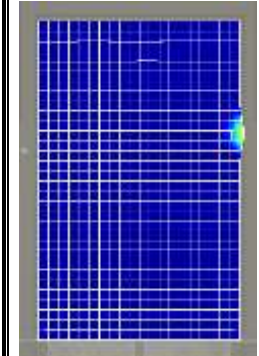 |
|---|---|---|
| 48 | 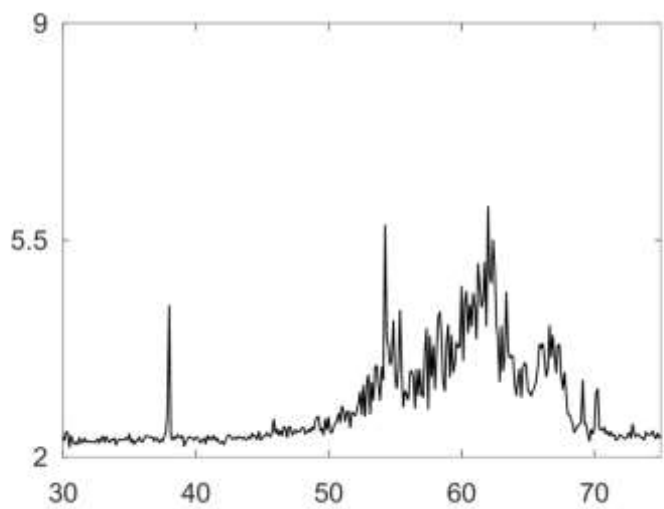 | 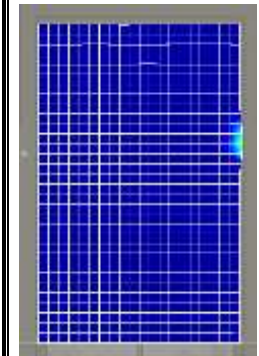 |
| 49 | 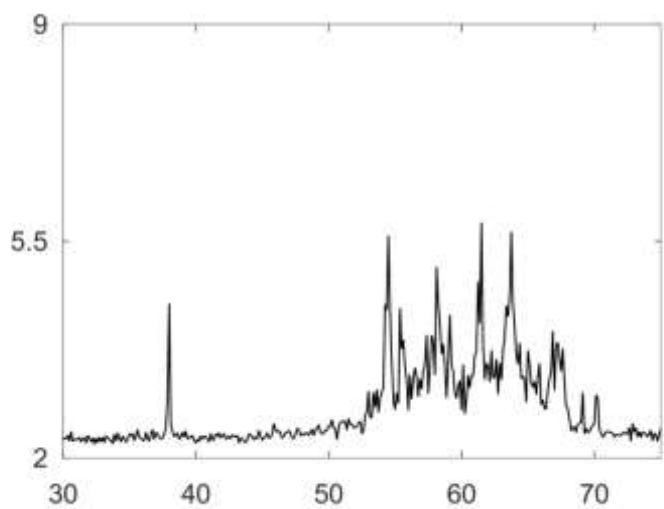 | 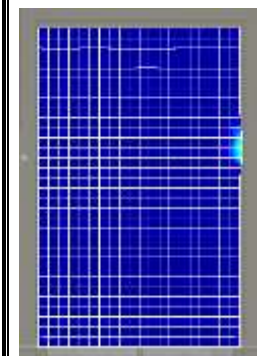 |

| 50 | 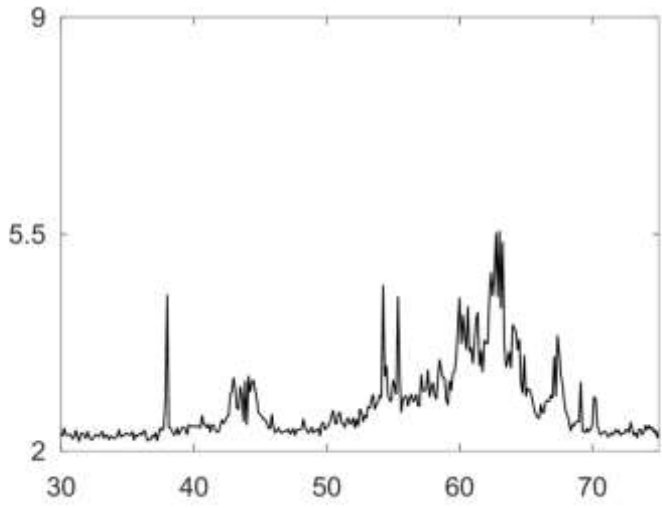 | 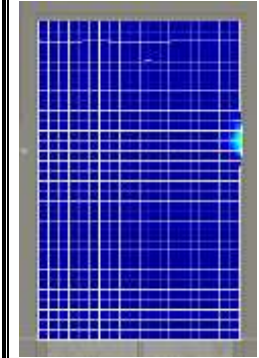 |
|---|---|---|
| 51 | 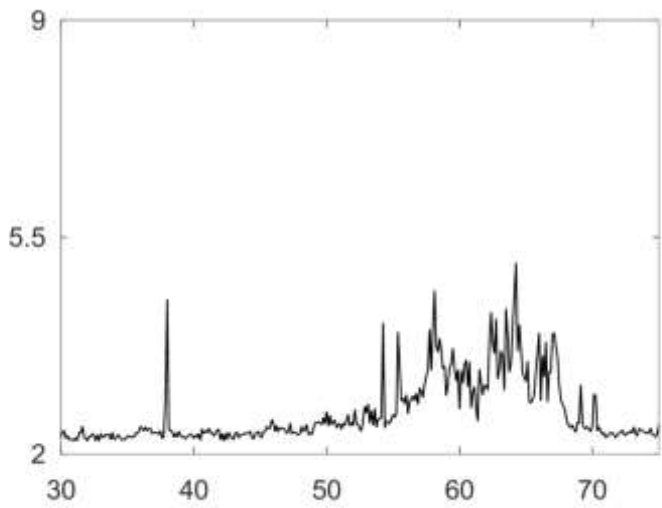 | 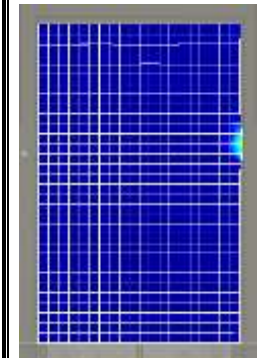 |
| 52 | 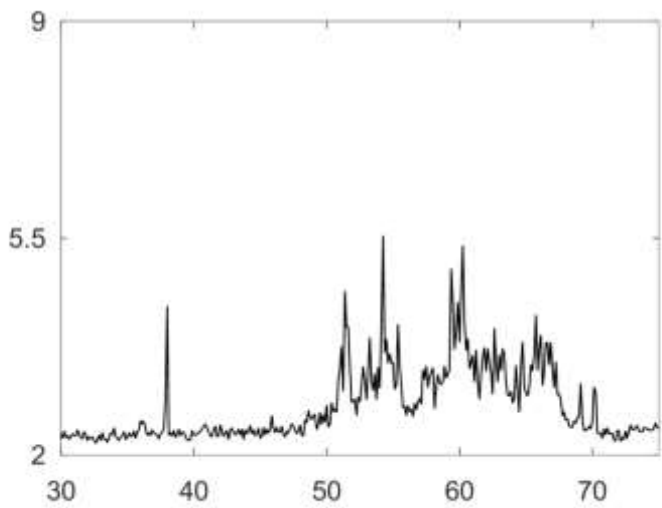 | 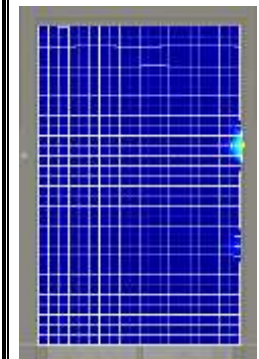 |

Run 3

| Time | Displacement (in pm) vs. Frequency (in kHz) | RMS Surface Displacment 1.2 nm — 0 nm |
|---|---|---|
| 1 | 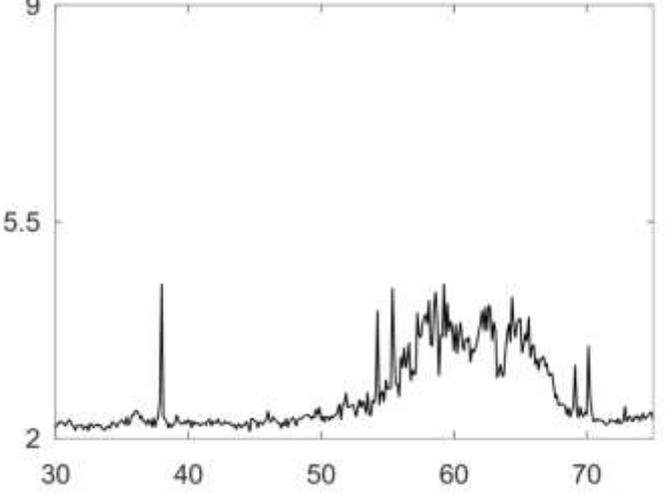 | 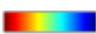 |
| 2 | 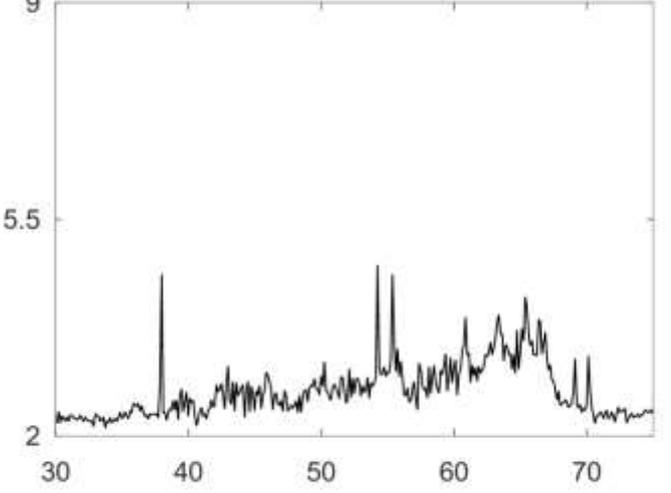 | 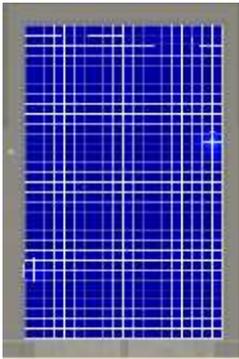 |

| 3 | 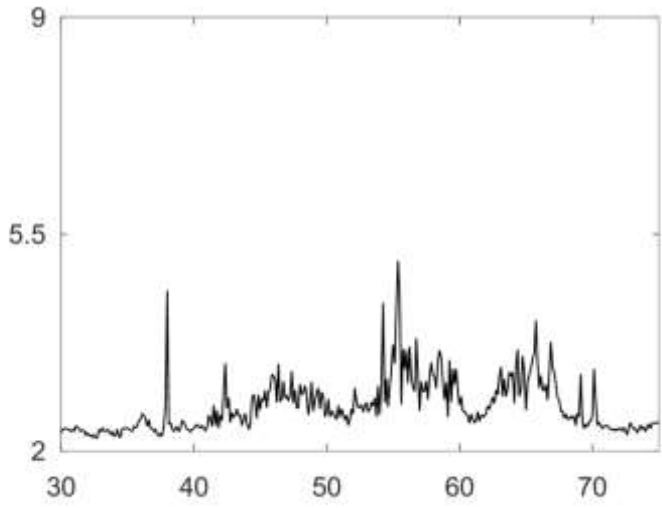 | 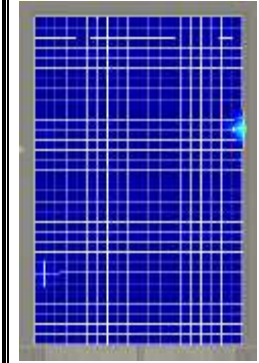 |
| 4 | 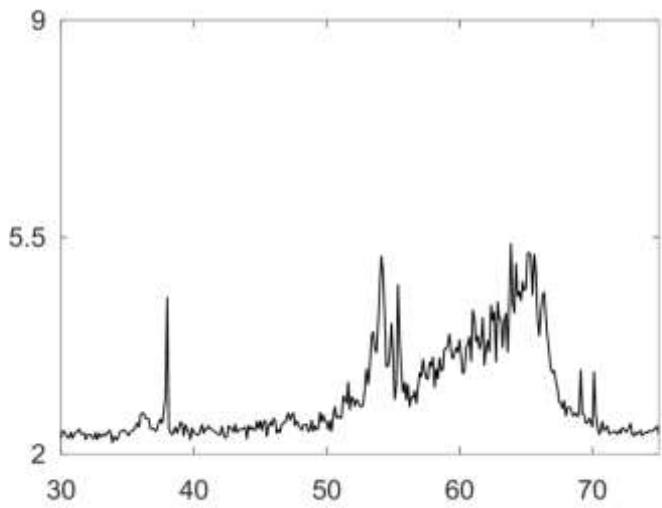 | 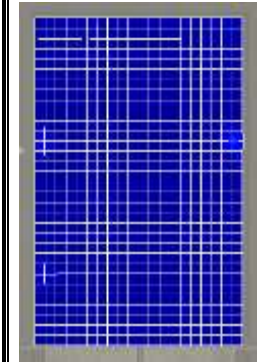 |
| 5 | 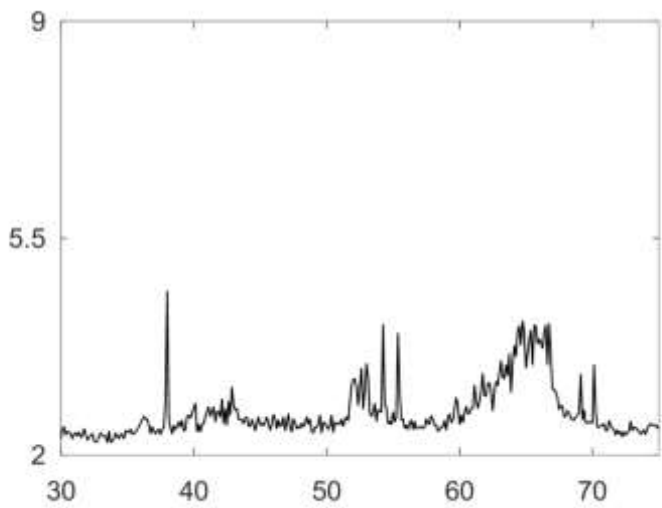 | 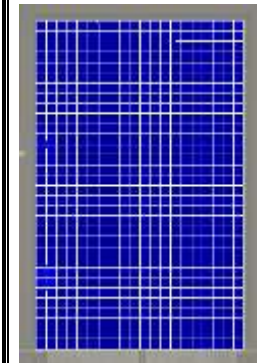 |

| 6 | 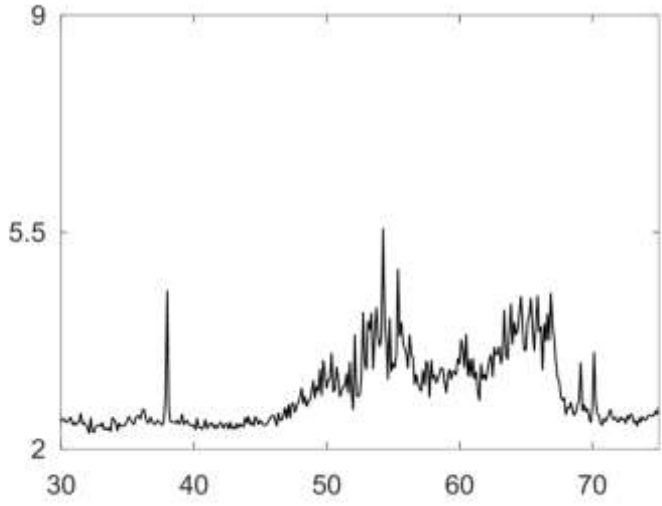 | 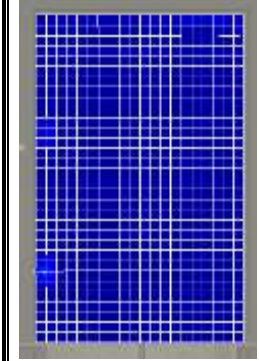 |
| --- | --- | --- |
| 7 | 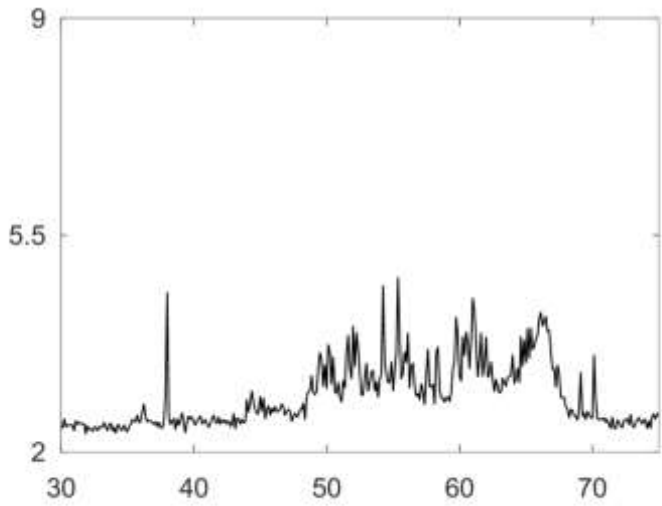 | 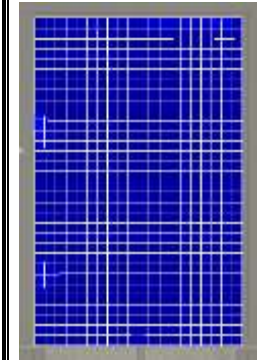 |
| 8 | 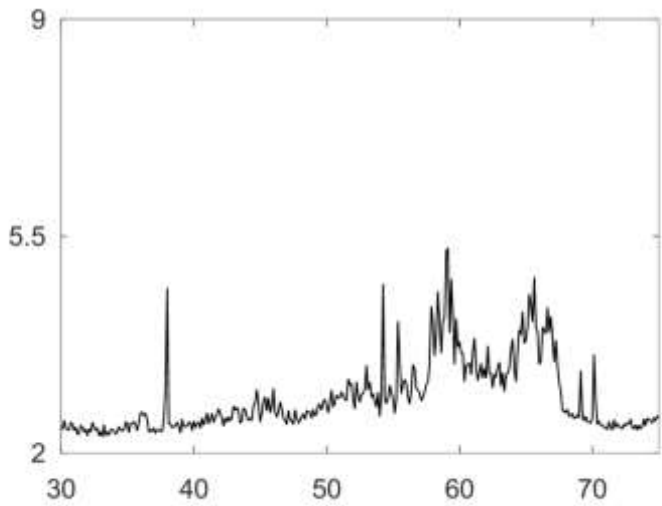 | 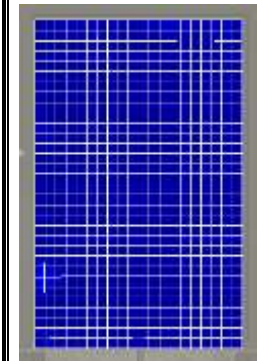 |

| 9 | 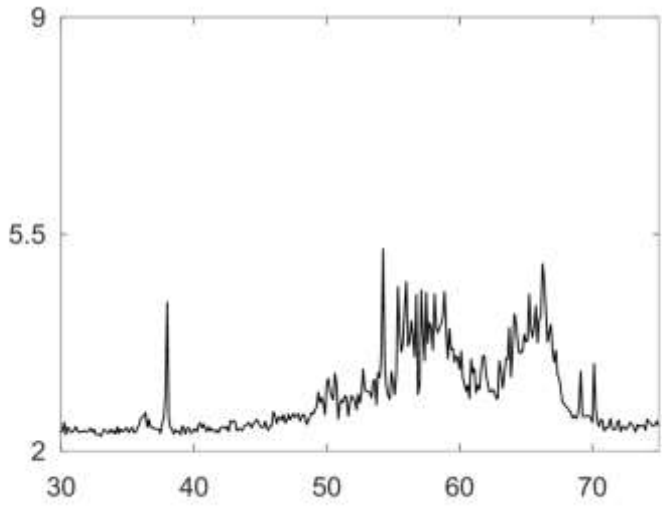 | 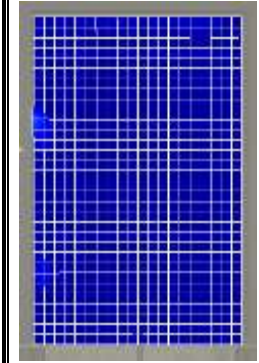 |
|---|---|---|
| 10 | 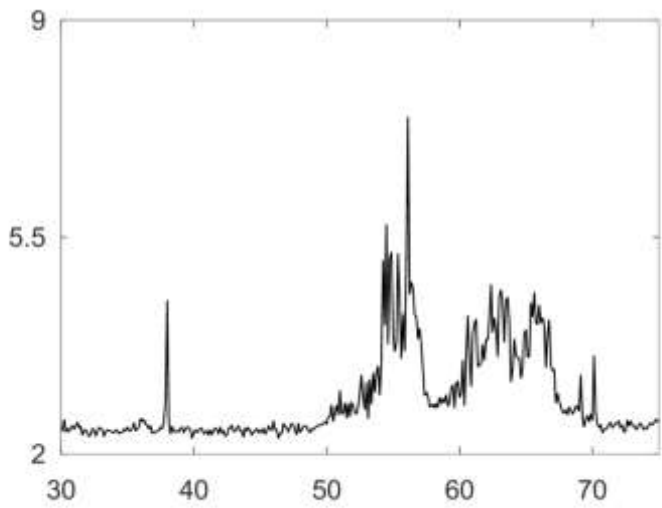 | 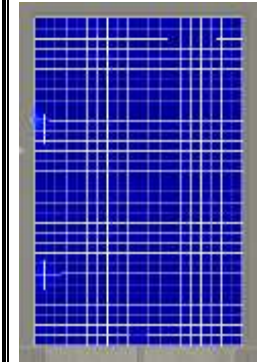 |
| 11 | 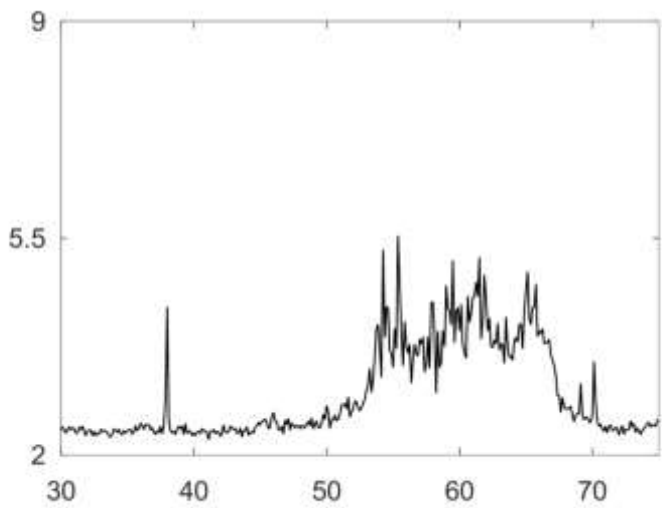 | 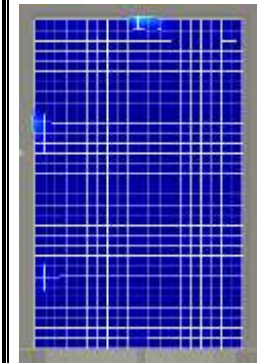 |

| 12 | 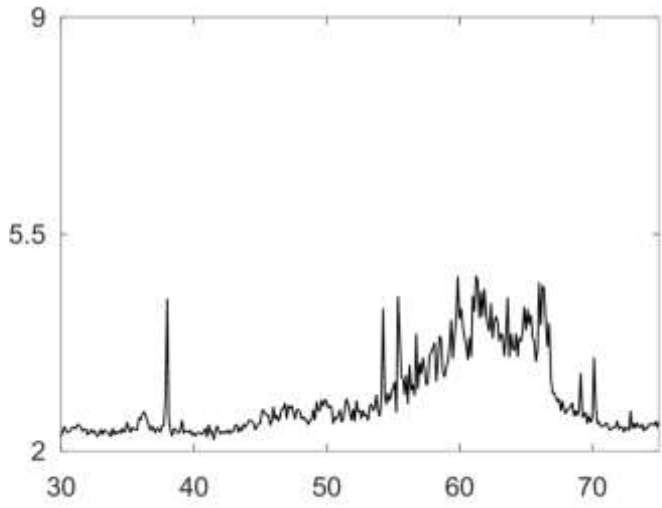 | 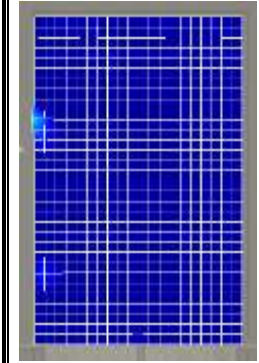 |
| --- | --- | --- |
| 13 | 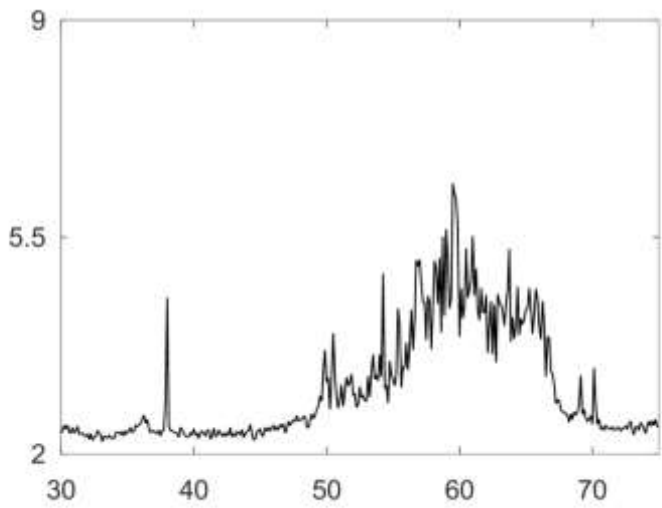 | 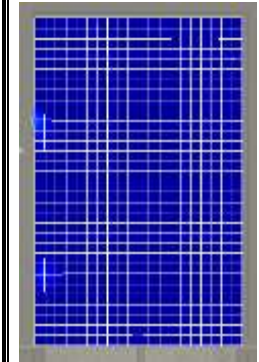 |
| 14 | 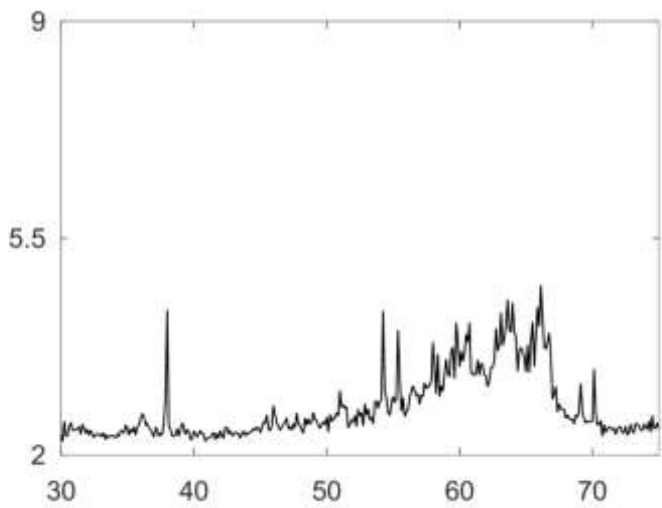 | 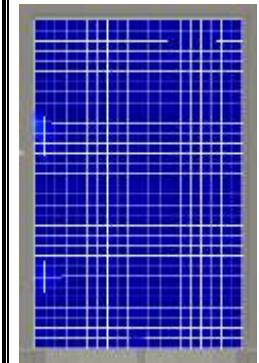 |

| 15 | 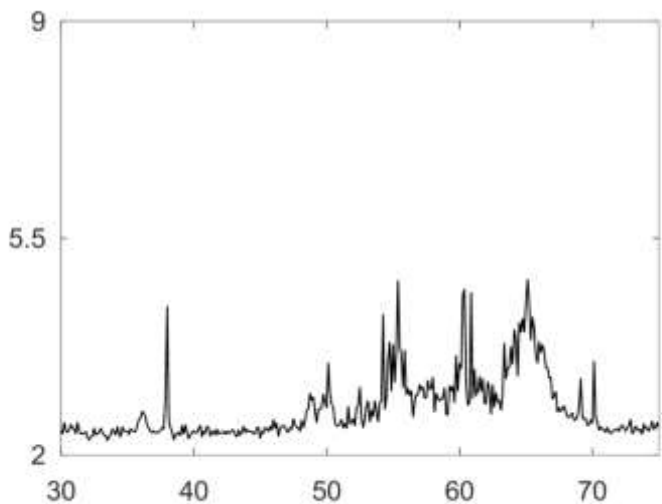 | 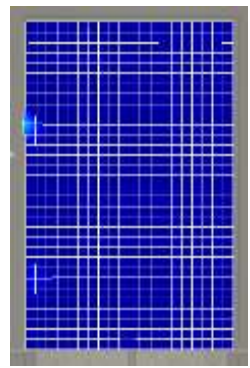 |
| --- | --- | --- |
| 16 | 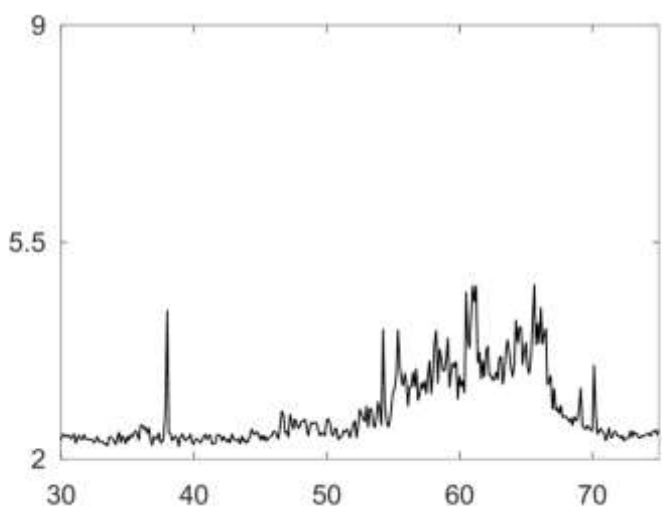 | 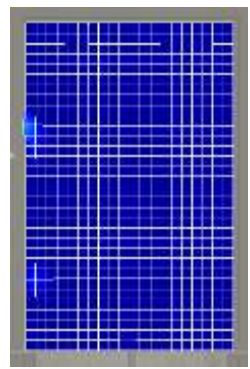 |
| 17 | 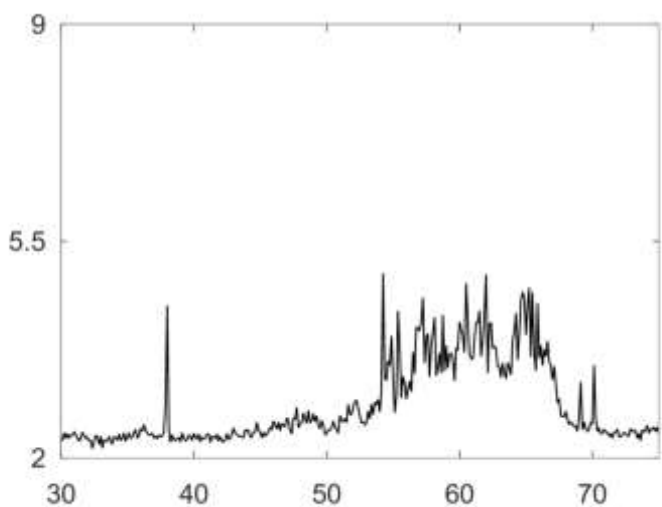 | 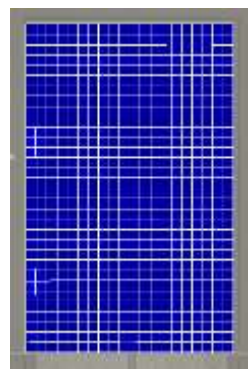 |

| 18 | 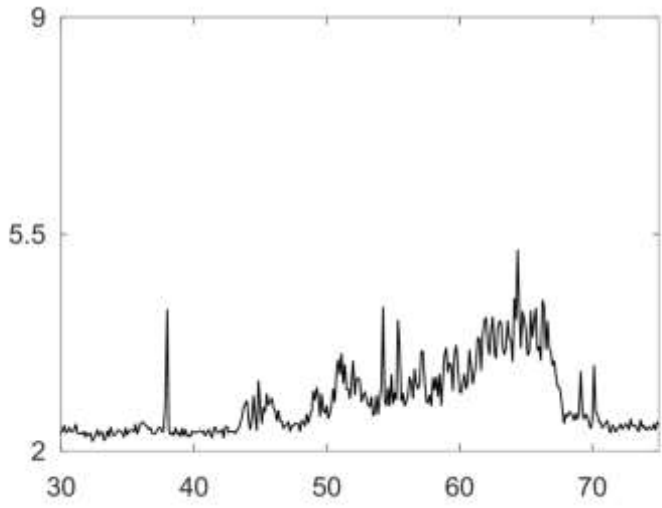 | 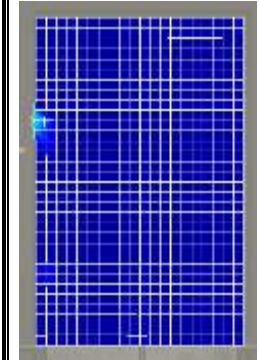 |
|---|---|---|
| 19 | 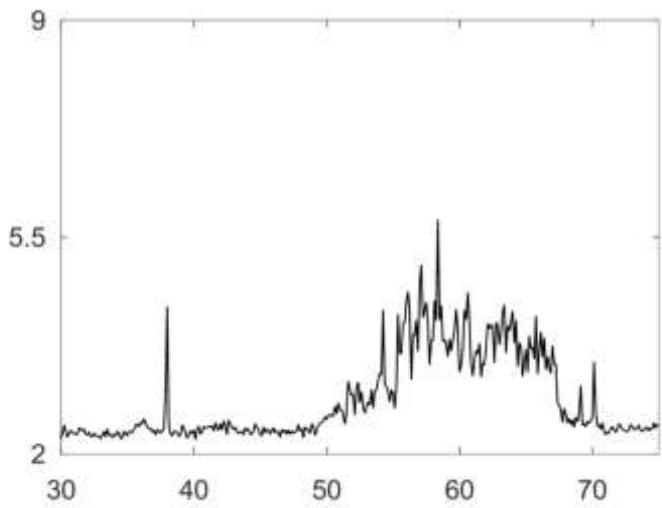 | 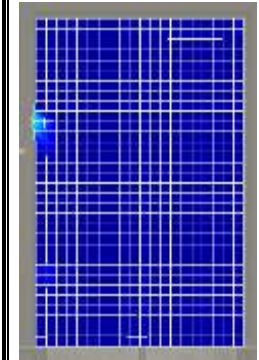 |
| 20 | 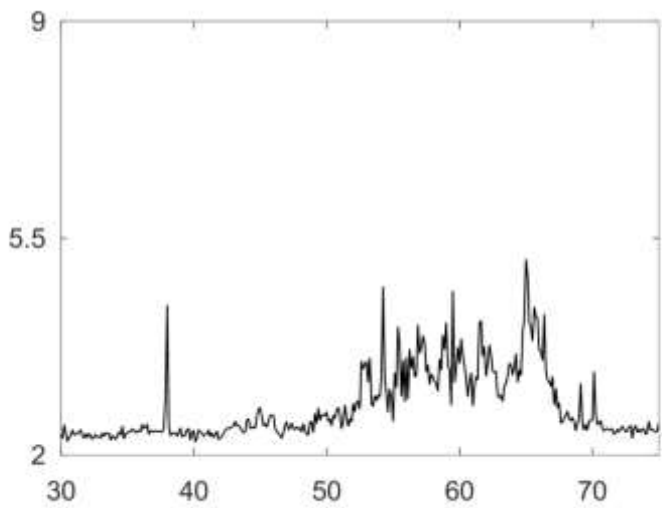 | 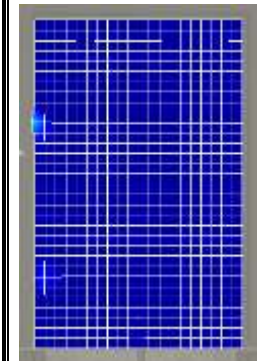 |

| 21 | 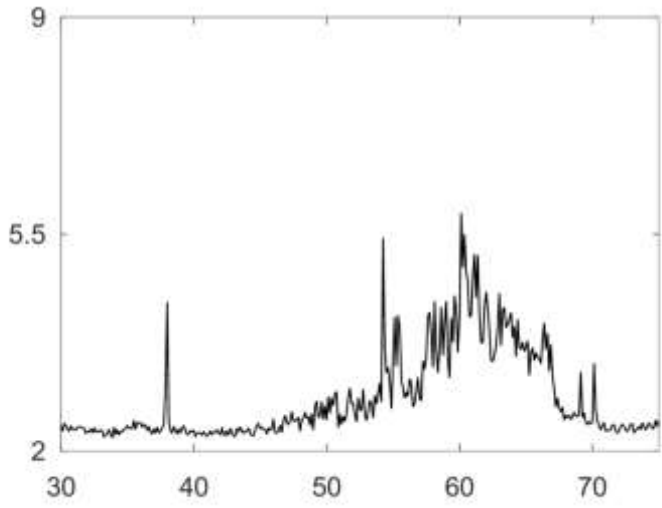 | 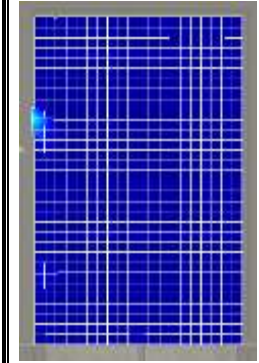 |
| 22 | 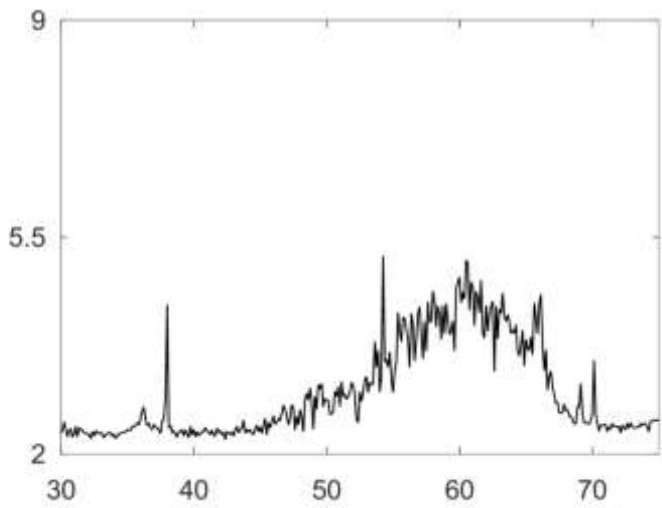 | 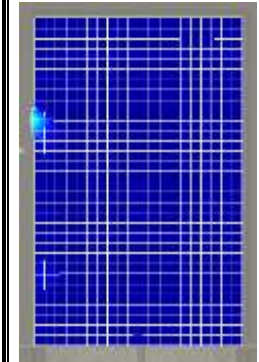 |
| 23 | 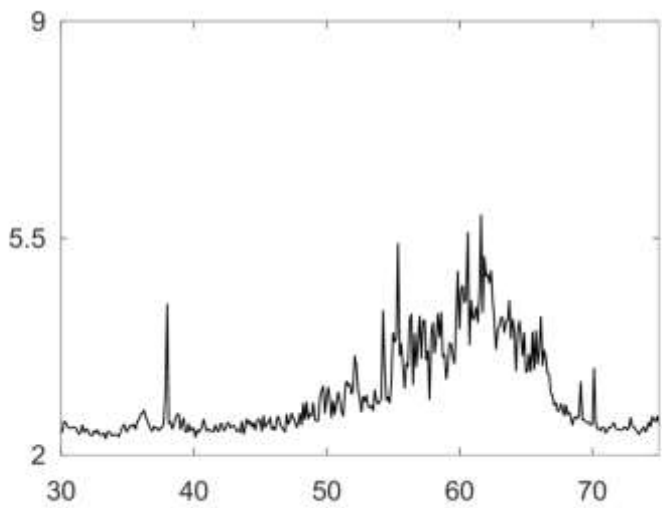 | 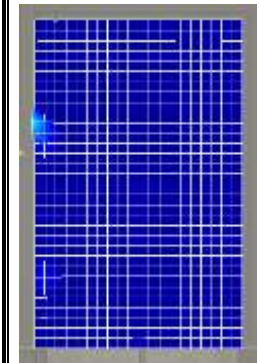 |

| 24 | 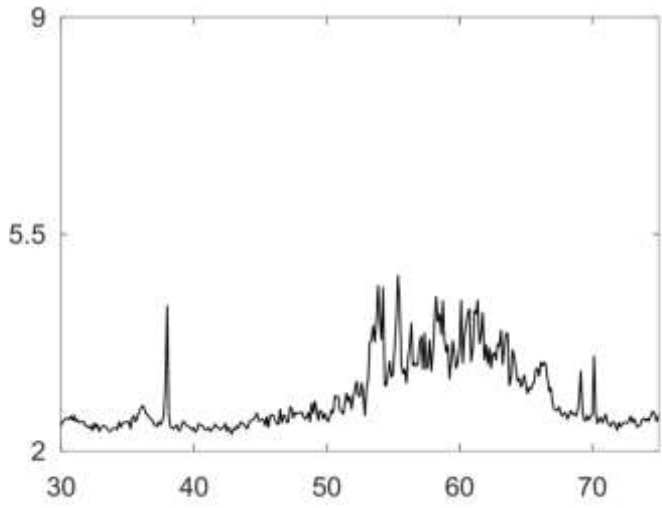 | 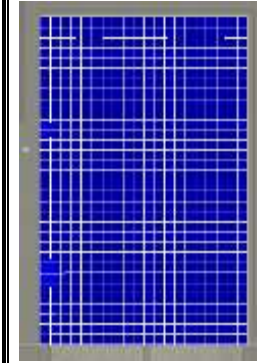 |
|---|---|---|
| 25 | 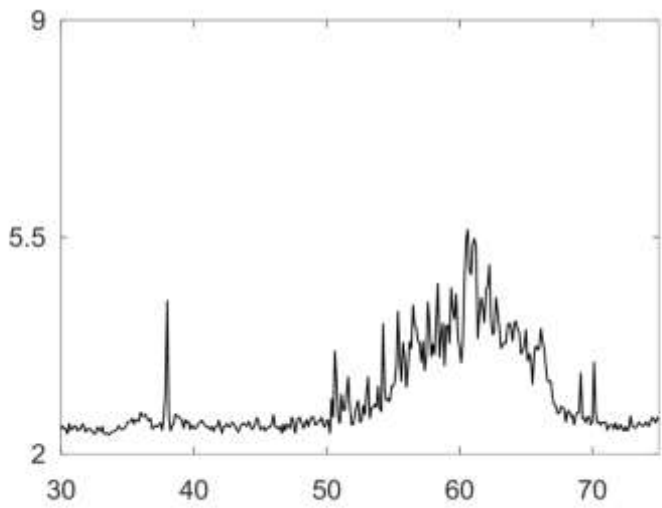 | 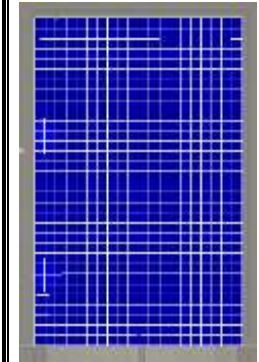 |
| 26 | 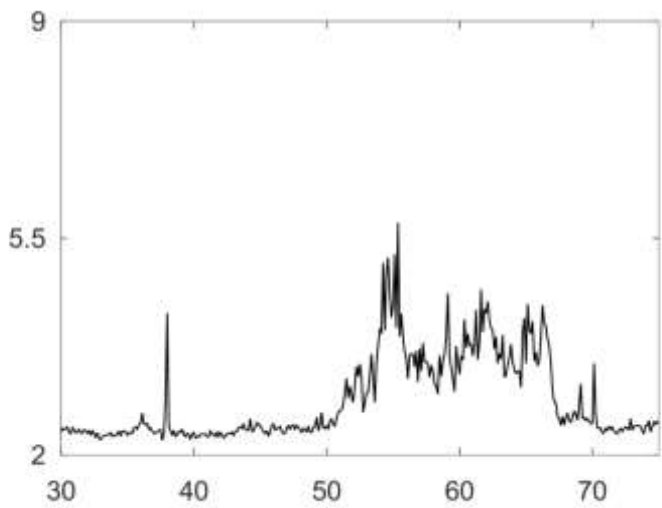 | 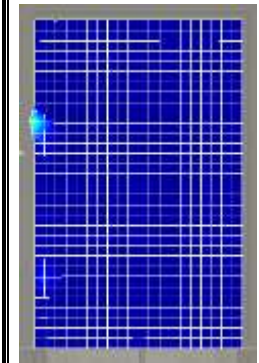 |

| 27 | 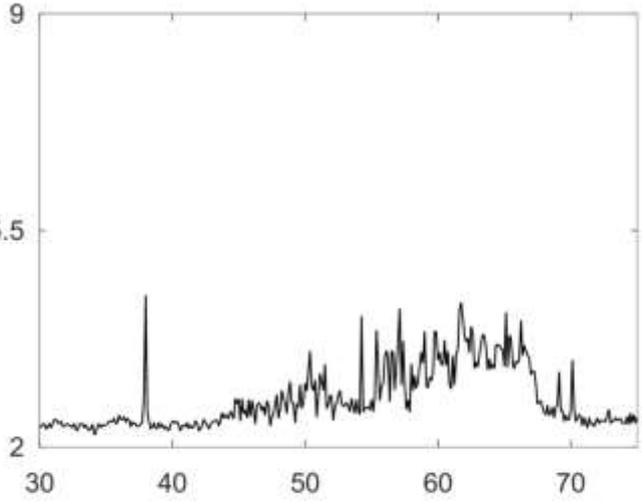 | 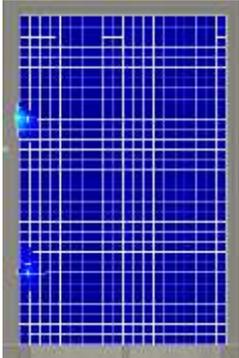 |
|---|---|---|
| 28 | 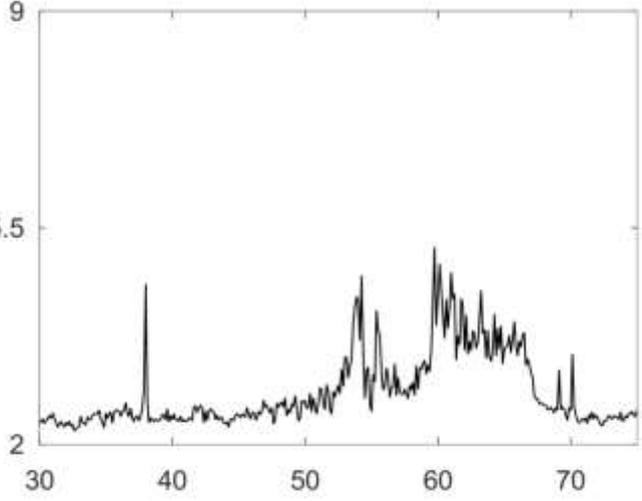 | 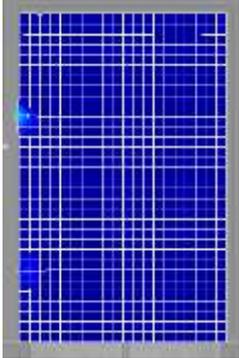 |
| 29 | 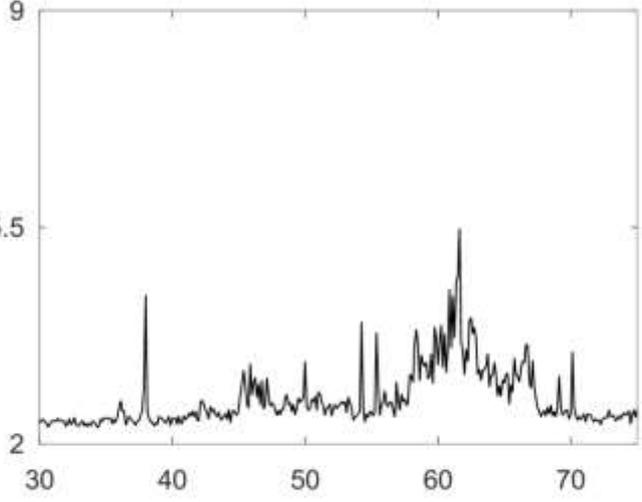 | 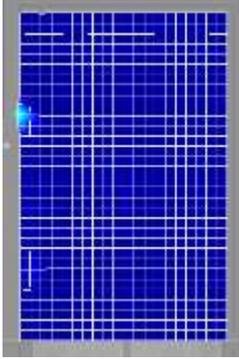 |

| 30 | 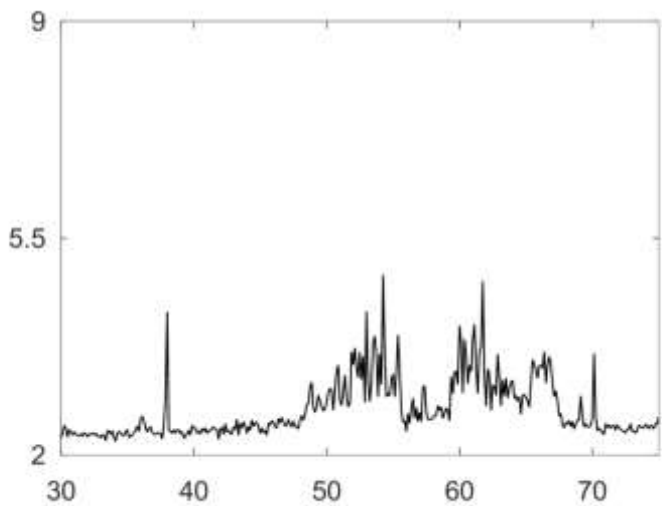 | 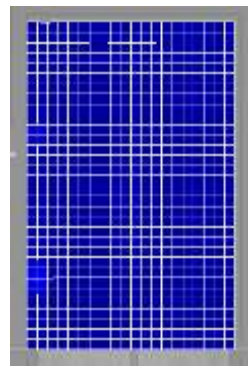 |
| --- | --- | --- |
| 31 | 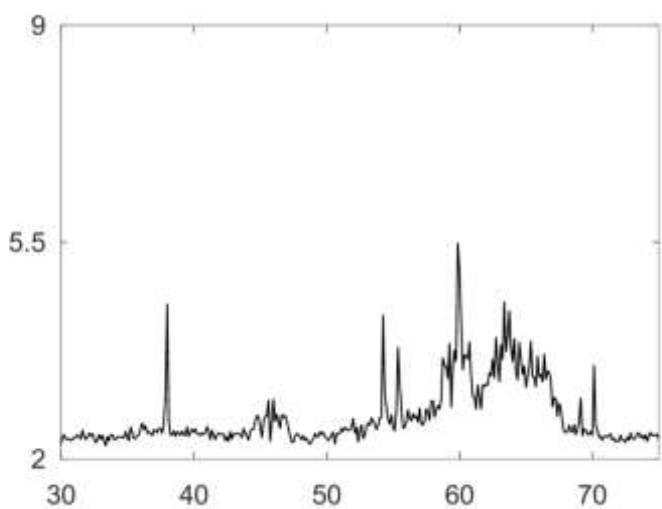 | 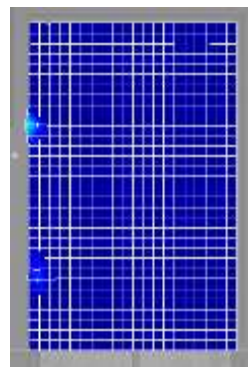 |
| 32 | 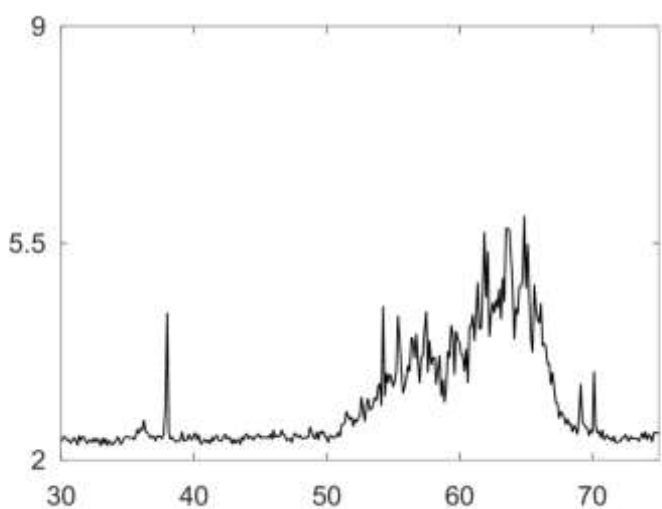 | 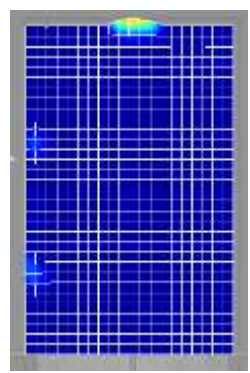 |

| 33 | 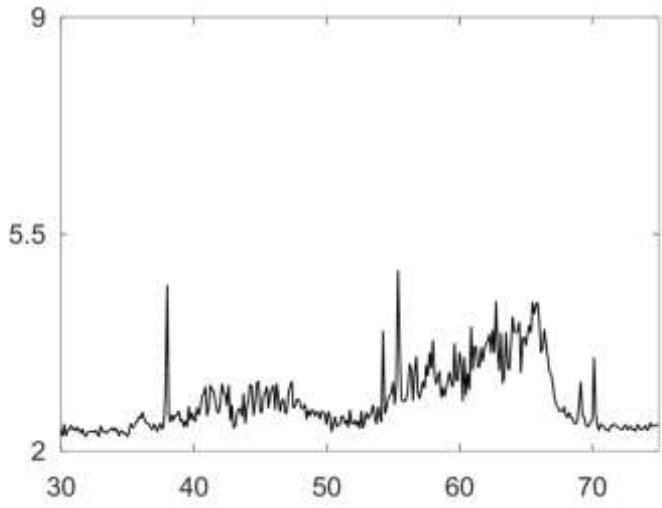 | 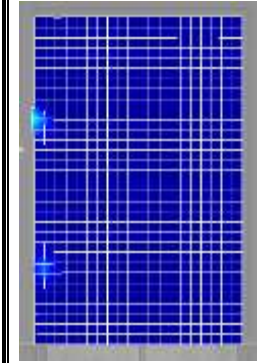 |
|---|---|---|
| 34 | 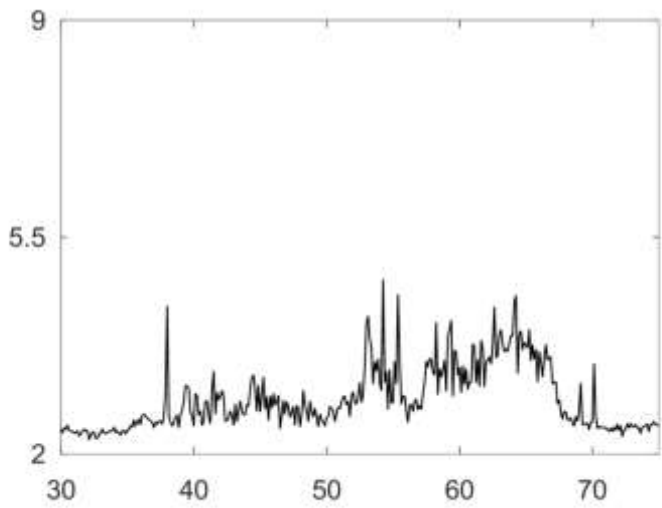 | 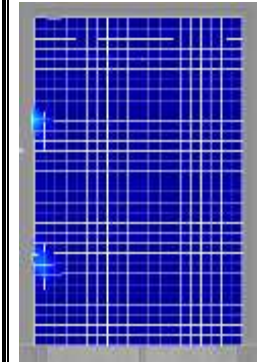 |
| 35 | 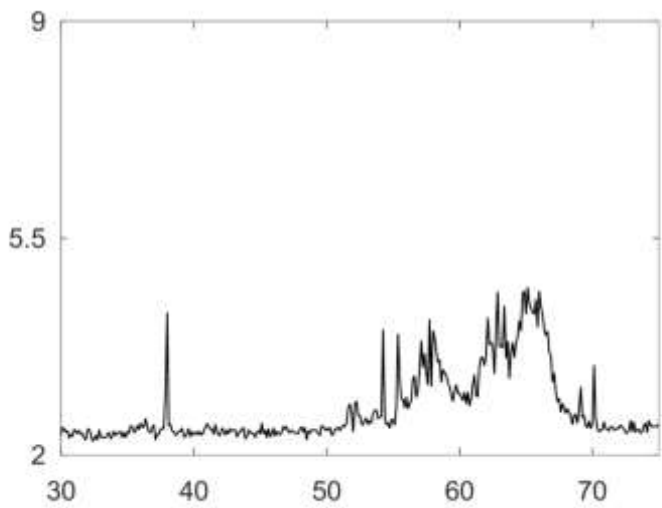 | 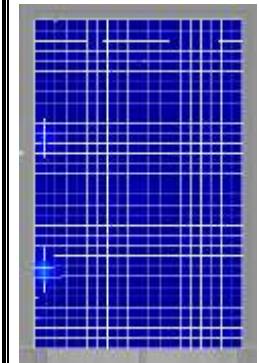 |

| 36 | 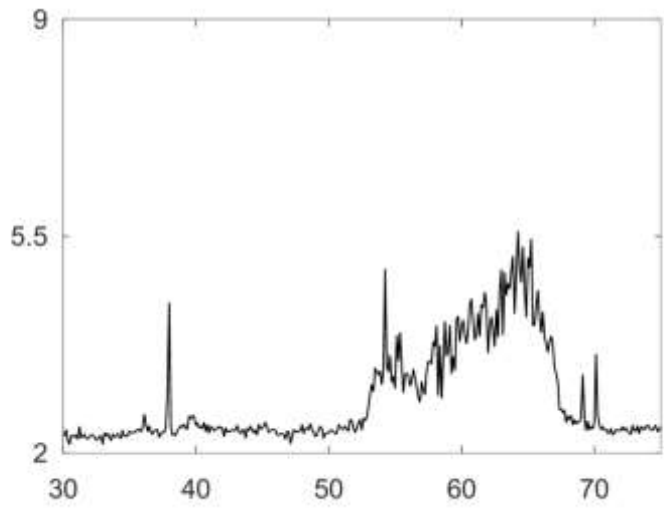 | 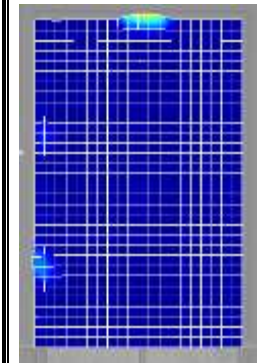 |
|---|---|---|
| 37 | 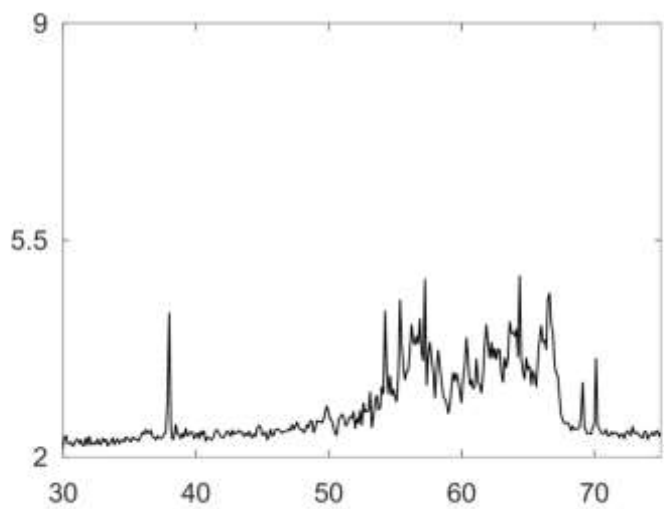 | 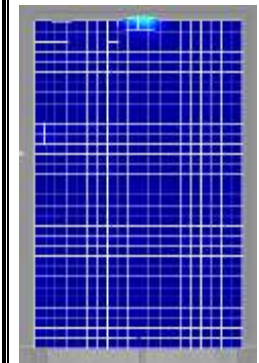 |
| 38 | 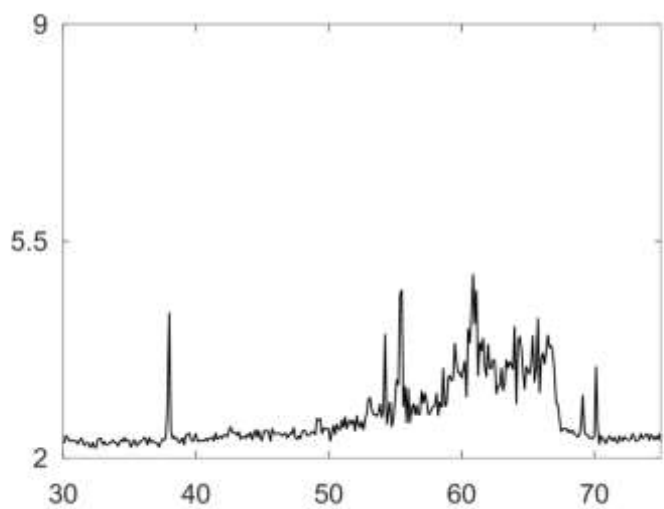 | 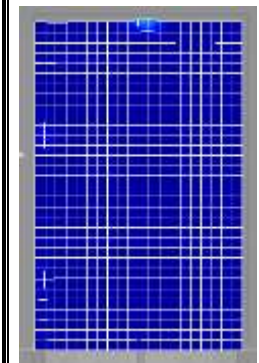 |

| 39 | 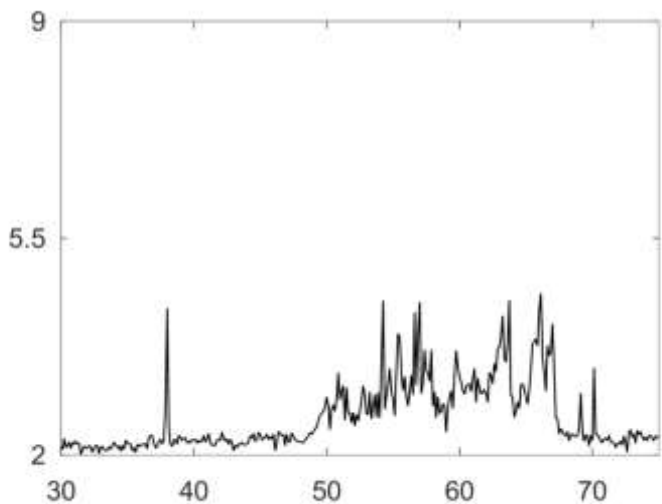 | 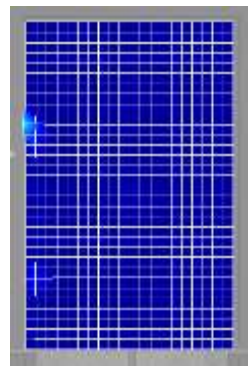 |
| --- | --- | --- |
| 40 | 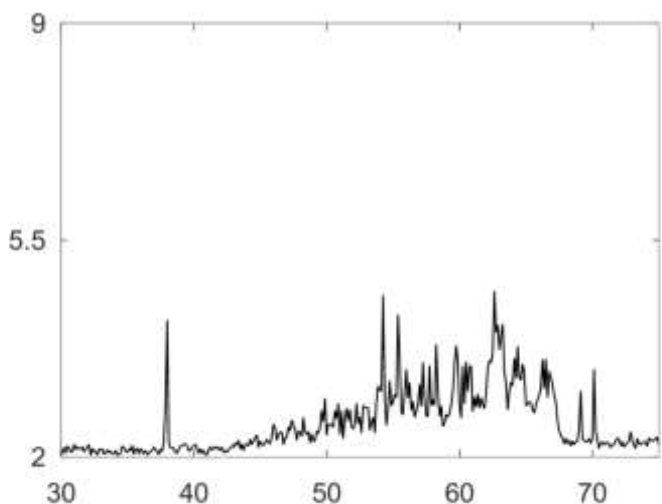 | 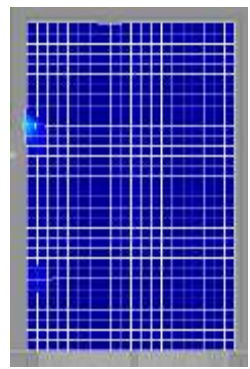 |
| 41 | 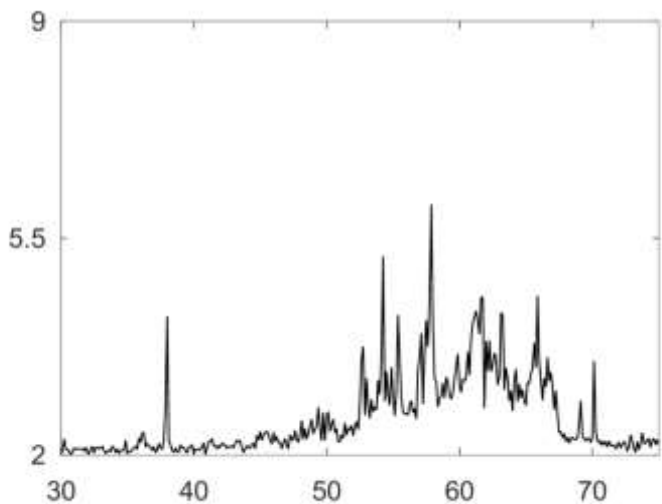 | 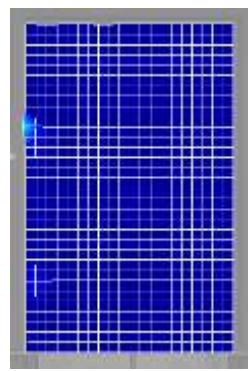 |
| 42 | 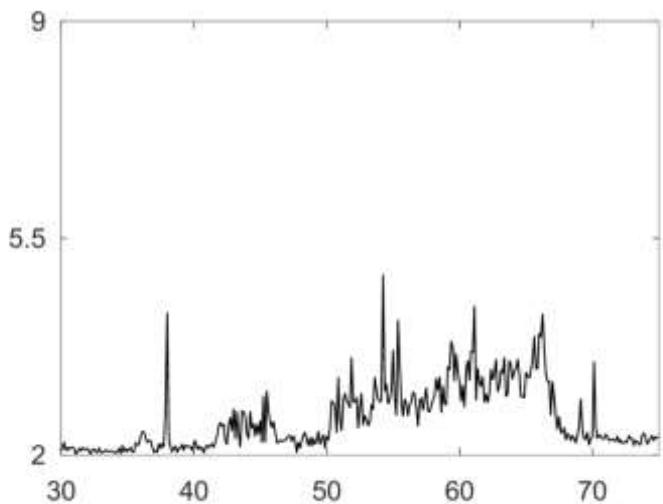 | 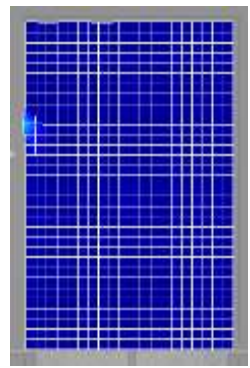 |
| 43 | 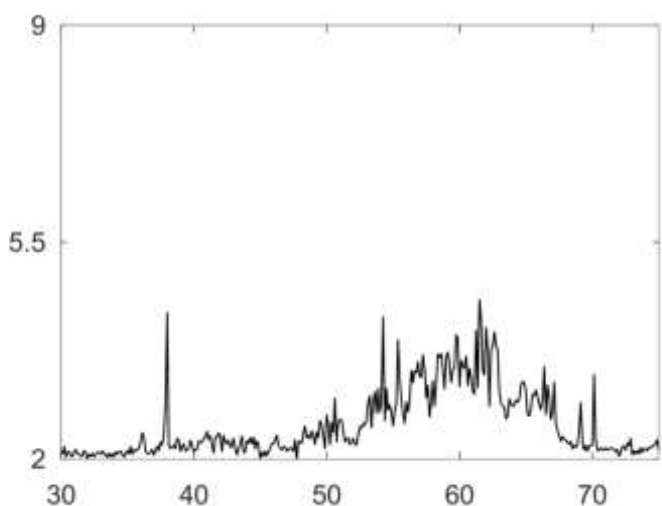 | 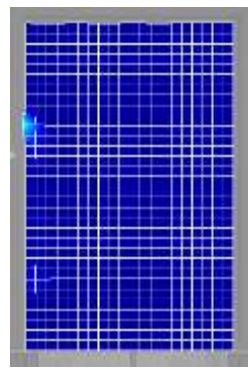 |
| 44 | 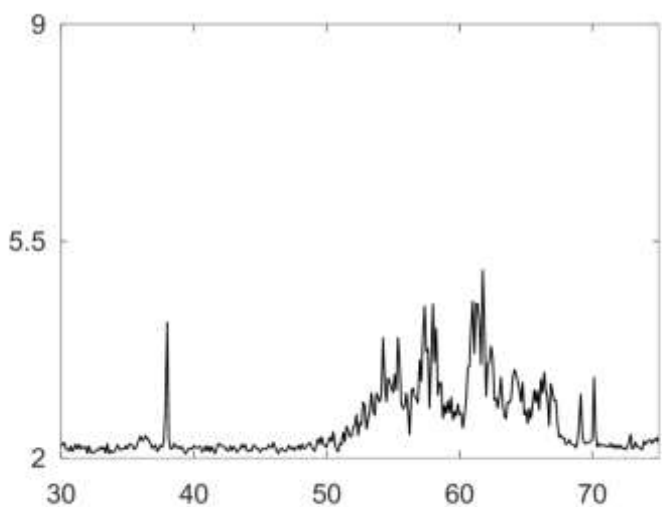 | 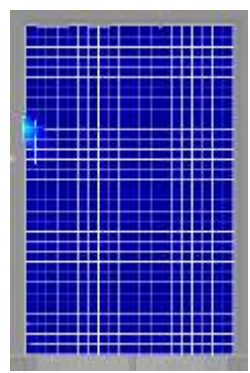 |

| 45 | 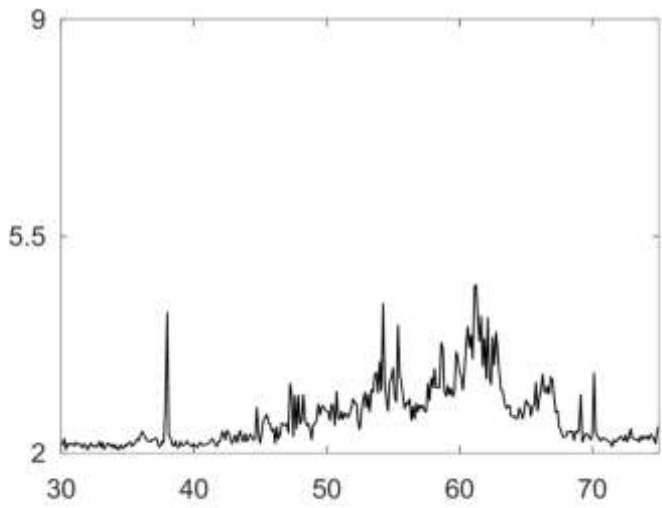 | 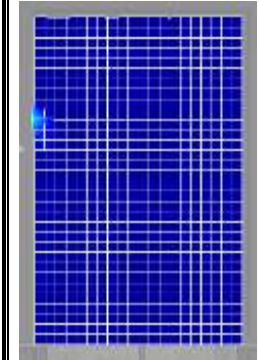 |
| --- | --- | --- |
| 46 | 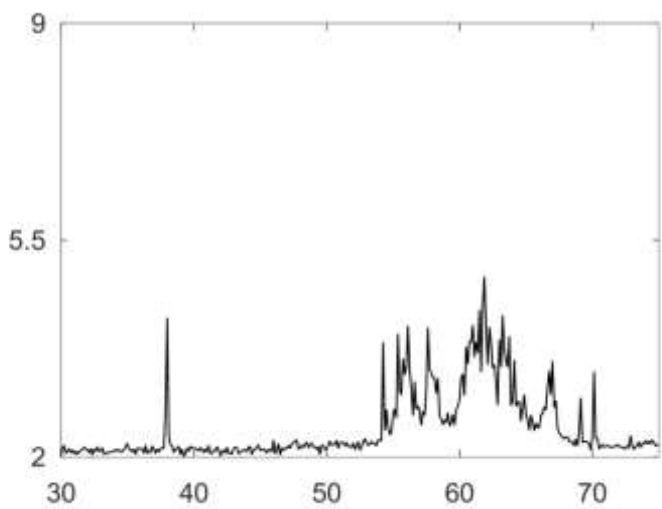 | 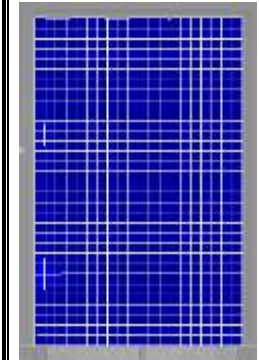 |
| 47 | 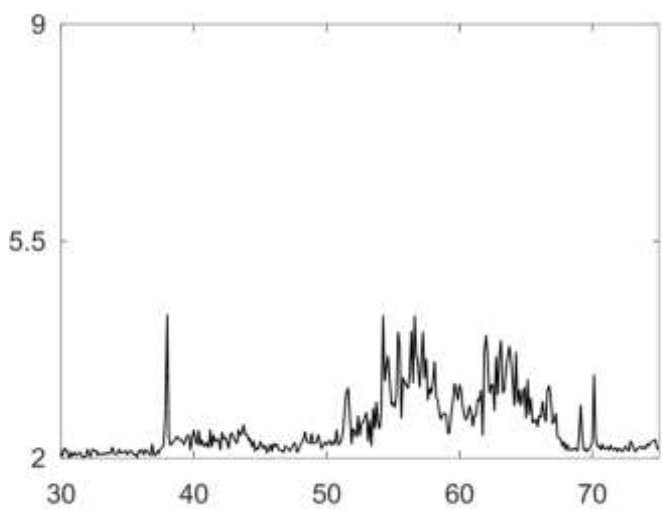 | 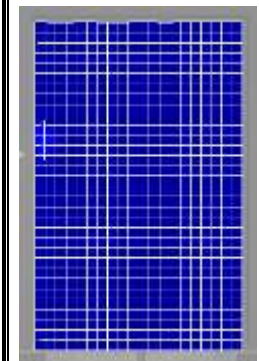 |

| 48 | 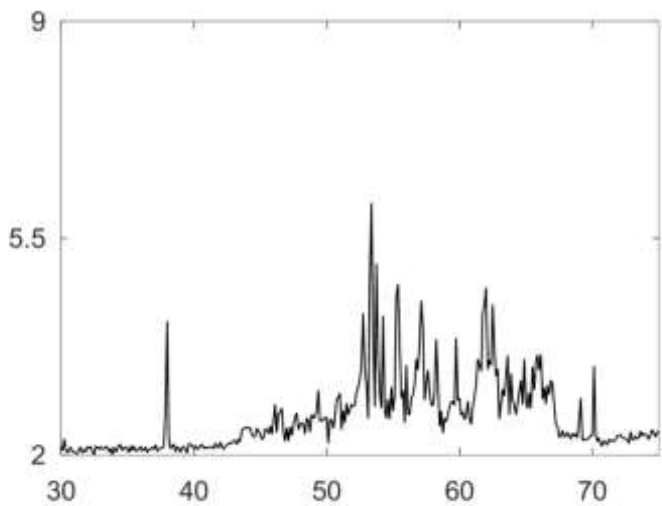 | 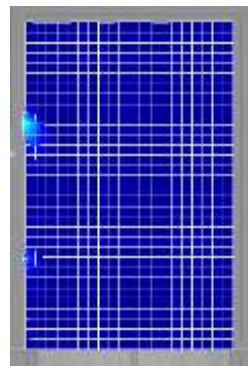 |
| --- | --- | --- |
| 49 | 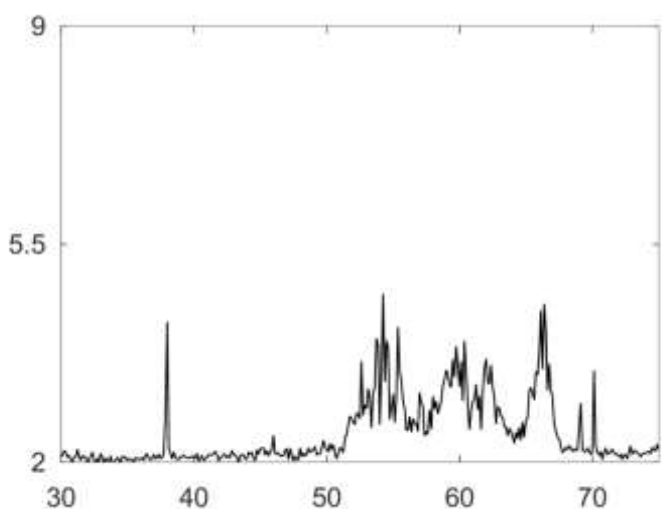 | 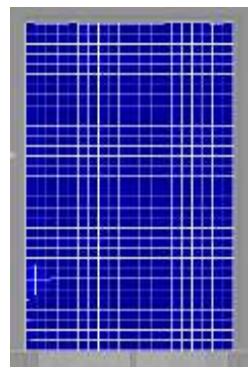 |
| 50 | 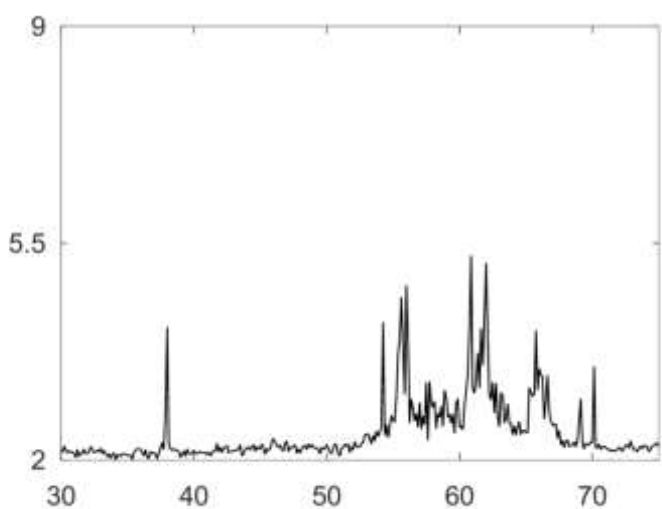 | 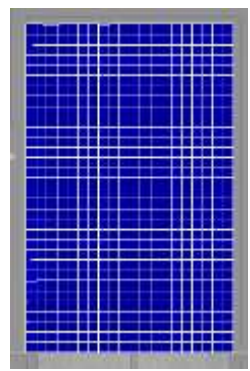 |

| 51 | 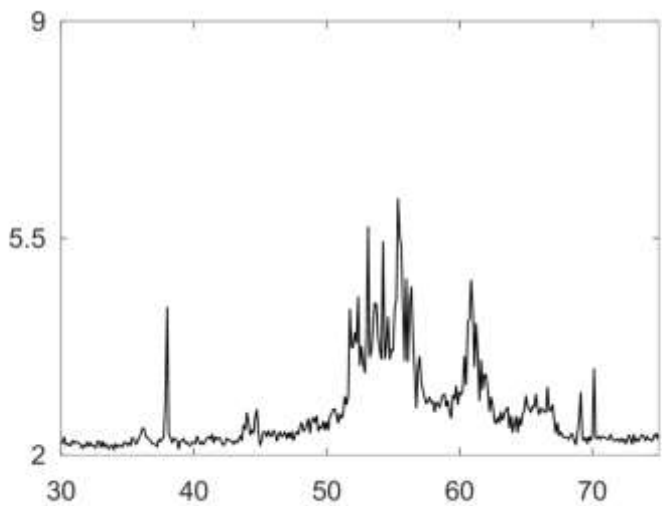 | 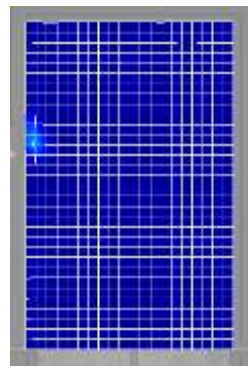 |
| --- | --- | --- |
| 52 | 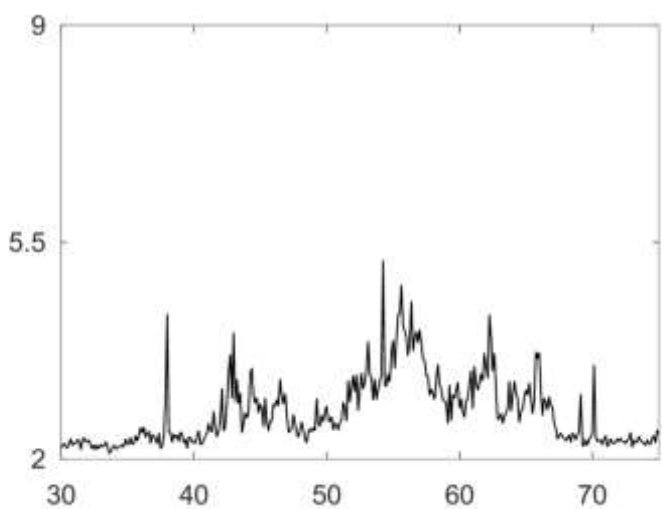 | 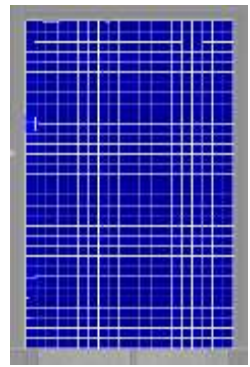 |
| 53 | 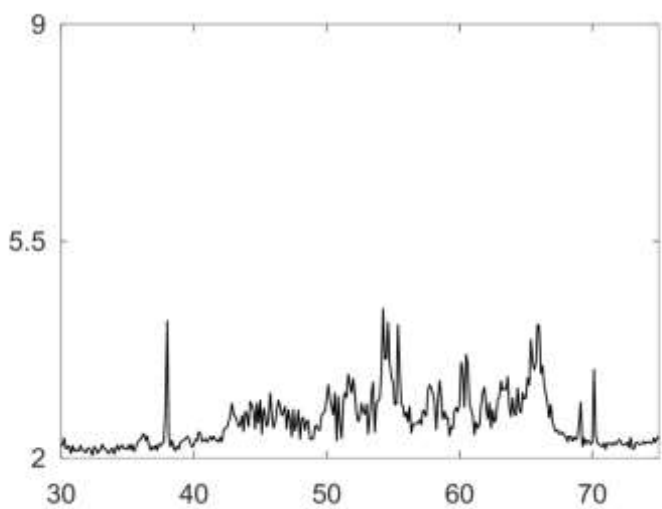 | 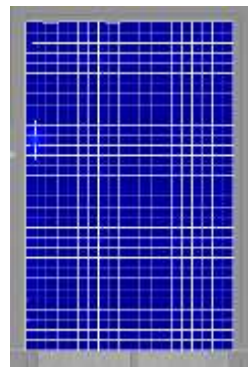 |

| 54 | 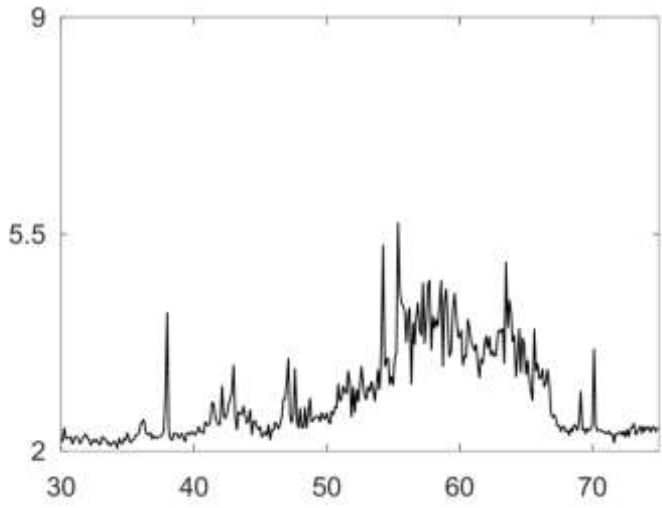 | 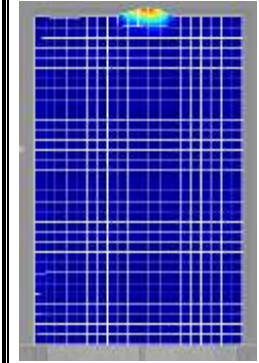 |
|---|---|---|
| 55 | 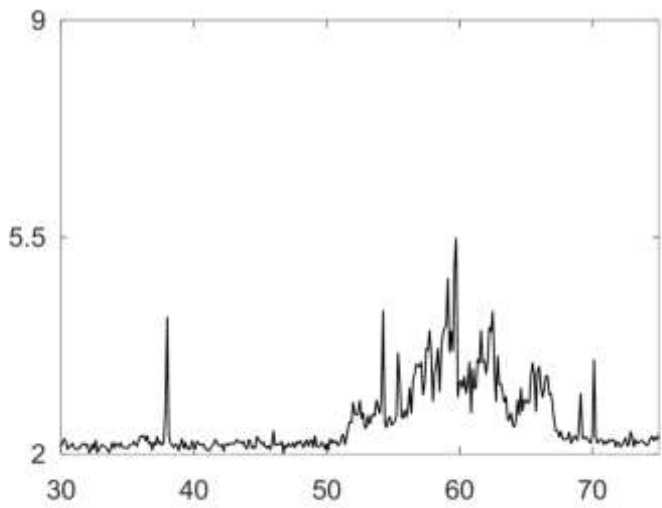 | 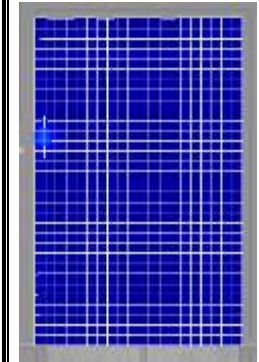 |
| 56 | 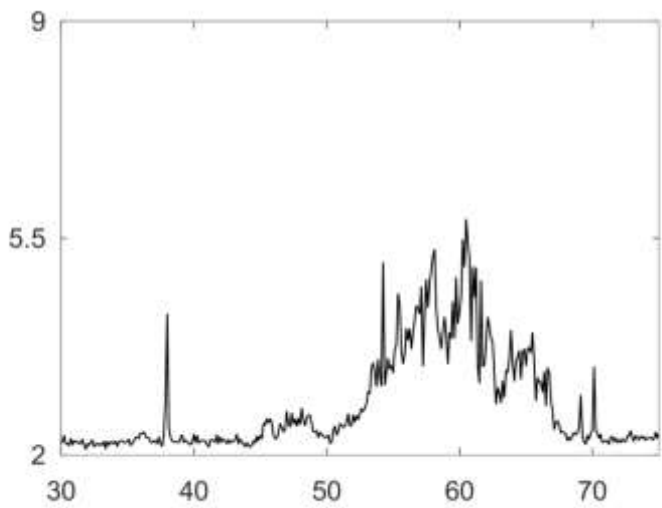 | 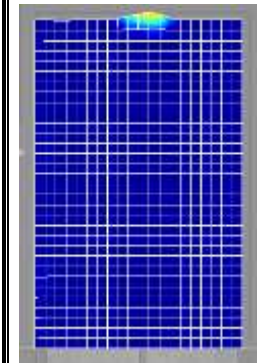 |

| 57 | 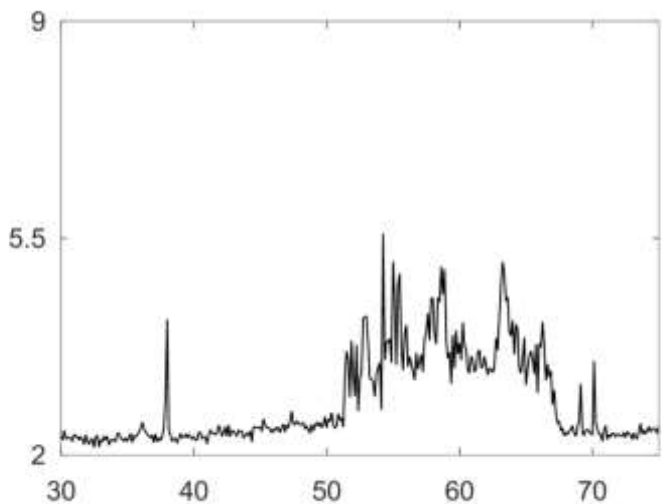 | 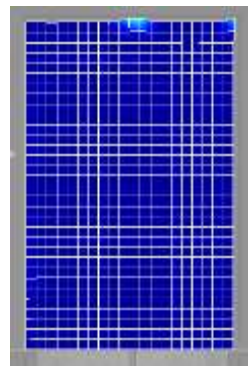 |
| 58 | 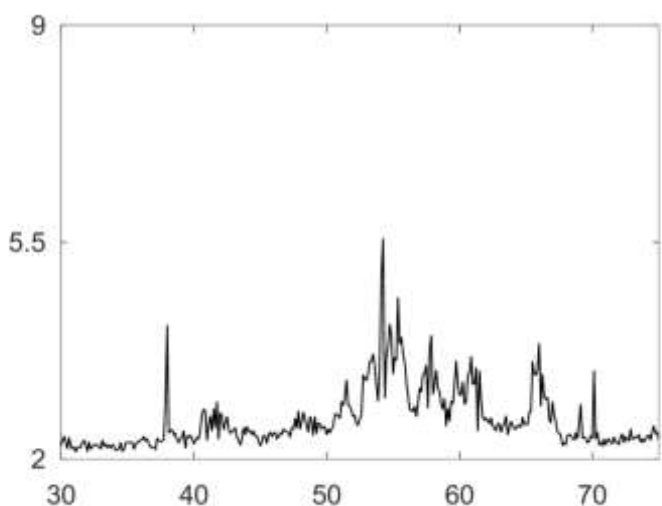 | 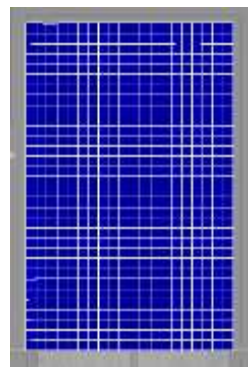 |
| 59 | 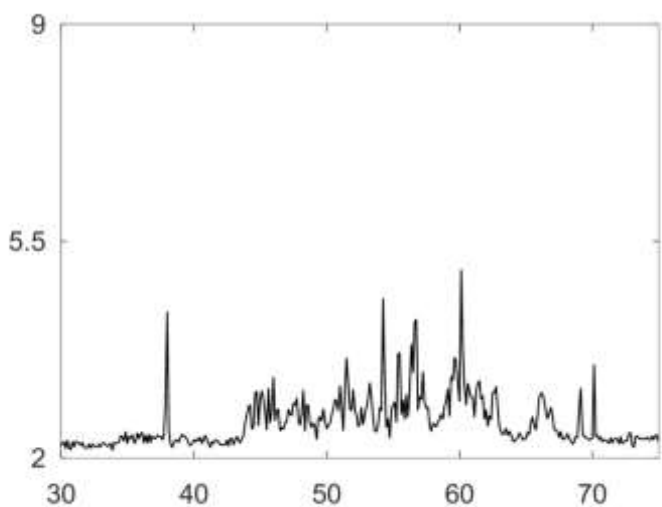 | 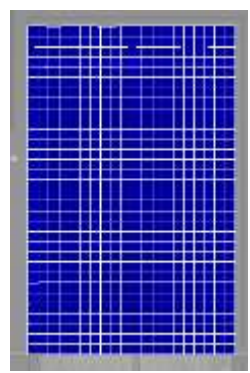 |

| 60 | 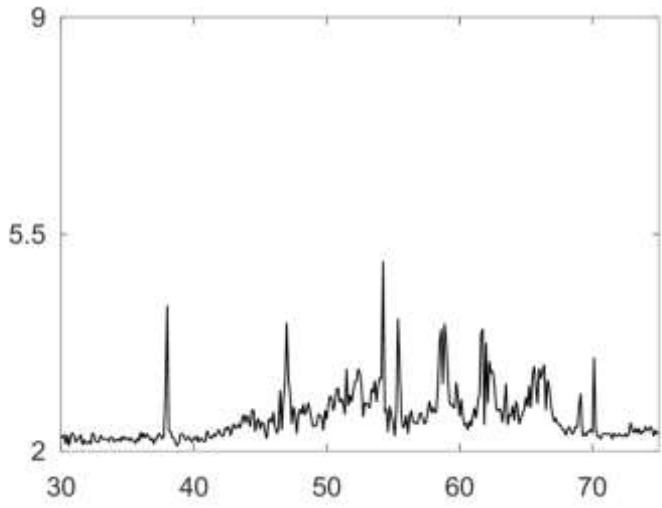 | 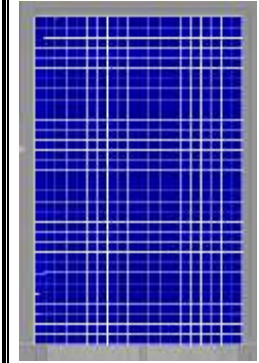 |
|---|---|---|
| 61 | 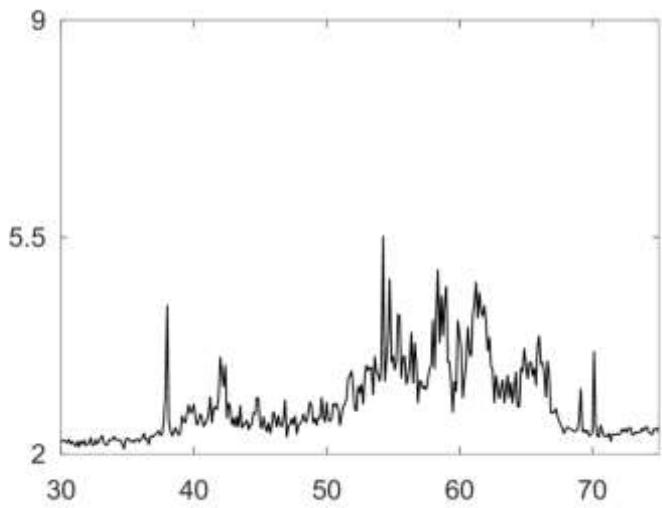 | 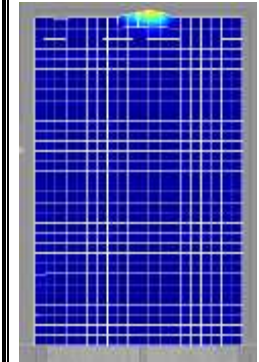 |
| 62 | 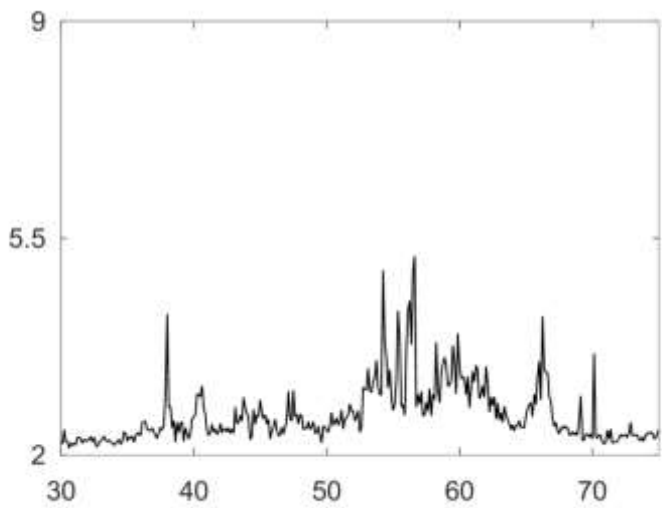 | 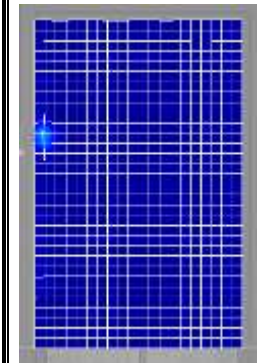 |

| 63 | 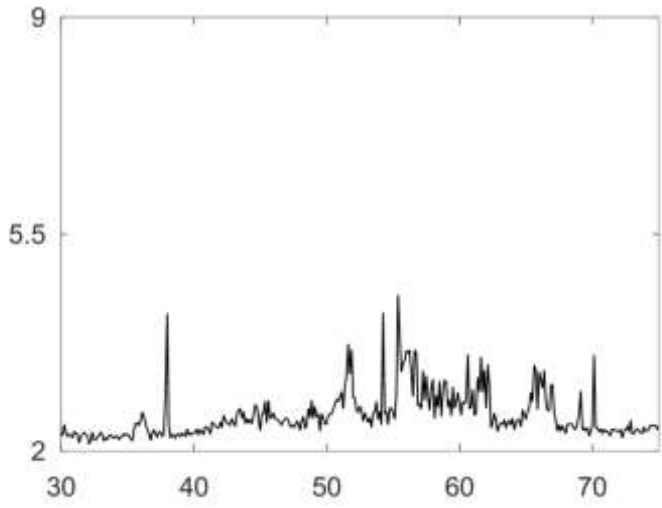 | 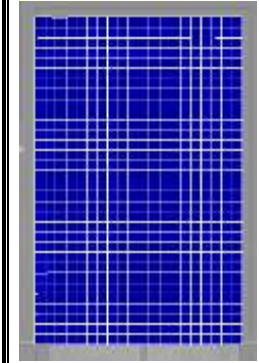 |
|---|---|---|
| 64 | 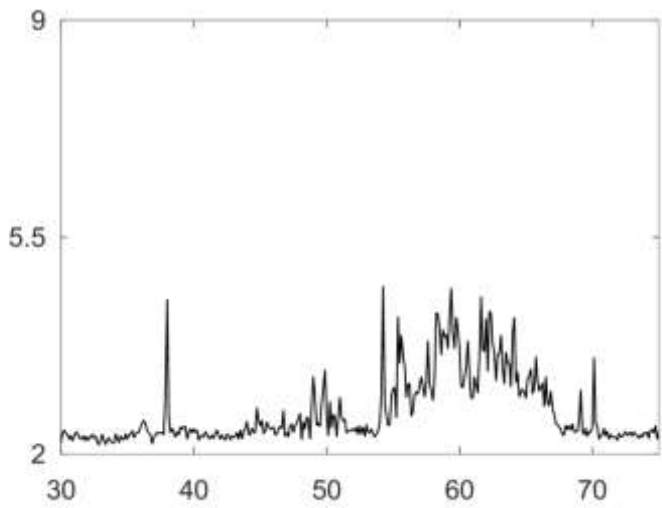 | 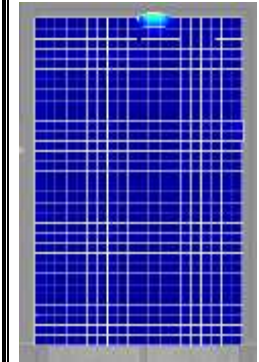 |
| 65 | 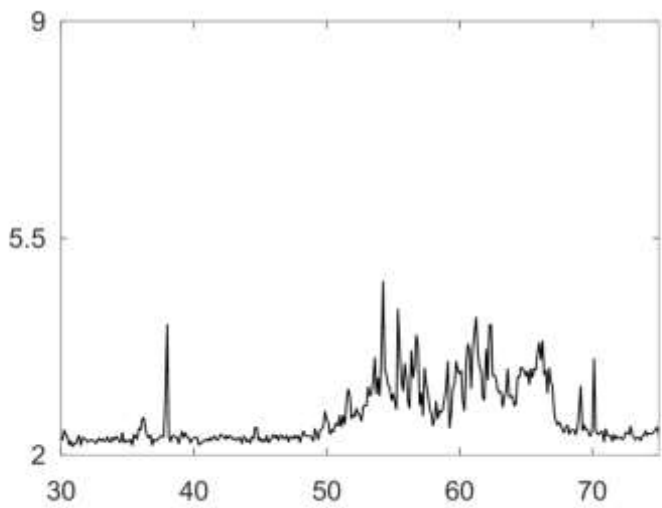 | 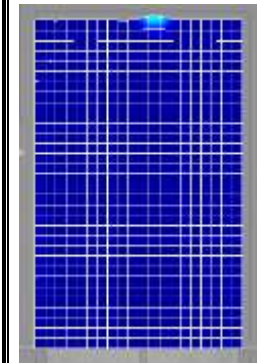 |

Run 4

| Time | Displacement (in pm) vs. Frequency (in kHz) | RMS Surface Displacment $1.2$ nm — $0$ nm |
|---|---|---|
| 1 | 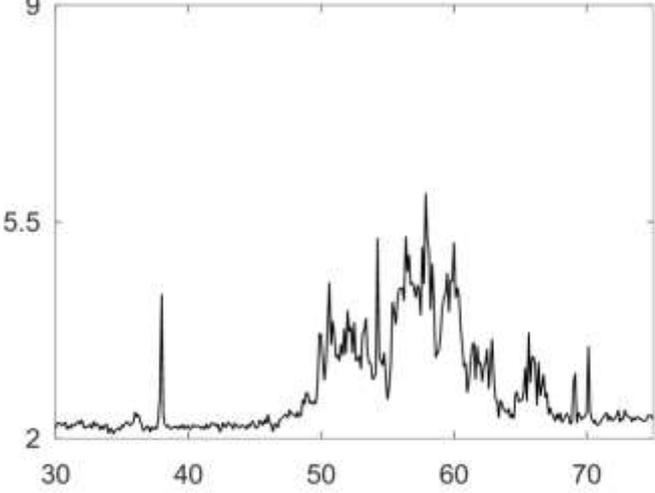 | 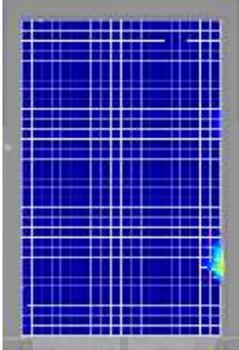 |
| 2 | 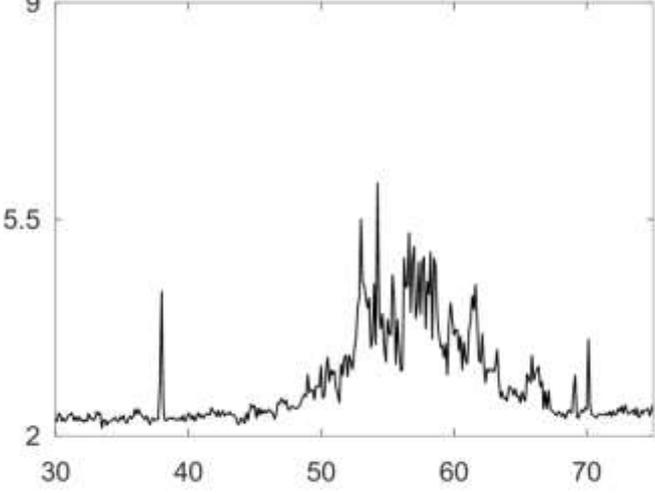 | 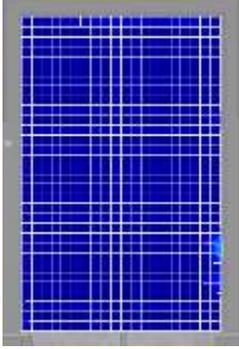 |

| 3 | 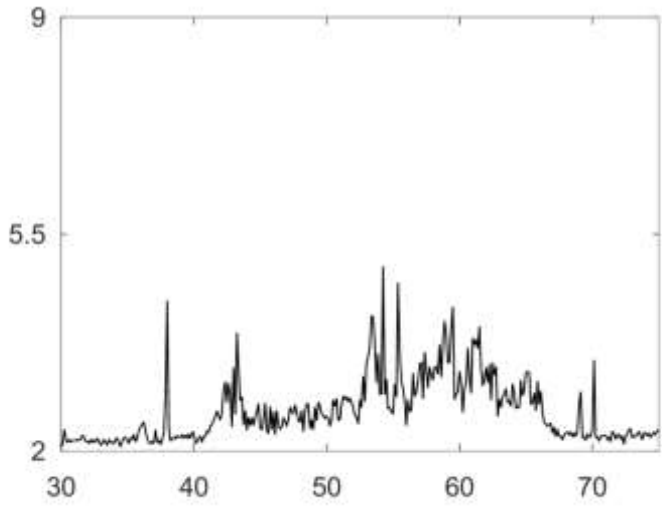 | 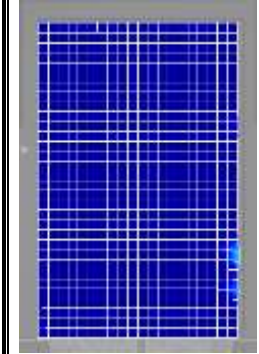 |
|---|---|---|
| 4 | 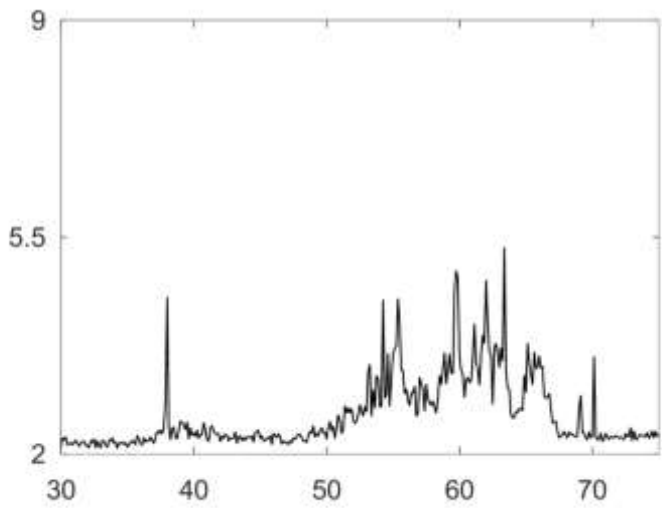 | 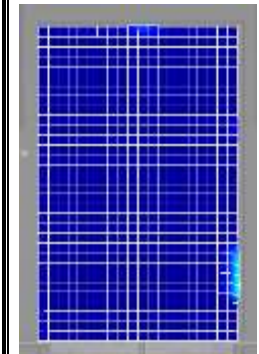 |
| 5 | 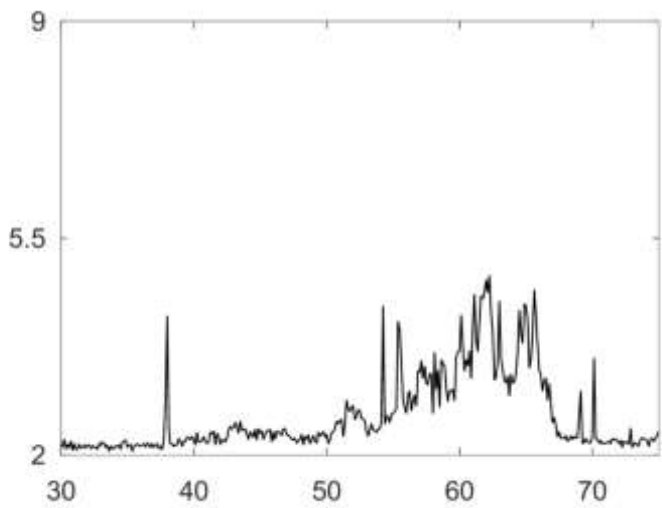 | 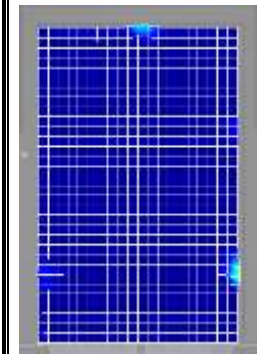 |

| 6 | 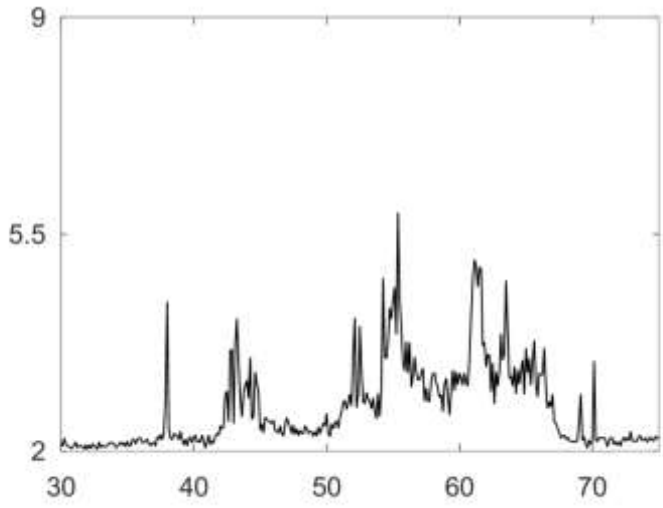 | 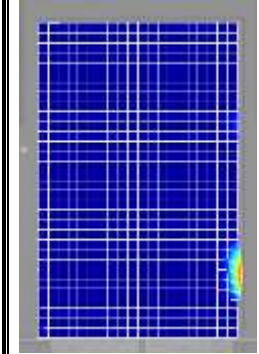 |
|---|---|---|
| 7 | 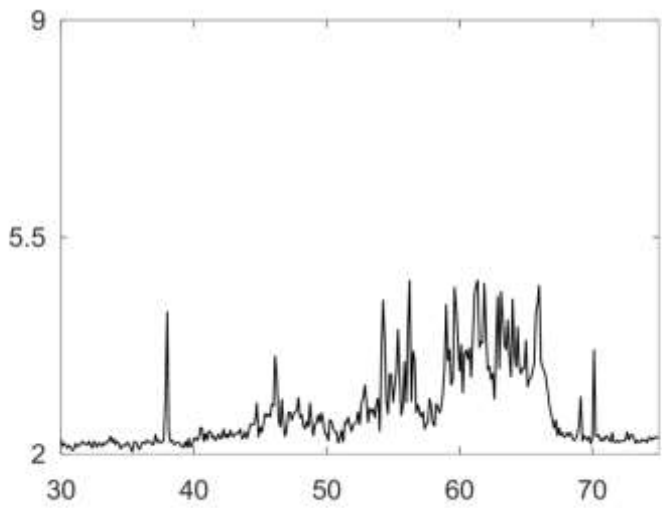 | 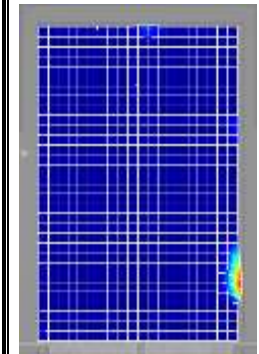 |
| 8 | 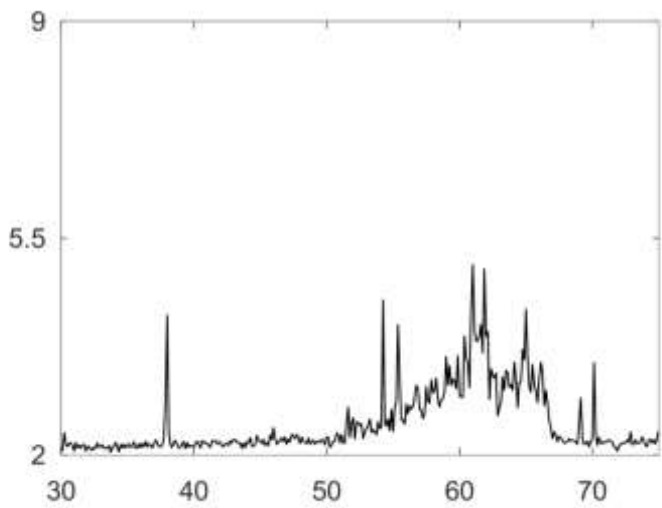 | 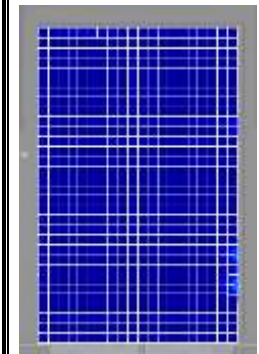 |

| 9 | 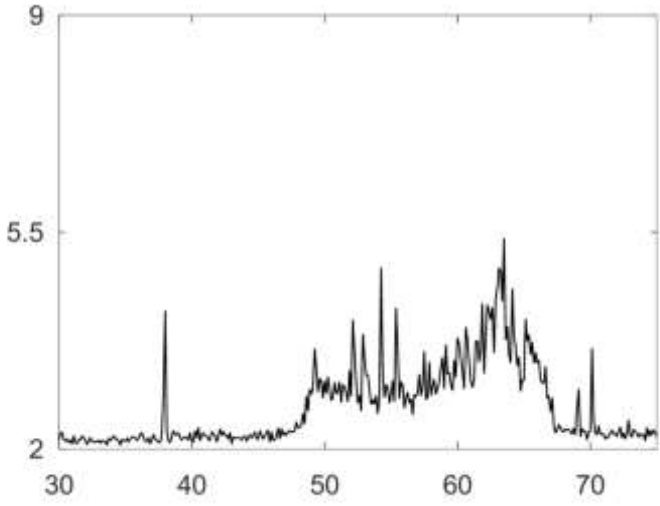 | 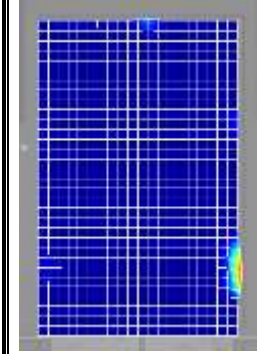 |
|---|---|---|
| 10 | 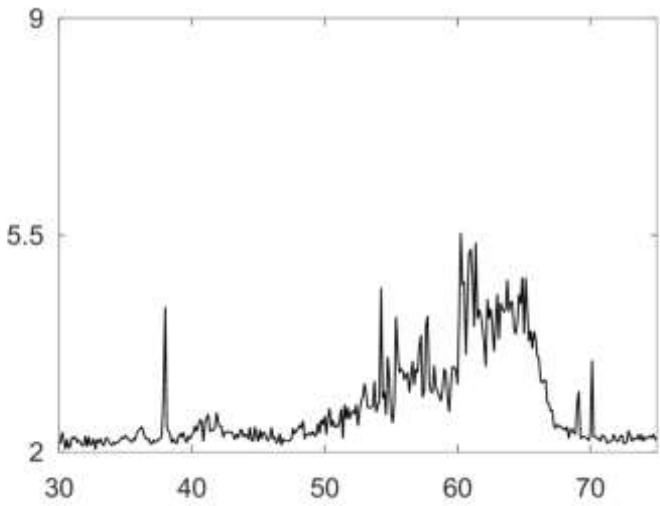 | 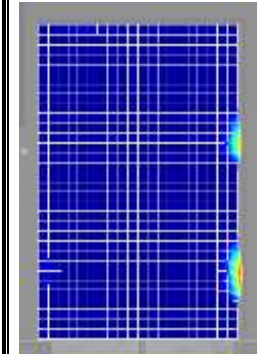 |
| 11 | 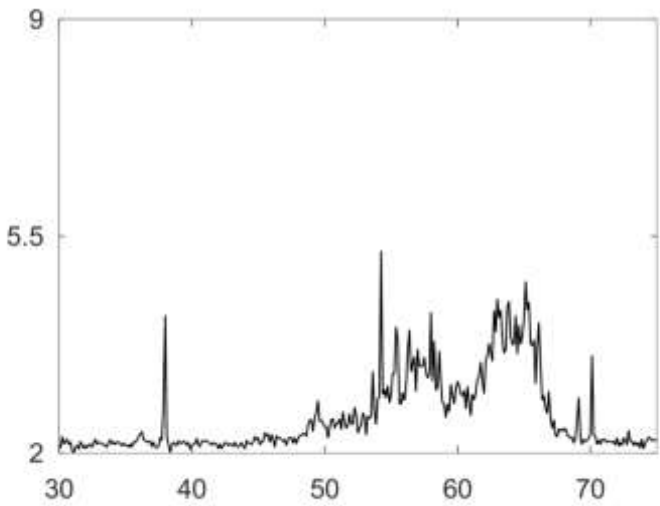 | 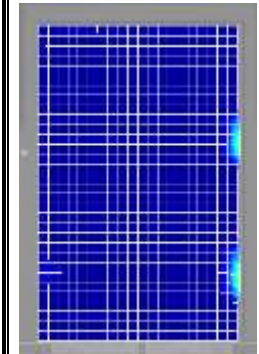 |

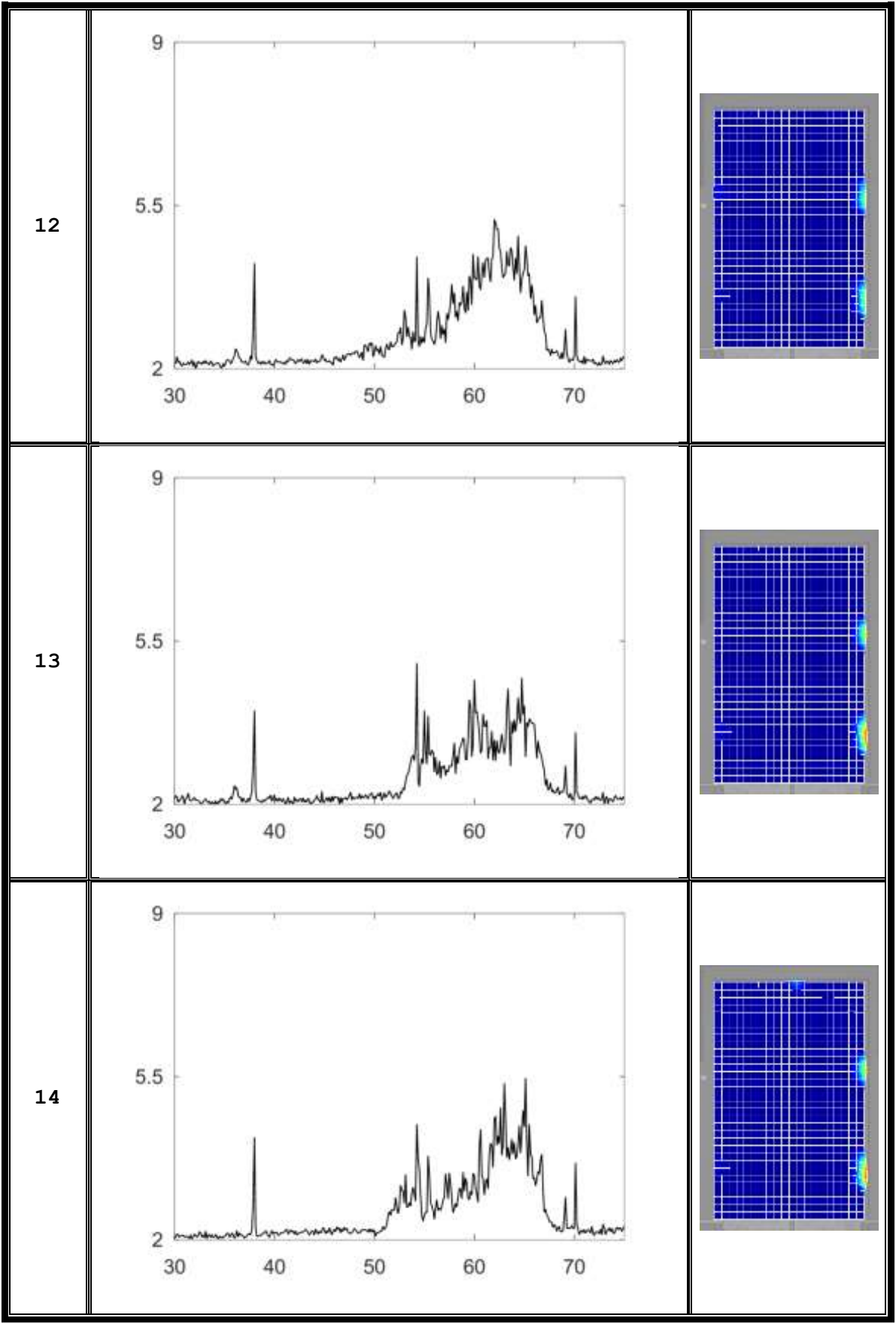

| 15 | 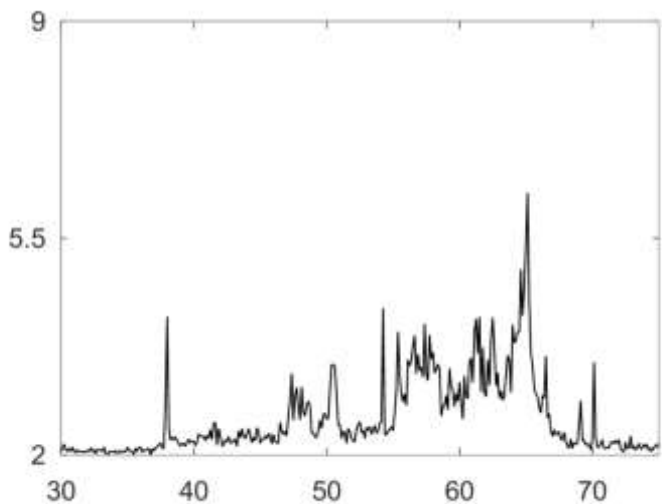 | 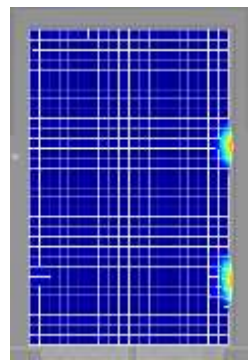 |
|---|---|---|
| 16 | 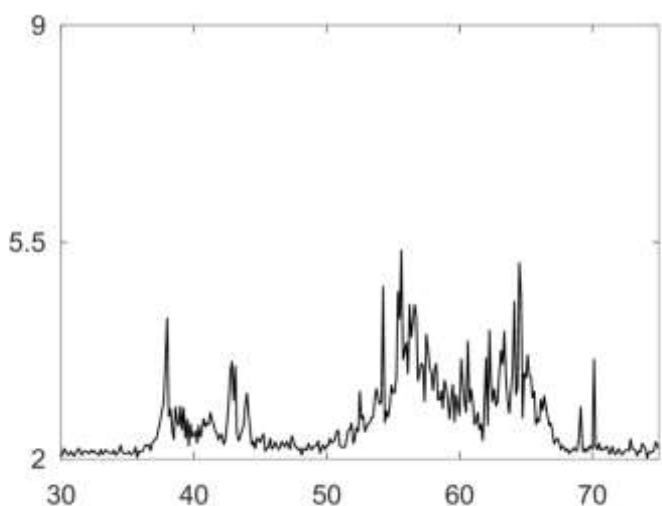 | 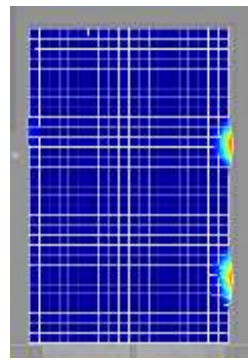 |
| 17 | 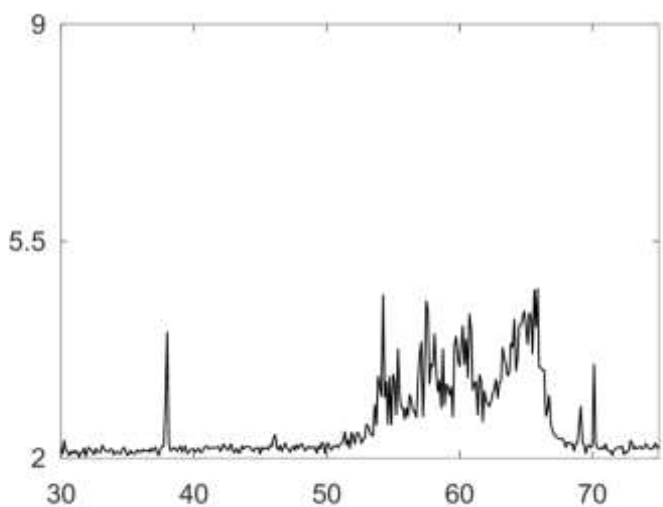 | 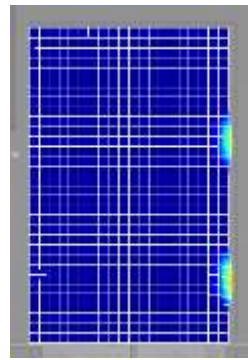 |

| 18 | 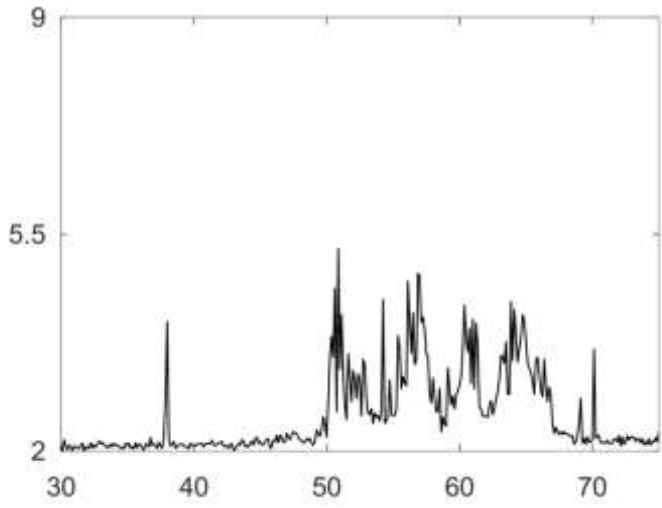 | 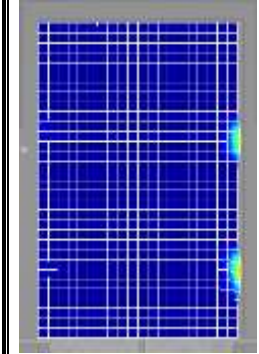 |
| --- | --- | --- |
| 19 | 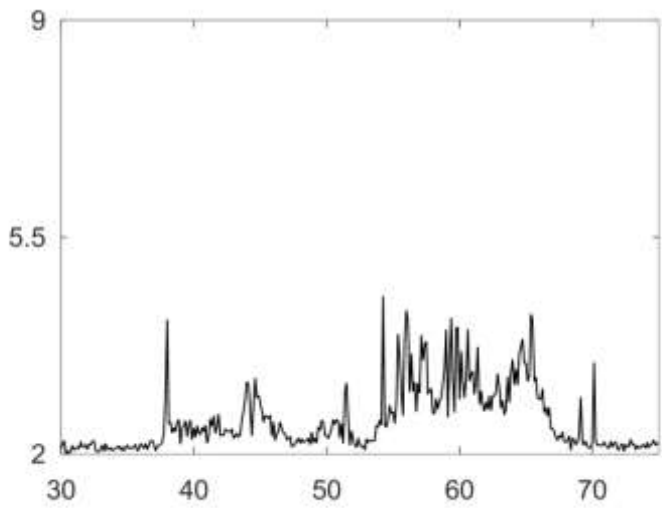 | 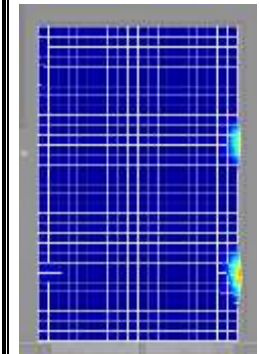 |
| 20 | 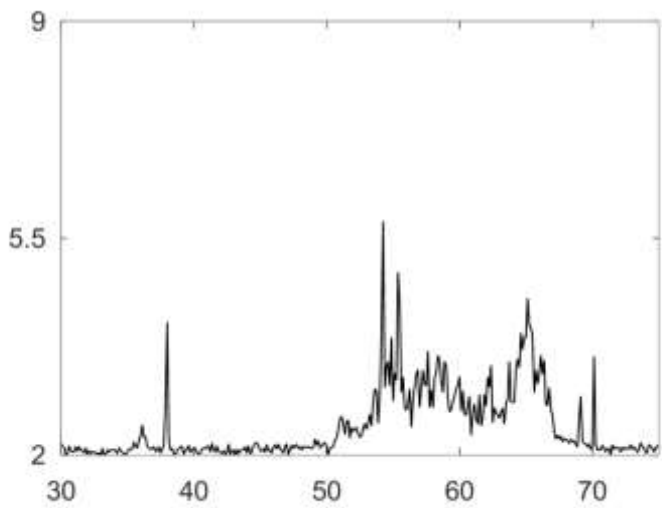 | 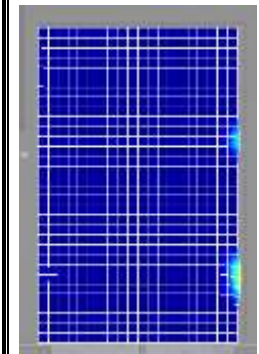 |

| 21 | 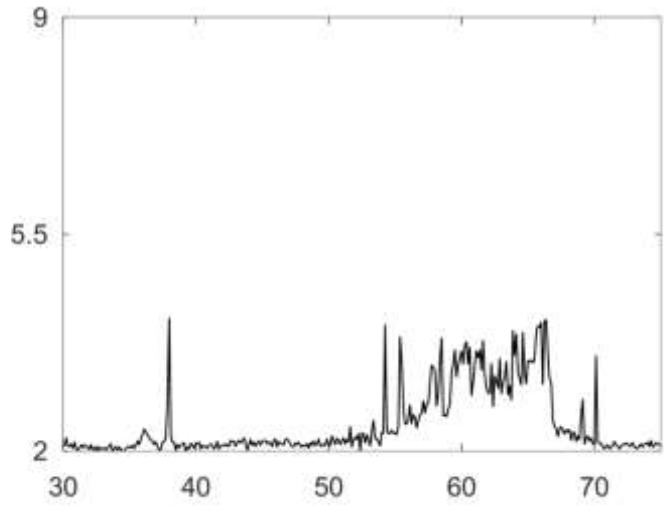 | 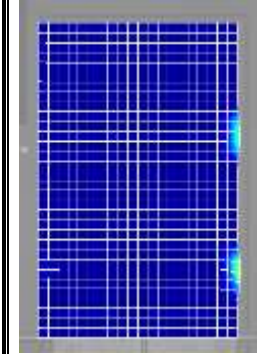 |
| --- | --- | --- |
| 22 | 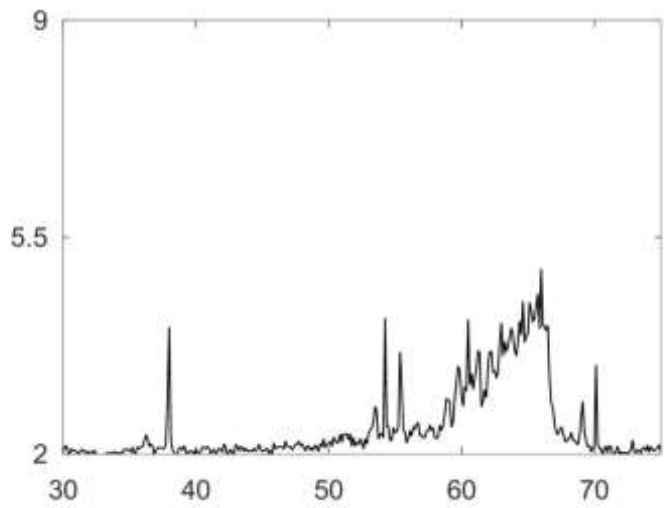 | 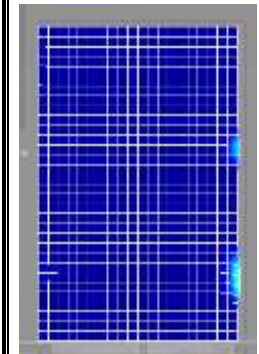 |
| 23 | 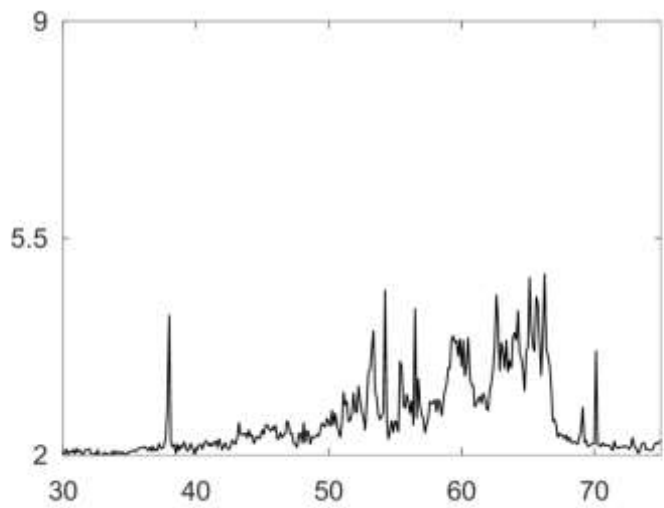 | 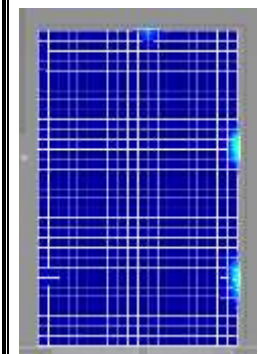 |

| 24 | 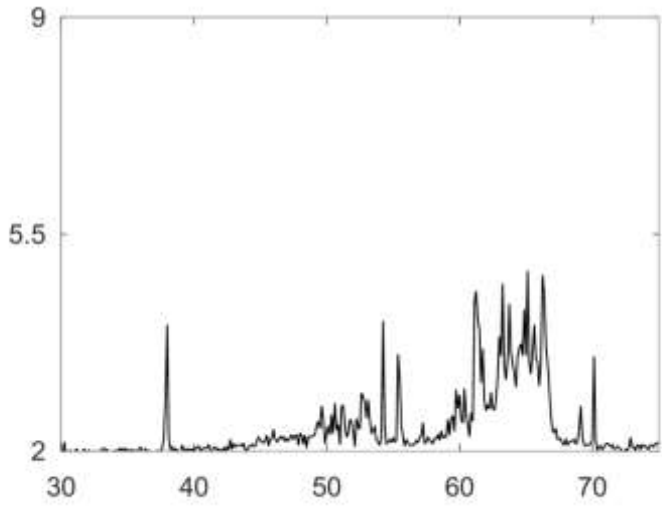 | 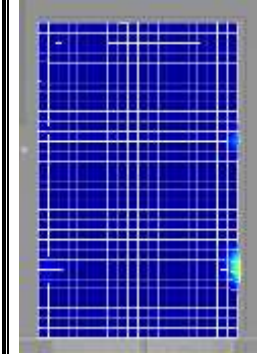 |
| --- | --- | --- |
| 25 | 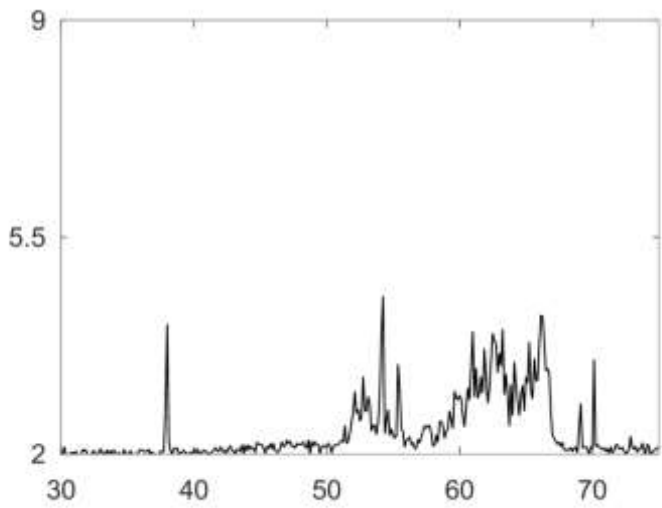 | 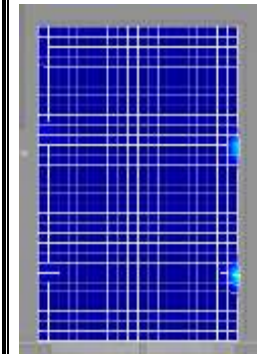 |
| 26 | 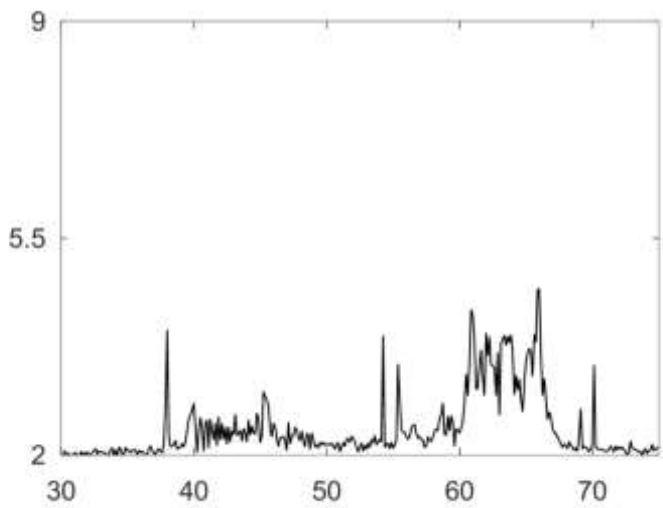 | 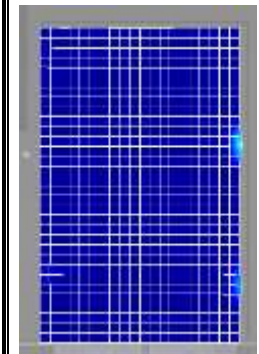 |

| 27 | 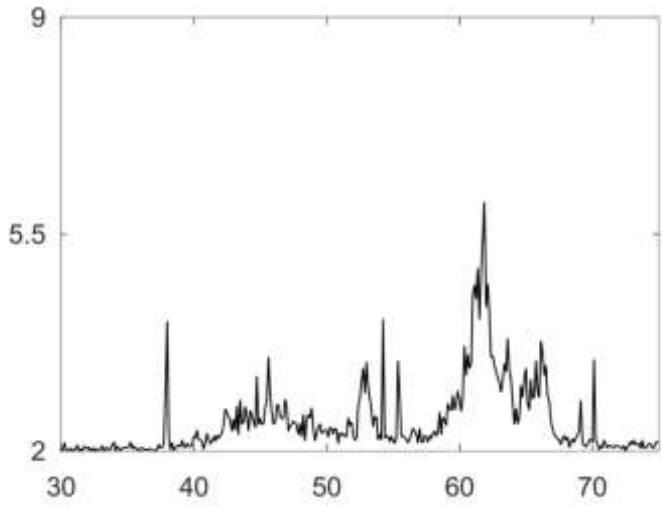 | 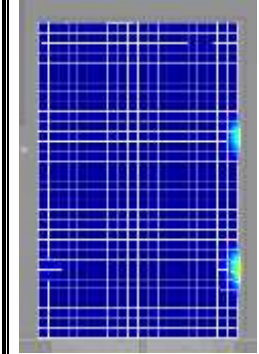 |
| --- | --- | --- |
| 28 | 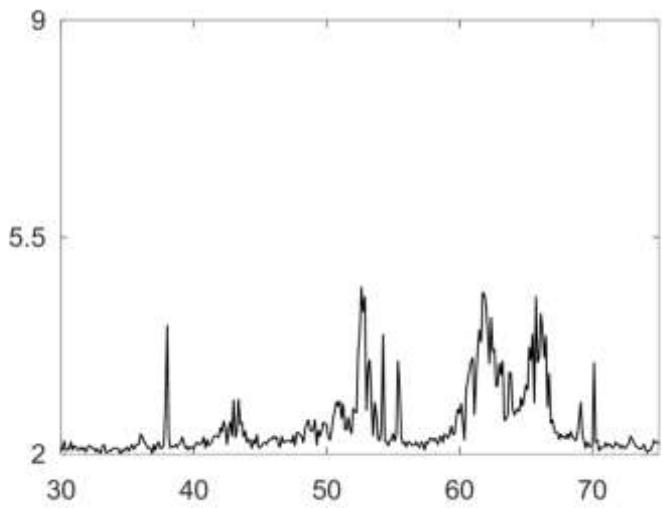 | 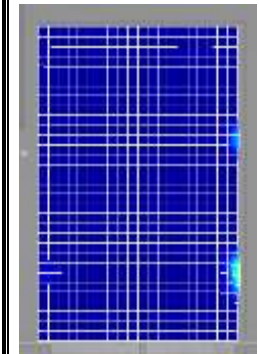 |
| 29 | 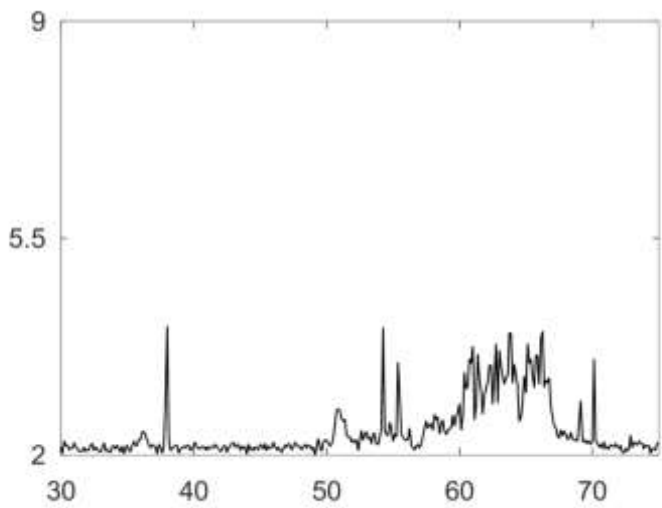 | 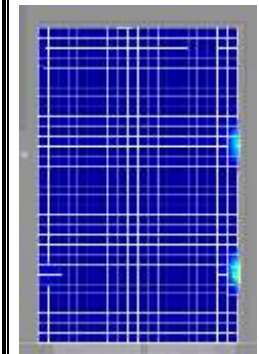 |

| 30 | 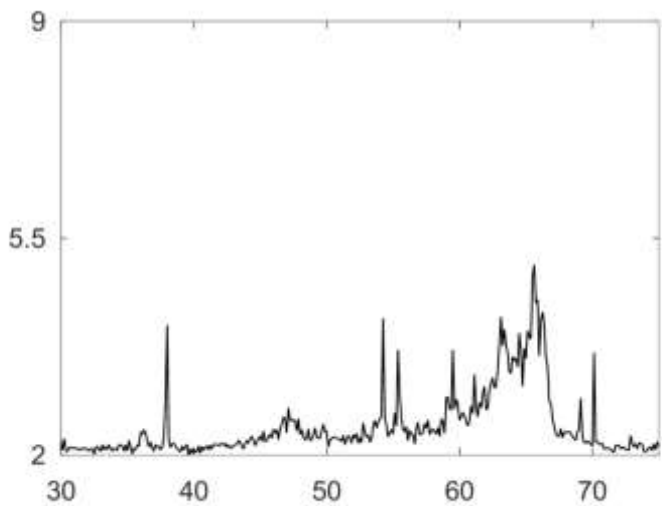 | 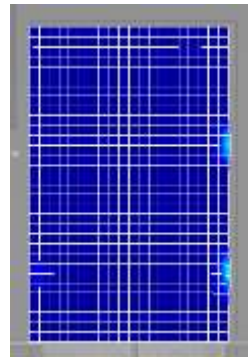 |
| --- | --- | --- |
| 31 | 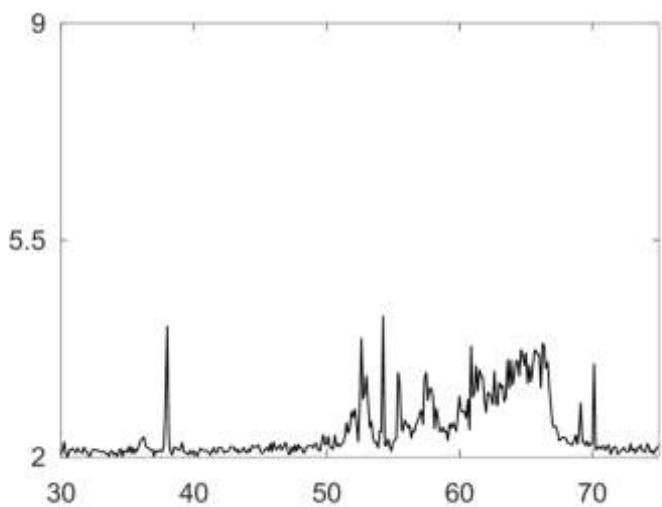 | 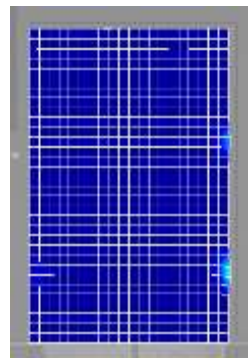 |